\documentclass[11pt]{article}
\usepackage{amsmath,amssymb,amsthm,array,cite,version,chngcntr}
\usepackage{graphicx,graphics,epsfig,subfigure,xcolor,color,float}
\allowdisplaybreaks

\theoremstyle{definition}

\newtheorem{sch}{Scheme}[section]
\theoremstyle{plain}
\newtheorem{thm}{Theorem}[section]
\newtheorem{lem}{Lemma}[section]

\theoremstyle{remark}
\newtheorem{rem}{Remark}[section]
\newtheoremstyle{notes}%
{3pt}
{3pt}
{}
{}
{\itshape\color{red}}
{.}
{.5em}
{}
\theoremstyle{notes}

\numberwithin{equation}{section}
\numberwithin{figure}{section}
\numberwithin{table}{section}
\numberwithin{footnote}{section}
\counterwithout{alg}{section}
\counterwithout{nts}{section}
\counterwithout{sch}{section}

\newcommand{\bena}{\begin{eqnarray}\begin{array}{l}}
\newcommand{\eena}{\end{array}\end{eqnarray}}
\newcommand{\ben}{\begin{eqnarray}}
\newcommand{\een}{\end{eqnarray}}
\newcommand{\bea}{\begin{array}}
\newcommand{\eea}{\end{array}}
\newcommand{\bes}{\begin{subequations}}
\newcommand{\ees}{\end{subequations}}
\newcommand{\bec}{\begin{cases}}
\newcommand{\eec}{\end{cases}}
\newcommand{\bef}{\begin{figure}[H]}
\newcommand{\eef}{\end{figure}}
\newcommand{\bet}{\begin{tikzpicture}}
\newcommand{\eet}{\end{tikzpicture}}
\newcommand{\beq}{\begin{equation}}
\newcommand{\eeq}{\end{equation}}
\newcommand{\bep}{\begin{proof}}
\newcommand{\eep}{\end{proof}}
\def\beg#1\eeg{\begin{align}#1\end{align}}
\def\besl#1\eesl{\begin{subequations}\begin{align}#1\end{align}\end{subequations}}

\newcommand{\parl}[2]{\ensuremath{\frac{\partial #1}{\partial #2}}}
\newcommand{\vparl}[2]{\ensuremath{\frac{\delta #1}{\delta #2}}}

\newcommand{\pr}{\ensuremath{\partial}}

\def\inc(#1){\includegraphics[height=3 cm]{pics/#1}}

\def\ba{\ensuremath{{\bf a}}}

\def\bq{\ensuremath{{\bf q}}}

\def\bx{\ensuremath{{\bf x}}}
\def\bn{\ensuremath{{\bf n}}}

\def\bv{\ensuremath{{\bf v}}}
\def\bM{\ensuremath{{\bf M}}}
\def\bI{\ensuremath{{\bf I}}}
\def\bD{\ensuremath{{\bf D}}}

\begin{document}
\title{Derivation and Efficient Entropy-Production-Rate-Preserving Algorithms for a Thermodynamically Consistent Nonisothermal Model of Incompressible Binary Fluids}


%


\author{
{Shouwen Sun}\footnote{ssw@sqnu.edu.cn, School of Mathematics and Statistics, Shangqiu Normal University, Shangqiu 476000, China.},
{Liangliang Lei}\footnote{leiliangliang@sqnu.edu.cn, School of Mathematics and Statistics, Shangqiu Normal University, Shangqiu 476000, China.},
{Qi Wang}\footnote{qwang@math.sc.edu, Department of Mathematics, University of South Carolina, Columbia, SC 29028, USA.}
}
\date{}
\maketitle
\begin{abstract}

We present a new hydrodynamic model for incompressible binary fluids that is thermodynamically consistent and non-isothermal. This model follows the generalized Onsager principle and Boussinesq approximation and preserves the volume of each fluid phase and the positive entropy production rate under consistent boundary conditions. To solve the governing partial differential equations in the model numerically, we  design a set of second-order, volume and entropy-production-rate preserving numerical algorithms.
Using an efficient adaptive time-stepping strategy, we conduct several numerical simulations. These simulations accurately simulate the Rayleigh-B\'{e}nard convection in binary fluids and the interfacial dynamics between two immiscible fluids under the effects of the temperature gradient, gravity, and interfacial forces. Our numerical results show roll cell patterns and thermally induced mixing of binary fluids in a rectangular computational domain with a set of specific boundary conditions: a zero velocity boundary condition all around, the insulation boundary condition at the lateral boundaries, and an imposed temperature difference vertically.
We also perform long-time simulations of interfacial dynamics, demonstrating the robustness of our new structure-preserving schemes and reveal interesting fluid mixing phenomena.

\vskip 12 pt

\noindent {\bf Keywords}:{ Nonisothermal binary incompressible viscous fluid flows; phase field; thermodynamical consistency; Rayleigh-B\'{e}nard convection; interfacial dynamics; structure-preserving; adaptive time-stepping.}

\end{abstract}

\section{Introduction}

\noindent \indent Multi-component material systems are ubiquitous in nature and industrial applications.
Modeling and simulating the hydrodynamics of such systems can be achieved through various methods, such as the traditional sharp-interface, volume-of-fluids, front-track methods, and the phase field method \cite{MORTON1996TWO, Du1991Numerical, Du2004A, Lowengrub1998Quasi, Teigen2011A, Yang2017Numerical, Yue2004A, Zhao2016Numerical,Boyer1999Mathematical,Cahn1958Free}. The latter has gained popularity in recent years due to its simplicity and effectiveness in describing multi-phase fluid flows. While most works on phase field models have focused on isothermal conditions, it is essential to consider nonisothermal conditions to faithfully describe real fluid systems.


The Rayleigh-B\'{e}nard convection is a common phenomenon for fluids  under non-isothermal conditions. The Rayleigh-B\'{e}nard convection takes place in a fluid system driven by a temperature gradient and gravity, which has been extensively documented in a single-phase fluid system. The classical example of the Rayleigh-B\'{e}nard convection is shown in the fluid confined between two parallel plates and heated from below, which is one of the well-known non-equilibrium, nonisothermal, hydrodynamic systems. Most works on Rayleigh-B\'{e}nard convection have only considered a single phase fluid however, where thermally induced hydrodynamic effects are primarily the results of gravity and the temperature gradient induced buoyancy effect \cite{favier_purseed_duchemin_2019,wen_goluskin_doering_2022,TWatanable_2004}. In multiphasic fluid systems, interactions among various fluid components introduce additional complications to the hydrodynamics of the systems.

In binary fluids with immiscible fluid components, the interfacial force between the two immiscible fluid phases adds an additional competing factor to the hydrodynamics, making the nonisothermal multiphasic fluid system more interesting. This motivates the current study. In 2015,  a non-isothermal, binary, incompressible, viscous fluid flow was proposed by Guo and Lin in \cite{Guo2015A}. A general framework for deriving transport equations involving heat flows through the energetic variational approach was proposed in  \cite{Pei2017Non,de2019non}. The author of the paper \cite{QingmingChang} showed a thermal lattice Boltzmann model for two-phase fluid flow with a double population distribution function. We took into account  the thermal-hydrodynamic coupling for an non-isothermal, incompressible binary material system to give a general thermodynamically consistent, nonisothermal, hydrodynamic model for binary viscous fluid flows using a phase-field approach recently \cite{SunShouwen}. Using a thermodynamically consistent, non-isothermal, hydrodynamic model of incompressible binary fluids to study Rayleigh-B\'{e}nard convection remains an active and less explored research area today.

To ensure the accuracy and efficiency of numerical approximations for thermodynamically consistent models, it is important to preserve their inherent structures and properties. Various numerical methods have been developed over the years to achieve this goal. Recently, several methods, including energy quadratization (EQ), scalar auxiliary variable (SAV), Lagrange multiplier SAV, and supplementary variable method (SVM), have been proposed to simplify the development of energy-stable and energy-dissipation-rate-preserving schemes, particularly to mitigate the nonlinearity in the chemical potential in phase-field models \cite{yang2018linear,Yang2017Numerical2,zhao2017numerical,KeWuandJieShen2022,ShenjieYangJiang2018,
TangTao2022,Yuhaijun2017,Chengqing2020Lagrange,DBLP:journals/moc/LiQW21}.
Numerous papers have been published on preserving the energy dissipation property, the energy dissipation rate or the entropy production rate for thermodynamically consistent models with adiabatic boundary conditions \cite{Gong2018Second,SunShouwen,Lijun2019Second}. However, for non-adiabatic boundaries in nonisothermal problems,  well-developed and fully implemented structure-preserving numerical approximations that preserve the entropy production rate for thermodynamically consistent non-isothermal hydrodynamical models are missing.


In this paper, we first derive a thermodynamically consistent, non-isothermal hydrodynamic model for incompressible binary fluids under the influence of gravity, simplifying our previous general model, by applying the Boussinesq approximation \cite{SunShouwen}. The resulting model consists of the coupled  Cahn-Hilliard equation, Navier-Stokes equation augmented by the buoyancy force and energy equation accounting for the thermal-hydrodynamic coupling and yields a positive entropy production rate. We remark that the interfacial force in this model is rooted in the Ericksen stress like in many hydrodynamical models developed previously \cite{Leslie1979,Wang&F&Z2004,Li&W2014,Lijun2019Second}. Thus, its contribution to the change of internal energy is through the "interfacial" heating analogous to the viscous heating \cite{DeGroot}.  Then, we present a modified model that allows weak compressibility due to the nonisothermal pressure effect. This modified model lays the foundation for us to design  entropy-production rate preserving numerical schemes using the projection strategy. For the simplified thermodynamically consistent model, we then devise a set of second-order schemes that preserve structure by utilizing the entropy quadratization (EQ) method in conjunction with a finite difference method on spatially staggered grids to address nonlinearity in chemical potential. We prove rigorously that  the proposed schemes preserve the  entropy-production-rate and fluid volume of each phase  in both the temporally semi-discrete system and the fully discrete system under thermodynamically consistent boundary conditions. With one of the developed,   second-order, fully discrete schemes,  we simulate the Rayleigh-B\'{e}nard convection of two-layered, superimposed viscous fluids in a rectangular 2-D domain with  adiabatic boundary conditions laterally and imposed Dirichlet boundary conditions of the temperature and velocity and no-flux boundary condition for the phase field variable vertically, where the binary immiscible fluid system is subject to a competing temperature gradient, interfacial force, and gravity effect. Our numerical results agree with the published results in resolving the roll cells at the onset of Rayleigh-B\'{e}nard convection and demonstrate quite significant fluid mixing in the long-time simulation.   Finally,  we simulate the dynamics of  drops in an immiscible binary viscous fluid system in the presence of temperature-gradient, gravity and interfacial forces to show how large scale roll cells form in the binary fluid system and how they interact with the interfacial dynamics of the two-phase fluid.

The second-order entropy-production-rate-preserving scheme is implemented in time with an adaptive time-step strategy to efficiently unleash the power of the EQ method in the numerical approximation \cite{zhang_qiao_2012}. This allows computations to continue with acceptable error tolerance for up to a very long time. We use the code to show the roll cell formation at the fluid interface and carry out the computation for a long time to explore fluid mixing after a long time. The code can be readily applied to other applications involving nonisothermal binary fluid flows subject to other thermodynamically consistent boundary conditions. However, in any other applications, the thermodynamical consistency of the fully discrete scheme needs to be established case-by-case depending on the spatial discretization and the boundary conditions involved.

The paper is organized as follows. In \S 2, we formulate the mathematical formulation of the simplified nonisothermal hydrodynamic model for incompressible binary fluids,  prove its thermodynamical consistency, and then extend it to a modified weakly compressible model. In \S3,  the EQ method is applied to derive semi-discrete numerical schemes in time. Subsequently,  the spatial discretization based on a finite difference method on staggered grids is carried out on semi-discrete schemes  to yield fully discrete schemes. Finally, we prove that the fully discrete numerical schemes preserve the properties of the entropy-production rate and the volume of each fluid phase under the given boundary conditions. In \S 4, we conduct two numerical simulations  to show the Rayleigh-B\'{e}nard convection phenomenon and simulate drop dynamics in a binary immiscible viscous fluid with an imposed temperature gradient, respectively. We give  a concluding remark in \S 5.

\section{Mathematical Formulation}

\noindent \indent To simulate the Rayleigh-B\'{e}nard convection phenomenon in two-phase fluid flows, we present a simplified thermodynamically consistent, non-isothermal hydrodynamic phase field model of incompressible binary viscous fluids. The model consists of a Cahn-Hilliard equation for the phase field,  the coupled Navier-Stokes equation for the velocity field and the energy conservation equation for the temperature. We call it the simplified Nonisothermal Cahn-Hilliard-Navier-Stokes  equation system. This simplified model is derived following the Onsager principle, analogous to the more general one in \cite{SunShouwen}, by postulating the fluid density is a prescribed function of the temperature and applying the Buossinesq approximation to the momentum balance equation.

\subsection{Balance equations}

\noindent \indent We consider that the mixture of binary fluid is composed of two viscous fluid components A and B, where $\phi$ represents the volume fraction of fluid A and $1-\phi$ represents that of  fluid B. \ben
\rho_1(T,\phi)=\hat{\rho_1}(T)\phi, \quad \rho_2(T,\phi)=\hat{\rho_2}(T)(1-\phi)
\een
 are their respective densities in the binary fluid, where $T$ is the absolute temperature, $\hat{\rho_1}(T)$ and $\hat{\rho_2}(T)$ are the intrinsic densities for fluid A and B at temperature $T$, respectively. The total mass density of the fluid mixture is defined as
 \ben
 \rho(T,\phi)=\rho_1(T,\phi)+\rho_2(T,\phi)=\hat{\rho_1}(T)\phi+\hat{\rho_2}(T)(1-\phi).
  \een
If the two constituents are of equal mass densities, i.e. $\hat{\rho_1}(T)=\hat{\rho_2}(T)=\rho_0(T)$, where $\rho_0$ is the fluid density under the reference temperature, independent of the phase variable. Then, $\rho=\rho_0(T)$. This is an approximation to the cases where the intrinsic density of the two fluid components are very close. We adopt this assumption in this paper.

We denote  $\bv$ the mass average velocity,  $e$  the internal energy per unit volume,  $s$ the entropy per unit volume,
$\Omega$  the material domain. Then, the total entropy of the fluid system is expressed as
\bena
S(e,\phi,\nabla\phi)=\int_\Omega s(e,\phi,\nabla \phi)dx=\int_\Omega [s_0(e,\phi)+s_1(\nabla \phi)]d \bx,
\label{system entropy}
\eena
where $s_0(e,\phi)$ is the bulk part of the entropy and $s_1(\nabla \phi)$ is the conformational entropy. Once again, we assume the conformational entropy is independent of the phase.
Mass conservation of the binary fluid system yields
\bena
\rho_t+\nabla\cdot(\rho\bv)=0.
\eena
It can be rewritten into
\bena
\nabla\cdot\bv=-\frac{1}{\rho}[\rho_t+\bv\cdot \nabla \rho].
\label{eq1.1.1}
\eena
This imposes a constraint on $\bv, T, \phi$ provided $\rho$ is prescribed.

In this study, we assume  the fluid density is  linearly dependent on the temperature\cite{PhysRevE.55.2780},
\bena
\rho(T)=\rho_0[1-\alpha(T-T_0)],
\eena
where $T_0$ is a reference temperature (i.e., the average value of the boundary temperature) and $\alpha$ is the fluid thermal expansion coefficient.
In the context of the Boussinesq approximation,  the buoyancy force of the system is given by
\bena
\textbf{b}=\rho_0\alpha(T-T_0)g\hat{\textbf{z}}
\eena
and
the approximate momentum balance equation is given by
\bena
\rho_0(\bv_t+\bv \cdot \nabla \bv)=\nabla \cdot\sigma_e+\eta\Delta\bv-\nabla p+\rho_0\alpha Tg\hat{\textbf{z}},
\label{eq1.1.2}
\eena
where $\sigma_e$ is the extra stress tensor, $p$ is the hydrostatic pressure, $\eta$ is the viscosity of the fluid, $g$ is the gravitational acceleration and $\hat{\textbf{z}}$ is the unit vector in the direction of gravity.
The mass conservation \eqref{eq1.1.1} is approximated by the following continuity equation
\ben
\nabla \cdot \bv=0.
\label{div-v}
\een

The approximate energy conservation equation is given by
\bena
(\frac{\rho_0}{2}|\bv|^2+e)_t+\nabla\cdot[(\frac{\rho_0}{2}|\bv|^2+e)\bv]=-\nabla\cdot\bq+\nabla\cdot[(\sigma_e+2\eta\bD-p\bI)\cdot\bv]+\rho_0\alpha Tg\hat{\textbf{z}}\cdot\bv,
\label{eq1.1.3}
\eena
where $\bq$ is the heat flux.
Considering \eqref{eq1.1.2} and \eqref{div-v}, we arrive at the transport equation for internal energy density $e$ as follows
\bena
e_t+\bv \cdot \nabla e = ({\sigma_e}+2\eta\bD-p\bI):\nabla \bv-\nabla \cdot \bq.
\label{eq1.1.5}
\eena

For the phase field, we postulate its transport equation  as follows
\bena
\phi_t +\nabla \cdot (\phi \bv) =j,
\label{eqpsiequation}
\eena
where $j=-\nabla\cdot J$ and $J$ is the excessive diffusive flux to be determined by the Onsager principle. 

\subsection{Constitutive equations}

\noindent \indent Note that $\frac{\delta S}{\delta e}=\frac{1}{T}$,
\bena
\nabla s=\frac{\delta S}{\delta e}\nabla e+\frac{\delta S}{\delta \phi}\nabla \phi+\nabla\cdot(\frac{\partial s}{\partial \nabla \phi}\nabla \phi),
\eena
and
\bena
-(\frac{\delta S}{\delta e}\nabla e+\frac{\delta S}{\delta \phi}\nabla \phi)\cdot \bv
=(s\bI-\frac{\partial s}{\partial \nabla \phi}\nabla \phi):\nabla\bv-\nabla\cdot[(s\bI-\frac{\partial s}{\partial \nabla \phi}\nabla \phi)\cdot\bv].
\eena
Considering the total entropy of system (\ref{system entropy}), then the corresponding entropy production rate is calculated as
\bena
\frac{d S}{dt}=\int_\Omega (\vparl{S}{e}e_t + \vparl{S}{\phi}\phi_t)d \bx +\int_{\pr \Omega}  \bn \cdot (\parl{s}{\nabla \phi} \phi_t ) d\ba  \\

=\int_\Omega \vparl{S}{e}[-\bv\cdot \nabla e+ ({\sigma_e}+2\eta\bD-p\bI) :\nabla \bv- \nabla \cdot \bq]
+ \vparl{S}{\phi}[-\nabla \cdot (\phi \bv)-\nabla\cdot J]d \bx\\
+\int_{\pr \Omega}  \bn \cdot (\parl{s}{\nabla \phi} \phi_t ) d\ba\\
=\int_\Omega [\frac{1}{ T} (\sigma_e-p\bI+2\eta\bD-T(\frac{\partial s}{\partial\nabla\phi}\nabla\phi-s\bI+\frac{\delta S}{\delta \phi}\phi\bI)) :\nabla \bv+\bq \cdot \nabla (\frac{1}{T})+\nabla\vparl{S}{\phi}\cdot J] d \bx\\
+\int_{\pr \Omega}  \bn \cdot [\parl{s}{\nabla \phi} \phi_t -(s\bI-\parl{s}{\nabla \phi} \nabla \phi)\cdot \bv-\frac{\bq}{T}-\frac{\delta S}{\delta \phi}J] d\ba,
\eena
where $\bn$ is the unit outward normal vector of $\pr \Omega$.
The bulk entropy production rate of system is then obtained as follows
\bena
\frac{d S_{gen}}{d t}=\int_\Omega [\frac{1}{ T}(\sigma_e-p\bI+2\eta\bD-T(\frac{\partial s}{\partial\nabla\phi}\nabla\phi-s\bI+\frac{\delta S}{\delta \phi}\phi\bI)) :\nabla \bv+\bq \cdot \nabla (\frac{1}{T})+\nabla\vparl{S}{\phi}\cdot J] d \bx.
\eena

According to the second law of thermodynamics, for non-equilibrium processes, the bulk entropy production rate of system should be non-negative. Therefore, based on the Onsager linear response theory, we put forward the following constitutive relation:
\bena
(J, \sigma_e-p\bI+2\eta\bD-T(\frac{\partial s}{\partial\nabla\phi}\nabla\phi-s\bI+\frac{\delta S}{\delta \phi}\phi\bI)+\pi \bI, \bq)^T = {\cal M} \cdot (\nabla\frac{\delta S}{\delta \phi},{\bf D},\nabla\frac{1}{T})^T,
\label{constitutive-relation}
\eena
where ${\cal M}\geq 0$ is the mobility operator and $\pi$ is an arbitrary function of $(\bx,t)$ owing to $\nabla \cdot \bv=0$.  This constitutive relation gives the necessary coupling between various hydrodynamical variables. The off-diagonal entries measure magnitudes of the cross-coupling between the stress, temperature and excessive volume fraction  flux. A special diagonal $\cal M$ and $\pi$ yields
\bena
J= \bM \nabla\frac{\delta S}{\delta \phi},\quad \sigma_e=T\frac{\partial s}{\partial\nabla\phi}\nabla\phi,\quad
\bq=D_e(T, \phi)\nabla\frac{1}{T}, \quad D_e=D_0(\phi)T^2,
\label{constitutive equations}
\eena
where ${\bM}>0$ is the mobility coefficient,   $D_0>0$ the thermal conductivity constant, and $\sigma_e$ is the well-known Ericksen stress. Hence, we deduce that the above non-isothermal hydrodynamic binary model with the non-negative bulk entropy production rate
\bena
\frac{d S_{gen}}{d t}=\int_\Omega (\bM|\nabla\vparl{S}{\phi}|^2 +2\eta\frac{1}{T}\bD:\bD+\frac{D_0}{T^2} |\nabla{T}|^2)d \bx\geq 0.
\eena

With the excessive flux $J$ given by \eqref{constitutive equations}, the entropy production rate involving the boundary is expressed in the following form
\ben
\bea{l}
\int_{\pr \Omega}  \bn \cdot [\parl{s}{\nabla \phi} \phi_t -(s\bI-\parl{s}{\nabla \phi} \nabla \phi)\cdot \bv-\frac{\bq}{T}-\frac{\delta S}{\delta \phi} \bM \nabla\frac{\delta S}{\delta \phi}] d\ba\\
=\int_{\pr \Omega} (\phi_t, \bv, \frac{1}{T},\frac{\delta S}{\delta \phi} )\cdot ( \bn \cdot \parl{s}{\nabla \phi},   -\bn \cdot (s\bI-\parl{s}{\nabla \phi} \nabla \phi), -\bn \cdot \bq ,-\bn \cdot \bM \nabla\frac{\delta S}{\delta \phi}) d\ba.
\eea
\label{entropy Boundary flux}
\een
We apply the Onsager principle to the boundary  entropy production rate to obtain
\bena
(\phi_t, \bv, \bn \cdot \bq,\bn \cdot \bM \nabla\frac{\delta S}{\delta \phi} )^T=
{\cal M}_s\cdot ( \bn \cdot \parl{s}{\nabla \phi},   -\bn \cdot (s\bI-\parl{s}{\nabla \phi} \nabla \phi), -\frac{1}{T}, -\frac{\delta S}{\delta \phi})^T,\label{TC-BC}
\eena
where ${\cal M}_s$ is the boundary mobility operator. If ${\cal M}_s \geq 0$, this boundary condition yields a positive contribution to the total entropy production rate from the boundary terms. The boundary conditions together with the bulk equations give a thermodynamically consistent partial differential equation system. We remark that in an open system, the entropy flux at the boundary may not be always positive. It depends on the property of the boundary and ambient property and  physics. This discussion is beyond the scope of this study.

To simulate the phenomenon related to   the Rayleigh-B\'{e}nard convection in the binary fluid flow in a rectangular domain, one often hold the temperature at the top and bottom boundary at different values to create a temperature gradient in the domain. It unavoidably creates entropy fluxes crossing boundaries of the physical domain that the binary fluid occupies. The Dirichlet boundary condition on the temperature does not warrant a positive entropy production at the boundary \eqref{TC-BC}. Namely, there could be entropy exchanges between the interior of the domain and the surrounding. This problem was not studied in our previous work \cite{SunShouwen}.

To study this phenomenon, we propose the following physical boundary conditions (shown in Figure \ref{Fig1}):
\ben
\bea{l}
\bv\mid_{\partial \Omega}=0, \quad \frac{\partial s}{\partial \nabla \phi}\cdot \bn\mid_{\partial \Omega}=0,\quad
(\bM \nabla\frac{\delta S}{\delta \phi})\cdot \bn\mid_{\partial \Omega}=0, \\
T\mid_{upper}=T_b, ~T\mid_{lower}=T_a, ~\bn \cdot \nabla T\mid_{left}=0,
~\bn \cdot \nabla T\mid_{right}=0.
\eea
\label{inhomogeneous BC1}
\een
Then, the entropy production rate involving the boundary (\ref{entropy Boundary flux}) reduces to
\bena
-\int_{\pr \Omega}  \bn \cdot \frac{\bq}{T} d\ba=\int_{\pr \Omega}  \bn \cdot D_0\frac{\nabla T}{T} d\ba.
\label{entropy Boundary flux1}
\eena
There is no theoretical  guarantee that this is positive although the overall entropy production rate is normally positive in our simulations.


The volume of fluid A is defined as
$
V=\int_\Omega \phi d\bx.
$
Under the physical condition (\ref{inhomogeneous BC1}), the rate of change of volume is calculated as
\bena
\frac{d V}{dt}=\int_\Omega \phi_t d\bx= -\int_\Omega  [\nabla\cdot \bM\nabla\frac{\delta S}{\delta \phi} + \nabla \cdot (\phi \bv)] d\bx=0,
\eena
indicating that the volume of each fluid component is conserved in the model.

We summarize the governing system of equations of the non-isothermal hydrodynamic model for incompressible binary fluids as follows:
\ben\bea{l}
\begin{cases}
\phi_t +\nabla \cdot (\phi \bv) =-\nabla\cdot \bM\nabla\frac{\delta S}{\delta \phi},\\
\nabla \cdot \bv = 0,\\
\rho_0(\bv_t+\bv \cdot \nabla \bv)=\nabla \cdot\sigma_e+{\eta}\Delta\bv-\nabla p+\rho_0 \alpha Tg\hat{\textbf{z}}, \\
e_t+\bv \cdot \nabla e = {\sigma_e}:\nabla \bv+{2\eta}\bD:\nabla \bv+ {D_0 } \nabla^2 T,
\end{cases}
\eea
\label{eq2.1}
\een
 The general thermodynamically consistent boundary conditions are given by (\ref{TC-BC}). Whereas, the physical boundary conditions we adopt for the Rayleigh-B\'{e}nard convection are given by \eqref{inhomogeneous BC1} which does not guarantee a positive entropy production.

 \begin{rem}
 This thermodynamically consistent model is an approximation of the thermodynamical consistent model derived in  \cite{SunShouwen} via the Boussinesq approximation. Here, we present the constitutive equations in a more general setting to allow potential coupling among the temperature, stress and phase dynamics, which was not given in \cite{SunShouwen}. Due to the Buossinesq approximation, this model is valid only when the density variation with respect to the temperature is small.
 \end{rem}

\subsection{Internal energy and entropy}

\noindent \indent  We recall that the bulk Helmholtz free energy is defined as follows
\bena
f(T,\phi)=e-Ts_0(e,\phi),
\label{entropy function 1}
\eena
where $\frac{\partial s_0}{\partial e}=\frac{1}{T}$.
We approximate the internal energy density by \cite{QingmingChang}
\bena
e=C_A T,
\label{entropy function 3}
\eena
where $C_A$ is a constant specific heat.

It follows from the derivation in \cite{SunShouwen} that
\bena
f(T,\phi)=T[-\int_{T_M}^T\frac{e(\xi)}{\xi^2}d\xi+F(\phi)]=-TC_A(\ln T-\ln T_M)+TF(\phi),
\label{entropy function 4}
\eena
where $T_M$ is a critical temperature and $F(\phi)=\gamma_2\phi^2(1-\phi)^2$, where $\gamma_2$
measures the strength of the repulsive potential.

Combining (\ref{entropy function 1}),(\ref{entropy function 3}) with (\ref{entropy function 4}), the bulk entropy is obtained as
\bena
s_0(T,\phi)=-\frac{1}{T}[f(T,\phi)-e]=C_A(\ln T-\ln T_M)-F(\phi)+C_A.
\label{entropy function 7}
\eena
Notice that the conformational entropy is expressed as
\bena
s_1(\nabla \phi)=-\frac{\gamma_1}{2}|\nabla\phi|^2,
\eena
where $\gamma_1$ is a parameter measuring the strength of the conformational entropy.
Therefore, the total entropy of the fluid system is expressed as
\bena
S(e,\phi,\nabla\phi)=\int_\Omega [-\frac{\gamma_1}{2}|\nabla\phi|^2+s_0(T,\phi)]dx,\quad e=C_A T.
\label{entropy function 10}
\eena

Using the relation between $e$ and $T$, we obtain the equivalent equations of (\ref{eq2.1}) in $(\bv, T, \phi)$ as follows
\ben\bea{l}
\begin{cases}
\phi_t +\nabla \cdot (\phi \bv) =-\nabla\cdot \bM\nabla\frac{\delta S}{\delta \phi},\\
\nabla \cdot \bv = 0,\\
\rho_0(\bv_t+\bv \cdot \nabla\bv)=\nabla \cdot\sigma_e+{\eta}\Delta\bv-\nabla p+\rho_0 \alpha Tg\hat{\textbf{z}}, \\
C_A(T_t+\bv \cdot \nabla T) = {\sigma_e}:\nabla \bv+{2\eta}\bD:\nabla \bv+{D_0 }\nabla^2 T.
\end{cases}
\eea
\label{equivalenteq2.1}
\een
The physical boundary conditions are given by \eqref{inhomogeneous BC1}.

\subsection{Weakly compressible model}

\noindent \indent We next extend the incompressible model to a weakly compressible model by stipulating the following relation between the density and the hydrostatic pressure:
\ben
\frac{d}{dt}\ln \rho=\epsilon \nabla^2 (\frac{p}{T})_t,
\een
where $\frac{d}{dt}$ denotes the material derivative and $\epsilon$ is a user-determined parameter.
Then, the continuity equation (\ref{eq1.1.1}) reduces to
\ben
\nabla \cdot \bv=- \epsilon \nabla^2 (\frac{p}{T})_t.
\een
The extra stress $\sigma_e$ is defined by
\bena
\sigma_e=T(\frac{\partial s}{\partial\nabla\phi}\nabla\phi-s\bI+\frac{\delta S}{\delta \phi}\phi\bI).
\label{the extra stress}
\eena

The entropy production rate is given by
\bena
\frac{d S}{dt}=\int_\Omega (\vparl{S}{e}e_t + \vparl{S}{\phi}\phi_t)d \bx +\int_{\pr \Omega}  \bn \cdot (\parl{s}{\nabla \phi} \phi_t ) d\ba  \\
=\int_\Omega [\frac{1}{ T} (-p\bI+2\eta\bD) :\nabla \bv+\bq \cdot \nabla (\frac{1}{T})+\nabla\vparl{S}{\phi}\cdot J] d \bx\\
+\int_{\pr \Omega}  \bn \cdot [\parl{s}{\nabla \phi} \phi_t -(s\bI-\parl{s}{\nabla \phi} \nabla \phi)\cdot \bv-\frac{\bq}{T}-\frac{\delta S}{\delta \phi}J] d\ba\\
=\int_\Omega [\frac{1}{ T} (2\eta\bD) :\nabla \bv+ \epsilon \frac{p}{T} \nabla^2 (\frac{p}{T})_t+\bq \cdot \nabla (\frac{1}{T})+\nabla\vparl{S}{\phi}\cdot J] d \bx\\
+\int_{\pr \Omega}  \bn \cdot [\parl{s}{\nabla \phi} \phi_t -(s\bI-\parl{s}{\nabla \phi} \nabla \phi)\cdot \bv-\frac{\bq}{T}-\frac{\delta S}{\delta \phi}J] d\ba\\
=\int_\Omega [\frac{1}{ T} (2\eta\bD) :\nabla \bv-\epsilon \nabla \frac{p}{T} \cdot \nabla (\frac{p}{T})_t+\bq \cdot \nabla (\frac{1}{T})+\nabla\vparl{S}{\phi}\cdot J] d \bx\\
+\int_{\pr \Omega}  \bn \cdot [\parl{s}{\nabla \phi} \phi_t -(s\bI-\parl{s}{\nabla \phi} \nabla \phi)\cdot \bv-\frac{\bq}{T}-\frac{\delta S}{\delta \phi}J +\epsilon\frac{p}{T} \nabla (\frac{p}{T})_t] d\ba.
\eena

We define the modified entropy
\bena
\hat{S}=S+\int_{\Omega} \frac{\epsilon}{2}|\nabla \frac{p}{T}|^2d\bx.
\eena
Then,
\bena
\frac{d \hat{S}}{dt}
=\int_\Omega [\frac{1}{ T} 2\eta\bD :\bD+D_0 T^2 \nabla (\frac{1}{T}) \cdot \nabla (\frac{1}{T})+\nabla\vparl{S}{\phi}\cdot M \nabla\vparl{S}{\phi}] d \bx\\
+\int_{\pr \Omega}  \bn \cdot [\parl{s}{\nabla \phi} \phi_t -(s\bI-\parl{s}{\nabla \phi} \nabla \phi)\cdot \bv-\frac{\bq}{T}-\frac{\delta S}{\delta \phi}J +{\epsilon} \frac{p}{T} \nabla (\frac{p}{T})_t] d\ba.
\eena
The bulk part of the entropy production is non-negative definite.  The additional boundary condition for $p$ besides \eqref{inhomogeneous BC1} is given by
\bena
\bn\cdot \nabla \frac{p}{T}\mid_{\partial \Omega}=0.\label{BC-p}
\eena
For both the homogeneous Neumann and constant Dirichlet boundary conditions in the temperature, \eqref{BC-p} implies $\bn\cdot \nabla p=0$.

The modified thermodynamically consistent weakly compressible hydrodynamical model is summarized as follows
\ben\bea{l}
\begin{cases}
\phi_t +\nabla \cdot (\phi \bv) =-\nabla\cdot \bM\nabla\frac{\delta S}{\delta \phi},\\
\nabla \cdot \bv =- \epsilon \nabla^2 (\frac{p}{T})_t,\\
\rho_0(\bv_t+\bv \cdot \nabla\bv)=\nabla \cdot\sigma_e+{\eta}\Delta\bv-\nabla p+\rho_0 \alpha Tg\hat{\textbf{z}}, \\
\sigma_e=T(\frac{\partial s}{\partial\nabla\phi}\nabla\phi-s\bI+\frac{\delta S}{\delta \phi}\phi\bI),\\
C_A(T_t+\bv \cdot \nabla T) = {\sigma_e}:\nabla \bv+{2\eta}\bD:\nabla \bv- p\bI:\nabla \bv+{D_0 }\nabla^2 T.
\end{cases}
\eea
\label{equivalenteq2.3}
\een
We note that this is an approximation to  the nonisothermal thermodynamically consistent incompressible model. We will show in the next section how we use this model to derive a family of numerical projection schemes for the incompressible model.

\subsection{Non-dimensionalization}

\noindent \indent  Using characteristic length scale $H$, temperature scale $\Delta T$ and velocity scale $U$, we achieve the corresponding dimensionless parameters and the physical variables:
\bena
\hat{\phi}=\phi, \quad \hat{x}=\frac{x}{H}, \quad \hat{y}=\frac{y}{H}, \quad \hat{t}=\frac{tU}{H},
\quad \hat{T} =\frac{T}{\Delta T}, \quad \hat{\bv}=\frac{\bv}{U}, \\
\hat{\frac{\delta S}{\delta \phi}}=\frac{\Delta T }{\rho_0 U^2}\frac{\delta S}{\delta \phi},
\quad \hat{\sigma}_e=\frac{\sigma_e}{ \rho_0 U^2},
\quad \hat{p}= \frac{p}{\rho_0 U^2},
\quad \hat{e}=\frac{e}{\rho_0U^2}, \quad \hat{\bM}=\frac{\rho_0 \bM U}{ \Delta T H },\quad
\hat{\gamma_1}=\frac{\Delta T \gamma_1}{\rho_0U^2H^2}, \\
\hat{\gamma_2}=\frac{\Delta T \gamma_2}{\rho_0U^2},
\quad \hat{\gamma_3}=\rho_0\Delta T U^2\gamma_3,
\quad \hat{C_A}=\frac{\Delta T C_A}{\rho_0U^2},\quad
 \hat{s_0}=\frac{\Delta T s_0}{\rho_0 U^2}, \quad\hat{S}=\frac{\Delta T S}{\rho_0 U^2H^2}.
\eena
In thermal convection, we express the maximum buoyancy-generated velocity as $U=\sqrt{\alpha g \Delta T H}$ and denote the Rayleigh number and the Prandtl number as follows
\bena
Ra=\frac{\alpha g \Delta T H^3}{\nu \xi}, \quad Pr=\frac{\nu}{\xi},
\eena
where $\nu=\frac{\eta}{\rho_0}$ is the kinematic viscosity and $\xi=\frac{D_0}{C_A}$ is the thermal diffusivity, the Rayleigh number indicates the strength of thermal forcing and is a measure of the ratio of buoyancy and dissipation, while the Prandtl number describes the relative importance of momentum diffusivity and thermal diffusivity.

For simplicity, after dropping the $\hat{}$s, we rewrite the dimensionless governing equations as below
\bena\bec
\phi_{t} +\nabla \cdot (\phi \bv) = -\nabla\cdot \bM\nabla\frac{\delta S}{\delta \phi},\\
\nabla \cdot \bv = 0,\\
\bv_t +\bv \cdot \nabla   \bv =\nabla \cdot \sigma_e+ \sqrt{\frac{Pr}{Ra}}\Delta \bv -\nabla p + T\hat{\textbf{z}}, \\
T_t+\bv \cdot \nabla T = \frac{1}{C_A}\sigma_e:\nabla \bv +\frac{2}{C_A}\sqrt{\frac{Pr}{Ra}} \bD:\nabla \bv + \frac{1}{\sqrt{PrRa}}\nabla^2 T,
\label{Non-dimensionalization equation1}
\eec\eena
where ${\sigma_e}=-\gamma_1 T\nabla\phi\otimes\nabla\phi$, the symbol $\otimes$ denotes the tensor product. And the corresponding non-dimensionalization total entropy rewrite as
\bena
S(e,\phi,\nabla \phi)=\int_V [-\frac{\gamma_1}{2}|\nabla\phi|^2+s_0(e,\phi)]dx.
\label{Non-dimensionalization entropy}
\eena

Similarly, we have the dimensionless governing equations of the modified model (\ref{equivalenteq2.3}) as follows
\bena\bec
\phi_{t} +\nabla \cdot (\phi \bv) = -\nabla\cdot \bM\nabla\frac{\delta S}{\delta \phi},\\
\nabla \cdot \bv = - \epsilon \nabla^2 (\frac{p}{T})_t,\\
\bv_t +\bv \cdot \nabla   \bv =\nabla \cdot \sigma_e+ \sqrt{\frac{Pr}{Ra}}\Delta \bv -\nabla p + T\hat{\textbf{z}}, \\
T_t+\bv \cdot \nabla T = \frac{1}{C_A}\sigma_e:\nabla \bv +\frac{2}{C_A}\sqrt{\frac{Pr}{Ra}} \bD:\nabla \bv - \frac{1}{C_A}p\bI:\nabla \bv + \frac{1}{\sqrt{PrRa}}\nabla^2 T,
\label{Non-dimensionalization equation1 of modified mode}
\eec\eena
where $\sigma_e=T(-\gamma_1\nabla\phi\otimes\nabla\phi-s\bI+\frac{\delta S}{\delta \phi}\phi\bI)$. And the modified non-dimensionalization entropy
\bena
\hat{S}=S+\int_{\Omega} \frac{\epsilon}{2}|\nabla \frac{p}{T}|^2d\bx.
\eena
where $S$ is the dimensionless total entropy (\ref{Non-dimensionalization entropy}).


\section{Numerical Approximations }


\noindent \indent  The non-isothermal, incompressible binary hydrodynamic model exhibits a positive entropy production rate and conserves the volume of each fluid phase when subjected to thermodynamically consistent physical or periodic boundary conditions. However, the thermodynamic consistency of the model's properties strongly depends on the specific boundary conditions employed. Notably, when using boundary conditions relevant to Rayleigh-B\'{e}nard convection, the entropy production rate cannot be guaranteed.

To address this issue, we will develop a set of second-order structure-preserving schemes in both time and space for the above nonisothermal model. These schemes aim to preserve the entropy production rate and conserve the volume of each fluid phase, regardless of whether the rate is positive definite or not. Furthermore, these schemes will maintain thermodynamic consistency at the discrete level when the boundary conditions warrant it, and be structure-preserving at all times.

\subsection{Model reformulation }

\noindent \indent Following the idea of energy quadratization method, we introduce a new variable to reformulate the governing system of equations to an equivalent form. Specifically, we set
\bena
q=\sqrt{-s_0-\gamma_2\phi^2-\gamma_3e^2+C_0},
\label{eq q}
\eena
where $C_0$ is a positive constant large enough to ensure that $-s_0-\gamma_2\phi^2-\gamma_3e^2+C_0>0$. Hence, the entropy of system (\ref{Non-dimensionalization entropy}) is express as a quadratic functional
\ben\bea{l}
S = \int_\Omega ( -\frac{\gamma_1}{2}|\nabla\phi|^2-q^2-\gamma_2\phi^2-\gamma_3e^2+C_0 )d{\bf x}.
\label{eq S}
\eea\een
Then, we have
\bena
q_{\phi}=\frac{\partial q}{\partial \phi}=\frac{-\frac{\partial s_0}{\partial \phi}-2\gamma_2\phi}{2\sqrt{-s_0-\gamma_2\phi^2-\gamma_3e^2+C_0}}
=\frac{F'(\phi)-2\gamma_2\phi}{2\sqrt{-s_0-\gamma_2\phi^2-\gamma_3e^2+C_0}},
\label{eq qpsi}
\eena
and
\bena
q_e=\frac{\partial q}{\partial e}=\frac{-\frac{\partial s_0}{\partial e}-2\gamma_3e}{2\sqrt{-s_0-\gamma_2\phi^2-\gamma_3e^2+C_0}}=\frac{-\frac{\partial s_0}{\partial T}\cdot\frac{\partial T}{\partial e}-2\gamma_3e}{2\sqrt{-s_0-\gamma_2\phi^2-\gamma_3e^2+C_0}}=-\frac{\frac{C_A}{e}+2\gamma_3e}{2\sqrt{-s_0-\gamma_2\phi^2-\gamma_3e^2+C_0}}.
\label{eq qe}
\eena
It  follows that
\bena
\frac{\delta S}{\delta e}=-2qq_e-2\gamma_3e=\frac{C_A}{e}=\frac{1}{T}.
\eena

The equations (\ref{Non-dimensionalization equation1}) can be written equivalently to the following EQ reformulated form
\ben\bea{l}
\begin{cases}
\phi_t +\nabla \cdot (\phi \bv) =- \nabla\cdot\bM\nabla(\gamma_1\Delta\phi-2qq_\phi-2\gamma_2\phi),\\
\nabla \cdot \bv = 0,\\
\bv_t + \bv \cdot \nabla \bv =\nabla \cdot \sigma_e+ \sqrt{\frac{Pr}{Ra}}\Delta \bv -\nabla p + T\hat{\textbf{z}}, \\
T_t+ \bv\cdot \nabla T = \frac{1}{C_A}\sigma_e:\nabla \bv +\frac{2}{C_A}\sqrt{\frac{Pr}{Ra}} \bD:\nabla \bv + \frac{1}{\sqrt{PrRa}}\nabla^2 T,\\
q_t=q_\phi\phi_t+q_ee_t.
\end{cases}
\eea
\label{eq1}
\een
where ${\sigma_e}=-\gamma_1 T\nabla\phi\otimes\nabla\phi, q_\phi=\frac{\partial q}{\partial \phi}, q_e=\frac{\partial q}{\partial e}, e=C_AT$ for a given $q(\bx, 0)$ calculated from \eqref{eq q}.

\subsection{Semi-discrete algorithms}

\noindent \indent Applying Crank-Nicolson method in time, we discretize the corresponding reformulated system given in \eqref{eq1}. We introduce the following notations:
\ben\bea{l}
(\cdot)^{n+\frac{1}{2}} =\frac{1}{2}( (\cdot)^{n+1} + (\cdot)^n), ~~\delta_t (\cdot)^{n+\frac{1}{2}} =\frac{1}{\Delta t}( (\cdot)^{n+1} - (\cdot)^n), \\
\overline{(\cdot)}^{n+\frac{1}{2}} = \frac{1}{2} (3(\cdot)^{n} - (\cdot)^{n-1}), ~~\tilde{(\cdot)}^{n+\frac{1}{2}} = \frac{1}{2} ((\tilde{\cdot})^{n+1} + (\cdot)^{n}).
\eea\een
We present two new, second order, semi-discrete algorithms in time for reformulated system (\ref{eq1}) below.

\begin{sch}[{\bf Semi-discrete entropy-production-rate-preserving scheme I}] Given $\bv^{n}$, $\phi^n$ and $T^n$, we update $\bv^{n+1}$, $\phi^{n+1}$ and $T^{n+1}$ as follows:
\ben\bea{l}
\begin{cases}
\delta_t\phi^{n+\frac{1}{2}} +\nabla \cdot (\phi^{n+\frac{1}{2}} \bv^{n+\frac{1}{2}}) =-\nabla\cdot\bM\nabla(\gamma_1\Delta\phi^{n+\frac{1}{2}}-2q^{n+\frac{1}{2}}q_\phi^{n+\frac{1}{2}}-2\gamma_2\phi^{n+\frac{1}{2}}),\\
\\
\nabla \cdot \bv^{n+\frac{1}{2}} = 0,\\
\\
\delta_t\bv^{n+\frac{1}{2}}+\bv^{n+\frac{1}{2}} \cdot \nabla\bv^{n+\frac{1}{2}}=\nabla\cdot\sigma_e^{n+\frac{1}{2}}+\sqrt{\frac{Pr}{Ra}}\Delta\bv^{n+\frac{1}{2}}
-\nabla p^{n+\frac{1}{2}}+T^{n+\frac{1}{2}}\hat{\textbf{z}},\\
\\
\delta_tT^{n+\frac{1}{2}}+\bv^{n+\frac{1}{2}}\cdot \nabla T^{n+\frac{1}{2}} = \frac{1}{C_A}\sigma_e^{n+\frac{1}{2}}:\nabla \bv^{n+\frac{1}{2}}+\frac{2}{C_A}\sqrt{\frac{Pr}{Ra}}\bD^{n+\frac{1}{2}}:\nabla \bv^{n+\frac{1}{2}}\\
~~~~~~~~~~~~~~~~~~~~~~~~~~~~~~~~~~+ \frac{1}{\sqrt{PrRa}}\nabla^2 T^{n+\frac{1}{2}}, \\
\\
\delta_tq^{n+\frac{1}{2}}=q^{n+\frac{1}{2}}_\phi\delta_t\phi^{n+\frac{1}{2}}+q^{n+\frac{1}{2}}_e\delta_te^{n+\frac{1}{2}},
\end{cases}
\eea
\label{Semi-discrete EQ Scheme-I}
\een
where Ericksen stress tensor $\sigma_e^{n+\frac{1}{2}}=-\gamma_1T^{n+\frac{1}{2}}\nabla\phi^{n+\frac{1}{2}}\otimes\nabla\phi^{n+\frac{1}{2}}$ and internal energy $e^{n+\frac{1}{2}}=C_AT^{n+\frac{1}{2}}.$
Furthermore, the corresponding boundary conditions as:
\ben
\left \{
\bea{l}
\bv^n|_{\partial \Omega} =0, \quad  \bn \cdot \frac{\partial s}{\partial \nabla \phi}^n|_{\partial \Omega}=0, \quad \bn \cdot \nabla \frac{\delta S}{\delta \phi}^n|_{\partial \Omega}=0, \quad
T^n\mid_{upper}=T_b,\\\\
T^n\mid_{lower}=T_a, ~\bn \cdot \nabla T^n\mid_{left}=0,
~\bn \cdot \nabla T^n\mid_{right}=0~(n=0,1,\cdots,N).
\eea\right.
\label{Semi-discrete EQ Scheme-I BC}
\een
\end{sch}
For this algorithm, we prove the following theorem.
\begin{thm}
Given boundary conditions (\ref{Semi-discrete EQ Scheme-I BC}), semi-discrete scheme-I preserves the volume conservation law: $V^{n+1}=V^n$,
and the entropy production rate
\bena
\frac{S^{n+1}-S^n}{\Delta t}=\int_\Omega [\bM(\nabla\frac{\delta S}{\delta \phi}^{n+\frac{1}{2}})^2+2\sqrt{\frac{Pr}{Ra}}\frac{1}{T^{n+\frac{1}{2}}}\bD^{n+\frac{1}{2}}: \bD^{n+\frac{1}{2}}+ \frac{C_A}{\sqrt{PrRa}}\frac{(\|\nabla T^{n+\frac{1}{2}}\|)^2}{(T^{n+\frac{1}{2}})^2}]d\bx\\
~~~~~~~~~~~+\frac{C_A}{\sqrt{PrRa}}\int_{\pr \Omega} \bn \cdot \frac{ \nabla T^{n+\frac{1}{2}}}{T^{n+\frac{1}{2}}} d\ba,
\eena
where
\bena
V^n=\int_\Omega \phi^{n}d\bx,\\
S^n=\int_\Omega[-|q^n|^2-\gamma_2|\phi^n|^2-\gamma_3|e^n|^2-\frac{\gamma_1}{2}|\nabla\phi^n|^2+C_0]d {\bx}.
\eena
\end{thm}

\noindent {\bf Proof.} Noting that $V^n$ and the prescribed boundary conditions,  we have $ \bv^n|_{\partial \Omega}=0$,  $\bn \cdot \nabla \frac{\delta S}{\delta \phi}^n|_{\partial \Omega}=0$, and
\bena
\frac{V^{n+1}-V^n}{\Delta t}
=\int_\Omega \frac{\phi^{n+1}-\phi^n}{\Delta t}d\bx\\
=-\int_\Omega \nabla\cdot\bM\nabla(\gamma_1\Delta\phi^{n+\frac{1}{2}}-2q^{n+\frac{1}{2}}q_\phi^{n+\frac{1}{2}}-2\gamma_2\phi^{n+\frac{1}{2}})+\nabla \cdot (\phi^{n+\frac{1}{2}} \bv^{n+\frac{1}{2}})d\bx=0.
\eena

It follows from (\ref{Semi-discrete EQ Scheme-I}-1), (\ref{Semi-discrete EQ Scheme-I}-4) and (\ref{Semi-discrete EQ Scheme-I}-5) that
\bena
\frac{S^{n+1}-S^n}{\Delta t}

=\int_\Omega [-(q^{n+1}+q^n)\frac{q^{n+1}-q^n}{\Delta t}-\gamma_2(\phi^{n+1}+\phi^n)\frac{\phi^{n+1}-\phi^n}{\Delta t}\\
 -\gamma_3(e^{n+1}+e^n)\frac{e^{n+1}-e^n}{\Delta t}-\frac{\gamma_1}{2}(\nabla\phi^{n+1}+\nabla\phi^n)\frac{\nabla\phi^{n+1}-\nabla\phi^n}{\Delta t} ]d\bx  \\

=\int_\Omega [-2\delta_t \phi^{n+\frac{1}{2}} q^{n+\frac{1}{2}}q_\phi^{n+\frac{1}{2}}+\delta_t \phi^{n+\frac{1}{2}} (\gamma_1\Delta\phi^{n+\frac{1}{2}})-2\gamma_2\phi^{n+\frac{1}{2}}(-\nabla\cdot\bM\nabla\frac{\delta S}{\delta \phi}^{n+\frac{1}{2}})\\

+2\gamma_2\phi^{n+\frac{1}{2}} \nabla \cdot(\phi^{n+\frac{1}{2}} \bv^{n+\frac{1}{2}})-2\delta_t e^{n+\frac{1}{2}} q^{n+\frac{1}{2}}q_e^{n+\frac{1}{2}}-2\gamma_3e^{n+\frac{1}{2}} \delta_t e^{n+\frac{1}{2}}] d\bx\\

=\int_\Omega [\delta_t \phi^{n+\frac{1}{2}}\frac{\delta S}{\delta \phi}^{n + \frac{1}{2}}-2\gamma_2\phi^{n+\frac{1}{2}} (-\nabla\cdot\bM\nabla\frac{\delta S}{\delta \phi}^{n+\frac{1}{2}}-\delta_t \phi^{n+\frac{1}{2}})\\
+2\gamma_2\phi^{n+\frac{1}{2}}\nabla \cdot(\phi^{n+\frac{1}{2}}\bv^{n+\frac{1}{2}})+\delta_t e^{n+\frac{1}{2}} \frac{\delta S}{\delta e}^{n + \frac{1}{2}}] d\bx\\
=\int_\Omega (\delta_t \phi^{n+\frac{1}{2}} \frac{\delta S}{\delta \phi}^{n + \frac{1}{2}}+\delta_t e^{n+\frac{1}{2}} \frac{\delta S}{\delta e}^{n + \frac{1}{2}}) d\bx.
\label{Theorem semi-discrete eq1}
\eena
Takeing into account
$\bI:\nabla\bv^{n+\frac{1}{2}}=0$, $\bn \cdot \bv^{n+\frac{1}{2}}|_{\partial \Omega}=0$ and
\bena
\nabla s^{n+\frac{1}{2}}=\frac{\delta S}{\delta \phi}^{n+\frac{1}{2}}\nabla \phi^{n+\frac{1}{2}}+\frac{\delta S}{\delta e}^{n+\frac{1}{2}}\nabla e^{n+\frac{1}{2}}+\nabla\cdot(\frac{\partial s}{\partial \nabla \phi}^{n+\frac{1}{2}}\nabla \phi^{n+\frac{1}{2}}),
\label{especial condition s}
\eena
we have
\bena
\int_\Omega (\delta_t \phi^{n+\frac{1}{2}} \frac{\delta S}{\delta \phi}^{n + \frac{1}{2}}+\delta_t e^{n+\frac{1}{2}} \frac{\delta S}{\delta e}^{n + \frac{1}{2}}) d\bx\\

=\int_\Omega [(-\nabla\cdot\bM\nabla\frac{\delta S}{\delta \phi}^{n + \frac{1}{2}}-\nabla\cdot(\phi^{n+\frac{1}{2}}\bv^{n+\frac{1}{2}}))\frac{\delta S}{\delta \phi}^{n + \frac{1}{2}}\\
+\big((\sigma_e^{n+\frac{1}{2}}+2\sqrt{\frac{Pr}{Ra}}\bD^{n+\frac{1}{2}}):\nabla \bv^{n+\frac{1}{2}}+\frac{C_A}{\sqrt{PrRa}}\nabla^2 T^{n+\frac{1}{2}}- \bv^{n+\frac{1}{2}}\cdot \nabla e^{n+\frac{1}{2}}\big)\frac{\delta S}{\delta e}^{n + \frac{1}{2}}] d\bx\\

=\int_\Omega [\bM(\nabla\frac{\delta S}{\delta \phi}^{n+\frac{1}{2}})^2+2\sqrt{\frac{Pr}{Ra}}\frac{1}{T^{n+\frac{1}{2}}}\bD^{n+\frac{1}{2}}:\nabla \bv^{n+\frac{1}{2}}+ \frac{C_A}{\sqrt{PrRa}}\frac{(\|\nabla T^{n+\frac{1}{2}}\|)^2}{(T^{n+\frac{1}{2}})^2}\\
+\frac{1}{T^{n+\frac{1}{2}}}\sigma_e^{n+\frac{1}{2}}:\nabla\bv^{n+\frac{1}{2}}
+(s^{n+\frac{1}{2}}\bI-\frac{\partial s}{\partial \nabla \phi}^{n+\frac{1}{2}}\nabla {\phi}^{n+\frac{1}{2}}):\nabla\bv^{n+\frac{1}{2}} ]d\bx\\
-\int_{\pr \Omega}  \bn \cdot (s^{n+\frac{1}{2}}\bI-\frac{\partial s}{\partial \nabla \phi}^{n+\frac{1}{2}}\nabla {\phi}^{n+\frac{1}{2}})\cdot \bv^{n+\frac{1}{2}} d\ba
+\frac{C_A}{\sqrt{PrRa}}\int_{\pr \Omega}  \bn \cdot \frac{ \nabla T^{n+\frac{1}{2}}}{T^{n+\frac{1}{2}}} d\ba\\

=\int_\Omega [\bM(\nabla\frac{\delta S}{\delta \phi}^{n+\frac{1}{2}})^2+2\sqrt{\frac{Pr}{Ra}}\frac{1}{T^{n+\frac{1}{2}}}\bD^{n+\frac{1}{2}}: \bD^{n+\frac{1}{2}}+ \frac{C_A}{\sqrt{PrRa}}\frac{(\|\nabla T^{n+\frac{1}{2}}\|)^2}{(T^{n+\frac{1}{2}})^2}]d\bx\\
+\frac{C_A}{\sqrt{PrRa}}\int_{\pr \Omega}  \bn \cdot \frac{ \nabla T^{n+\frac{1}{2}}}{T^{n+\frac{1}{2}}} d\ba.
\label{conclusion of S}
\eena
Thus,
\bena
\frac{S^{n+1}-S^n}{\Delta t}=\int_\Omega [\bM(\nabla\frac{\delta S}{\delta \phi}^{n+\frac{1}{2}})^2+2\sqrt{\frac{Pr}{Ra}}\frac{1}{T^{n+\frac{1}{2}}}\bD^{n+\frac{1}{2}}: \bD^{n+\frac{1}{2}}+ \frac{C_A}{\sqrt{PrRa}}\frac{(\nabla T^{n+\frac{1}{2}})^2}{(T^{n+\frac{1}{2}})^2}]d\bx\\
~~~~~~~~~~~+\frac{C_A}{\sqrt{PrRa}}\int_{\pr \Omega}  \bn \cdot \frac{ \nabla T^{n+\frac{1}{2}}}{T^{n+\frac{1}{2}}} d\ba.
\eena
This completes the proof.

\rem{To simulate Rayleigh-B\'{e}nard convection phenomena, we adopt boundary conditions \eqref{Semi-discrete EQ Scheme-I BC}, which may lead to a boundary entropy flux. When the entire boundary is isothermal, i.e.,
\bena
\bn \cdot \nabla T\mid_{\partial \Omega}=0,
\label{Temper inhomogeneous BC111}
\eena
the Semi-discrete EQ Scheme-I yields a positive entropy production rate at the semidiscrete level:
\bena
\frac{S^{n+1}-S^n}{\Delta t}=\int_\Omega [\bM(\nabla\frac{\delta S}{\delta \phi}^{n+\frac{1}{2}})^2+2\sqrt{\frac{Pr}{Ra}}\frac{1}{T^{n+\frac{1}{2}}}\bD^{n+\frac{1}{2}}: \bD^{n+\frac{1}{2}}+ \frac{C_A}{\sqrt{PrRa}}\frac{(\nabla T^{n+\frac{1}{2}})^2}{(T^{n+\frac{1}{2}})^2}]d\bx\geq 0.
\eena
}

Notice that this algorithm is fully coupled. To decouple the pressure from the velocity,
 we apply the pressure-correction method \cite{GUERMOND20066011} to the momentum and the continuity equation of   reformulated system (\ref{eq1}). Then, we arrive at the Semi-discrete  Scheme-II as follows.

\begin{sch}[{\bf Semi-discrete entropy-production-rate-preserving scheme-II}] Given $\bv^{n}$, $\phi^n$, $T^n$ and $p^n$, we update $\bv^{n+1}$, $\phi^{n+1}$, $T^{n+1}$ and $p^{n+1}$ as follows:
\ben\bea{l}
\textit{Step 1.}
\begin{cases}
\frac{1}{\Delta t}(\tilde{\bv}^{n+1}-\bv^{n})+\bar{\bv}^{n+\frac{1}{2}}\cdot\nabla\tilde{\bv}^{n+\frac{1}{2}}=\nabla\cdot\sigma_e^{n+\frac{1}{2}}+\sqrt{\frac{Pr}{Ra}}\Delta\tilde{\bv}^{n+\frac{1}{2}}
-\nabla p^{n}+T^{n+\frac{1}{2}}\hat{\textbf{z}},\\
\\
\frac{1}{\Delta t}(\bv^{n+1}-\tilde{\bv}^{n+1})=-\frac{1}{2}\nabla(p^{n+1}-p^n),\\
\\
\nabla \cdot \bv^{n+1} = 0 \quad \Rightarrow ~\nabla^2 (p^{n+1}-p^n)=\frac{2}{\Delta t}\nabla \cdot \tilde{\bv}^{n+1}, 
\end{cases}
\\where ~\tilde{\bv}^{n+1}\mid_{\partial \Omega}=0;
\\
\textit{Step 2.}
\begin{cases}
\frac{1}{\Delta t}(\phi^{n+1}-\phi^n)+\nabla \cdot (\bar{\phi}^{n+\frac{1}{2}} \bv^{n+\frac{1}{2}}) =-\nabla\cdot\bM\nabla(\gamma_1\Delta\phi^{n+\frac{1}{2}}-2q^{n+\frac{1}{2}}\bar{q}_\phi^{n+\frac{1}{2}}-2\gamma_2\phi^{n+\frac{1}{2}}),\\
\\
\frac{1}{\Delta t}(T^{n+1}-T^n)+\nabla \cdot (\bar{T}^{n+\frac{1}{2}}\bv^{n+\frac{1}{2}}) = \frac{1}{C_A}\sigma_e^{n+\frac{1}{2}}:\nabla \bv^{n+\frac{1}{2}}+\frac{2}{C_A}\sqrt{\frac{Pr}{Ra}}\bD^{n+\frac{1}{2}}:\nabla \bv^{n+\frac{1}{2}}\\
~~~~~~~~~~~~~~~~~~~~~~~~~~~~~~~~~~~~~~~~~~~~~~+ \frac{1}{\sqrt{PrRa}}\nabla^2 T^{n+\frac{1}{2}}, \\
\\
\frac{1}{\Delta t}(q^{n+1}-q^n)=\bar{q}^{n+\frac{1}{2}}_\phi\frac{1}{\Delta t}(\phi^{n+1}-\phi^n)+\bar{q}^{n+\frac{1}{2}}_e\frac{1}{\Delta t}(e^{n+1}-e^n),
\end{cases}
\eea
\label{scheme1}
\een
where   $\sigma_e^{n+\frac{1}{2}}=-\gamma_1T^{n+\frac{1}{2}}\nabla\phi^{n+\frac{1}{2}}\otimes\nabla\bar{\phi}^{n+\frac{1}{2}}, e^{n+\frac{1}{2}}=C_AT^{n+\frac{1}{2}}.$
Moreover, the corresponding boundary conditions as:
\ben
\left \{
\bea{l}
\bv^n|_{\partial \Omega} =0, \quad \bn \cdot \nabla p^n|_{\partial \Omega}=0, \quad \bn \cdot \nabla \phi^n|_{\partial \Omega}=0, \quad \bn \cdot \nabla \frac{\delta S}{\delta \phi}^n|_{\partial \Omega}=0, \\\\
T^n\mid_{upper}=T_b, ~T^n\mid_{lower}=T_a, ~\bn \cdot \nabla T^n\mid_{left}=0,
~\bn \cdot \nabla T^n\mid_{right}=0~(n=0,1,\cdots,N).
\label{semi-discrete BC}
\eea\right.
\een
\end{sch}

In the next theorem, we prove this semi-discrete entropy-production-rate-preserving scheme-II preserves the volume and the entropy production rate as well.

\begin{thm}
Given boundary conditions (\ref{semi-discrete BC}), semi-discrete scheme-II preserves the volume conservation law: $V^{n+1}=V^n$,
and  the entropy production rate
\bena
\frac{S^{n+1}-S^n}{\Delta t}=\int_\Omega [\bM(\nabla\frac{\delta S}{\delta \phi}^{n+\frac{1}{2}})^2+2\sqrt{\frac{Pr}{Ra}}\frac{1}{T^{n+\frac{1}{2}}}\bD^{n+\frac{1}{2}}: \bD^{n+\frac{1}{2}}+ \frac{C_A}{\sqrt{PrRa}}\frac{(\nabla T^{n+\frac{1}{2}})^2}{(T^{n+\frac{1}{2}})^2}]d\bx\\
~~~~~~~~~~+\frac{C_A}{\sqrt{PrRa}}\int_{\pr \Omega}  \bn \cdot \frac{ \nabla T^{n+\frac{1}{2}}}{T^{n+\frac{1}{2}}} d\ba,
\eena
where
\bena
V^n=\int_\Omega \phi^{n}d\bx,\\
S^n=\int_\Omega[-|q^n|^2-\gamma_2|\phi^n|^2-\gamma_3|e^n|^2-\frac{\gamma_1}{2}|\nabla\phi^n|^2+C_0]d {\bx}.
\eena
\end{thm}
\noindent {\bf Proof.} The proof is basically similar to the proof of Theorem 4.1 except for one fine detail, which we highlight here. Following the equations in $Step~2$ of scheme (\ref{scheme1}), we obtain
\bena
\frac{S^{n+1}-S^n}{\Delta t}
=\int_\Omega (\delta_t \phi^{n+\frac{1}{2}} \frac{\delta S}{\delta \phi}^{n + \frac{1}{2}}+\delta_t e^{n+\frac{1}{2}} \frac{\delta S}{\delta e}^{n + \frac{1}{2}}) d\bx.
\label{Theorem semi-discrete proj scheme eq1}
\eena
Next, we introduce an auxiliary variable $s_*^{n+\frac{1}{2}}$ satisfying
\bena
\nabla s_*^{n+\frac{1}{2}}=\frac{\delta S}{\delta \phi}^{n+\frac{1}{2}}\nabla \bar{\phi}^{n+\frac{1}{2}}+\frac{\delta S}{\delta e}^{n+\frac{1}{2}}\nabla \bar{e}^{n+\frac{1}{2}}+\nabla\cdot(\frac{\partial s}{\partial \nabla \phi}^{n+\frac{1}{2}}\nabla \bar{\phi}^{n+\frac{1}{2}}),
\label{especial condition s11}
\eena
subject to a proper boundary condition, for example, a Dirichlet boundary condition. In fact, $s_*^{n+\frac{1}{2}}$ satisfies a Poisson equation with a Dirichlet boundary condition. So, its existence of the function is warranted. We then
employ the following relations:
$\bI:\nabla\bv^{n+\frac{1}{2}}=0$ and $\bn \cdot \bv^{n+\frac{1}{2}}|_{\partial \Omega}=0$ to obtain
\bena
\int_\Omega (\delta_t \phi^{n+\frac{1}{2}} \frac{\delta S}{\delta \phi}^{n + \frac{1}{2}}+\delta_t e^{n+\frac{1}{2}} \frac{\delta S}{\delta e}^{n + \frac{1}{2}}) d\bx\\

=\int_\Omega [(-\nabla\cdot\bM\nabla\frac{\delta S}{\delta \phi}^{n + \frac{1}{2}}-\nabla\cdot(\bar{\phi}^{n+\frac{1}{2}}\bv^{n+\frac{1}{2}}))\frac{\delta S}{\delta \phi}^{n + \frac{1}{2}}\\
+\big((\sigma_e^{n+\frac{1}{2}}+2\sqrt{\frac{Pr}{Ra}}\bD^{n+\frac{1}{2}}):\nabla \bv^{n+\frac{1}{2}}+\frac{C_A}{\sqrt{PrRa}}\nabla^2 T^{n+\frac{1}{2}}- \nabla\cdot ( \bar{e}^{n+\frac{1}{2}}\bv^{n+\frac{1}{2}})\big)\frac{\delta S}{\delta e}^{n + \frac{1}{2}}] d\bx\\

=\int_\Omega [\bM(\nabla\frac{\delta S}{\delta \phi}^{n+\frac{1}{2}})^2+2\sqrt{\frac{Pr}{Ra}}\frac{1}{T^{n+\frac{1}{2}}}\bD^{n+\frac{1}{2}}:\nabla \bv^{n+\frac{1}{2}}+ \frac{C_A}{\sqrt{PrRa}}\frac{(\nabla T^{n+\frac{1}{2}})^2}{(T^{n+\frac{1}{2}})^2}\\
+\frac{1}{T^{n+\frac{1}{2}}}\sigma_e^{n+\frac{1}{2}}:\nabla\bv^{n+\frac{1}{2}}
+(s_*^{n+\frac{1}{2}}\bI-\frac{\partial s}{\partial \nabla \phi}^{n+\frac{1}{2}}\nabla \bar{\phi}^{n+\frac{1}{2}}):\nabla\bv^{n+\frac{1}{2}} ]d\bx\\
-\int_{\pr \Omega}  \bn \cdot (s_*^{n+\frac{1}{2}}\bI-\frac{\partial s}{\partial \nabla \phi}^{n+\frac{1}{2}}\nabla \bar{\phi}^{n+\frac{1}{2}})\cdot \bv^{n+\frac{1}{2}} d\ba
+\frac{C_A}{\sqrt{PrRa}}\int_{\pr \Omega}  \bn \cdot \frac{ \nabla T^{n+\frac{1}{2}}}{T^{n+\frac{1}{2}}} d\ba\\

=\int_\Omega [\bM(\nabla\frac{\delta S}{\delta \phi}^{n+\frac{1}{2}})^2+2\sqrt{\frac{Pr}{Ra}}\frac{1}{T^{n+\frac{1}{2}}}\bD^{n+\frac{1}{2}}: \bD^{n+\frac{1}{2}}+ \frac{C_A}{\sqrt{PrRa}}\frac{(\nabla T^{n+\frac{1}{2}})^2}{(T^{n+\frac{1}{2}})^2}]d\bx\\
+\frac{C_A}{\sqrt{PrRa}}\int_{\pr \Omega}  \bn \cdot \frac{ \nabla T^{n+\frac{1}{2}}}{T^{n+\frac{1}{2}}} d\ba.
\label{conclusion of proj S}
\eena
 The theorem is hence proved.

\rem{When the entire boundary is isothermal, i.e. $\bn \cdot \nabla T\mid_{\partial \Omega}=0$, semi-discrete scheme-II also yields a positive entropy production rate at the semidiscrete level:
\bena
\frac{S^{n+1}-S^n}{\Delta t}=\int_\Omega [\bM(\nabla\frac{\delta S}{\delta \phi}^{n+\frac{1}{2}})^2+2\sqrt{\frac{Pr}{Ra}}\frac{1}{T^{n+\frac{1}{2}}}\bD^{n+\frac{1}{2}}: \bD^{n+\frac{1}{2}}+ \frac{C_A}{\sqrt{PrRa}}\frac{(\nabla T^{n+\frac{1}{2}})^2}{(T^{n+\frac{1}{2}})^2}]d\bx\geq 0.
\eena}

Next, we propose a family of alternative projection algorithms to decouple the pressure from the velocity based on a modified thermodynamically consistent model.
We note that the modified model (\ref{Non-dimensionalization equation1 of modified mode}) can also be reformulated in the EQ form as follows
\ben\bea{l}
\begin{cases}
\phi_t +\nabla \cdot (\phi \bv) =- \nabla\cdot\bM\nabla(\gamma_1\Delta\phi-2qq_\phi-2\gamma_2\phi),\\
\nabla \cdot \bv = - \epsilon \nabla^2 (\frac{p}{T})_t,\\
\bv_t +\bv \cdot \nabla \bv =\nabla \cdot \sigma_e+ \sqrt{\frac{Pr}{Ra}}\Delta \bv -\nabla p + T\hat{\textbf{z}}, \\
T_t+\bv \cdot \nabla T = \frac{1}{C_A}\sigma_e:\nabla \bv +\frac{2}{C_A}\sqrt{\frac{Pr}{Ra}} \bD:\nabla \bv - \frac{1}{C_A}p\bI:\nabla \bv + \frac{1}{\sqrt{PrRa}}\nabla^2 T,\\
q_t=q_\phi\phi_t+q_ee_t.
\end{cases}
\eea
\label{The modified mode EQ form}
\een
where $\sigma_e=T(-\gamma_1\nabla\phi\otimes\nabla\phi-s\bI+\frac{\delta S}{\delta \phi}\phi\bI), q_\phi=\frac{\partial q}{\partial \phi}, q_e=\frac{\partial q}{\partial e}, e=C_AT.$
Based on modified model (\ref{The modified mode EQ form}), we next devise a family of schemes using the pressure-correction strategy.

In order to derive the projection method, we recast (\ref{The modified mode EQ form}) in a suitable form. Specifically, we replace $\bv$ by $\tilde{\bv}$ and define a new divergence free velocity field by
\bena
\bv=\tilde{\bv}+ \epsilon \nabla (\frac{p}{T})_t.
\eena
Namely, the modified reformulated system (\ref{The modified mode EQ form}) is recast in as follows
\ben\bea{l}
\begin{cases}
\tilde{\bv}_t +\bar{\bv} \cdot \nabla\tilde{\bv}=\nabla \cdot \sigma_e+ \sqrt{\frac{Pr}{Ra}}\Delta \tilde{\bv} -\nabla p + T\hat{\textbf{z}}, \\
\bv=\tilde{\bv}+ \epsilon \nabla (\frac{p}{T})_t,\\
\nabla \cdot \bv = 0,\\
\phi_t +\nabla \cdot (\phi \tilde{\bv}) =- \nabla\cdot\bM\nabla(\gamma_1\Delta\phi-2qq_\phi-2\gamma_2\phi),\\
T_t+\tilde{\bv} \cdot \nabla T = \frac{1}{C_A}\sigma_e:\nabla \tilde{\bv} +\frac{2}{C_A}\sqrt{\frac{Pr}{Ra}} \tilde{\bD}:\nabla \tilde{\bv} - \frac{1}{C_A}p\bI:\nabla \tilde{\bv} + \frac{1}{\sqrt{PrRa}}\nabla^2 T,\\
q_t=q_\phi\phi_t+q_ee_t,
\end{cases}
\eea
\label{Recast The modified mode EQ form}
\een
where $\sigma_e=T(-\gamma_1\nabla\phi\otimes\nabla\phi-s\bI+\frac{\delta S}{\delta \phi}\phi\bI), q_\phi=\frac{\partial q}{\partial \phi}, q_e=\frac{\partial q}{\partial e}, e=C_AT.$

For $\epsilon=O(\Delta t^k), k\geq 4$, we apply the Crank-Nicolson method to recast modified systems (\ref{Recast The modified mode EQ form}) to arrive at a second-order, semi-discrete scheme as  follows:
\begin{sch}[{\bf Semi-discrete entropy-production-rate-preserving scheme III}] Given $\tilde{\bv}^n$, $\phi^n$, $T^n$ and $p^n$, we update $\tilde{\bv}^{n+1}$, $\phi^{n+1}$, $T^{n+1}$ and $p^{n+1}$ as follows:
\ben\bea{l}
\textit{Step 1.}
\begin{cases}
\frac{1}{\Delta t}(\tilde{\bv}^{n+1}-\bv^{n})+\bar{\bv}^{n+\frac{1}{2}}\cdot \nabla\tilde{\bv}^{n+\frac{1}{2}}=\nabla\cdot\sigma_e^{n+\frac{1}{2}}+\sqrt{\frac{Pr}{Ra}}\Delta\tilde{\bv}^{n+\frac{1}{2}}
-\nabla p^{n+\frac{1}{2}}+T^{n+\frac{1}{2}}\hat{\textbf{z}},\\
\\
\frac{1}{\Delta t}(\bv^{n+1}-\tilde{\bv}^{n+1})=\frac{2\epsilon}{(\Delta t)^2}\nabla((\frac{p}{T})^{n+1}-(\frac{p}{T})^n)\Leftrightarrow\bv^{n+\frac{1}{2}}=\tilde{\bv}^{n+\frac{1}{2}}+ \epsilon \nabla (\frac{p}{T})^{n+\frac{1}{2}}_t,\\
\\
\nabla \cdot \bv^{n+1} = 0 \quad \Rightarrow ~\nabla^2 ((\frac{p}{T})^{n+1}-(\frac{p}{T})^n)=-\frac{\Delta t}{2 \epsilon}\nabla \cdot \tilde{\bv}^{n+1}, 
\end{cases}
\\where ~\tilde{\bv}^{n+1}\mid_{\partial \Omega}=0;
\\
\textit{Step 2.}
\begin{cases}
\frac{1}{\Delta t}(\phi^{n+1}-\phi^n)+\nabla \cdot (\phi^{n+\frac{1}{2}} \tilde{\bv}^{n+\frac{1}{2}}) =-\nabla\cdot\bM\nabla(\gamma_1\Delta\phi^{n+\frac{1}{2}}-2q^{n+\frac{1}{2}}\bar{q}_\phi^{n+\frac{1}{2}}-2\gamma_2\phi^{n+\frac{1}{2}}),\\
\\
\frac{1}{\Delta t}(T^{n+1}-T^n)+\tilde{\bv}^{n+\frac{1}{2}} \cdot \nabla T^{n+\frac{1}{2}} = \frac{1}{C_A}\sigma_e^{n+\frac{1}{2}}:\nabla \tilde{\bv}^{n+\frac{1}{2}}+\frac{2}{C_A}\sqrt{\frac{Pr}{Ra}}\tilde{\bD}^{n+\frac{1}{2}}:\nabla \tilde{\bv}^{n+\frac{1}{2}}\\
~~~~~~~~~~~~~~~~~~~~~~~~~~~~~~~~~~~~~~~~~~~-\frac{1}{C_A}p^{n+\frac{1}{2}}\bI:\nabla \tilde{\bv}^{n+\frac{1}{2}}+ \frac{1}{\sqrt{PrRa}}\nabla^2 T^{n+\frac{1}{2}}, \\
\\
\frac{1}{\Delta t}(q^{n+1}-q^n)=\bar{q}^{n+\frac{1}{2}}_\phi\frac{1}{\Delta t}(\phi^{n+1}-\phi^n)+\bar{q}^{n+\frac{1}{2}}_e\frac{1}{\Delta t}(e^{n+1}-e^n),
\end{cases}
\eea
\label{schemeIII}
\een
where
$\sigma_e^{n+\frac{1}{2}}=T^{n+\frac{1}{2}}(-\gamma_1\nabla\phi^{n+\frac{1}{2}}\otimes\nabla\phi^{n+\frac{1}{2}}-s^{n+\frac{1}{2}}\bI+\frac{\delta S}{\delta \phi}^{n+\frac{1}{2}}\phi^{n+\frac{1}{2}}\bI), e^{n+\frac{1}{2}}=C_AT^{n+\frac{1}{2}}.$
In addition, the corresponding physical boundary conditions as:
\ben
\left \{
\bea{l}
\tilde{\bv}^n|_{\partial \Omega} =0, ~ \bn \cdot \nabla p^n|_{\partial \Omega}=0, ~\bn \cdot \nabla \phi^n|_{\partial \Omega}=0, ~\bn \cdot \nabla \frac{\delta S}{\delta \phi}^n|_{\partial \Omega}=0, \\\\
T^n\mid_{upper}=T_b, ~T^n\mid_{lower}=T_a, ~\bn \cdot \nabla T^n\mid_{left}=0,
~\bn \cdot \nabla T^n\mid_{right}=0~(n=0,1,\cdots,N).
\eea\right.
\label{semi-discrete schemeIII BC}
\een
\end{sch}
Next, we present a theorem to show the semi-discrete entropy-production-rate-preserving scheme III also preserves the volume and the entropy production rate.

\begin{thm}
Given boundary conditions (\ref{semi-discrete schemeIII BC}), semi-discrete scheme III preserves the volume conservation law: $V^{n+1}=V^n$,
and the entropy production rate
\bena
\frac{\hat{S}^{n+1}-\hat{S}^n}{\Delta t}=\int_\Omega [\bM(\nabla\frac{\delta S}{\delta \phi}^{n+\frac{1}{2}})^2+2\sqrt{\frac{Pr}{Ra}}\frac{1}{T^{n+\frac{1}{2}}}\tilde{\bD}^{n+\frac{1}{2}}: \tilde{\bD}^{n+\frac{1}{2}}+ \frac{C_A}{\sqrt{PrRa}}\frac{(\nabla T^{n+\frac{1}{2}})^2}{(T^{n+\frac{1}{2}})^2}]d\bx\\
~~~~~~~~~~~+\frac{C_A}{\sqrt{PrRa}}\int_{\pr \Omega}  \bn \cdot \frac{ \nabla T^{n+\frac{1}{2}}}{T^{n+\frac{1}{2}}} d\ba
+\epsilon\int_{\pr \Omega}  \bn \cdot (\frac{p}{T})^{n+\frac{1}{2}} \nabla (\frac{p}{T})^{n+\frac{1}{2}}_t d\ba,
\eena
where $V^n=\int_\Omega \phi^{n}d\bx$ and the modified entropy
\bena
\hat{S}^n=S^n+\int_\Omega \frac{\epsilon}{2}|\nabla (\frac{p}{T})^{n}|^2d {\bx},\\
S^n=\int_\Omega[-|q^n|^2-\gamma_2|\phi^n|^2-\gamma_3|e^n|^2-\frac{\gamma_1}{2}|\nabla\phi^n|^2+C_0]d {\bx}.
\eena
\end{thm}

\noindent {\bf Proof.} Based on the definition of $V^n$, we readily prove $V^{n+1}=V^n$. Apply the phase field equation and temperature field equation in $Step~2$ of scheme (\ref{schemeIII}), we have
\bena
\frac{\hat{S}^{n+1}-\hat{S}^n}{\Delta t}=\frac{S^{n+1}-S^n}{\Delta t}+\int_\Omega [\frac{\epsilon}{2}(\nabla(\frac{p}{T})^{n+1}+\nabla(\frac{p}{T})^n)\frac{\nabla(\frac{p}{T})^{n+1}-\nabla(\frac{p}{T})^n}{\Delta t} ]d\bx,
\eena
where
\bena
\frac{S^{n+1}-S^n}{\Delta t}
=\int_\Omega [-(q^{n+1}+q^n)\frac{q^{n+1}-q^n}{\Delta t}-\gamma_2(\phi^{n+1}+\phi^n)\frac{\phi^{n+1}-\phi^n}{\Delta t}\\
 -\gamma_3(e^{n+1}+e^n)\frac{e^{n+1}-e^n}{\Delta t}-\frac{\gamma_1}{2}(\nabla\phi^{n+1}+\nabla\phi^n)\frac{\nabla\phi^{n+1}-\nabla\phi^n}{\Delta t} ]d\bx
 \\

=\int_\Omega [-2\delta_t \phi^{n+\frac{1}{2}} q^{n+\frac{1}{2}}\bar{q}_\phi^{n+\frac{1}{2}}+\delta_t \phi^{n+\frac{1}{2}} (\gamma_1\Delta\phi^{n+\frac{1}{2}})-2\gamma_2\phi^{n+\frac{1}{2}}(-\nabla\cdot\bM\nabla\frac{\delta S}{\delta \phi}^{n+\frac{1}{2}})\\

+2\gamma_2\phi^{n+\frac{1}{2}} \nabla \cdot(\phi^{n+\frac{1}{2}} \tilde{\bv}^{n+\frac{1}{2}})-2\delta_t e^{n+\frac{1}{2}} q^{n+\frac{1}{2}}\bar{q}_e^{n+\frac{1}{2}}-2\gamma_3e^{n+\frac{1}{2}} \delta_t e^{n+\frac{1}{2}}] d\bx\\

=\int_\Omega [\delta_t \phi^{n+\frac{1}{2}}\frac{\delta S}{\delta \phi}^{n + \frac{1}{2}}-2\gamma_2\phi^{n+\frac{1}{2}} (-\nabla\cdot\bM\nabla\frac{\delta S}{\delta \phi}^{n+\frac{1}{2}}-\delta_t \phi^{n+\frac{1}{2}})\\
+2\gamma_2\phi^{n+\frac{1}{2}}\nabla \cdot(\phi^{n+\frac{1}{2}}\tilde{\bv}^{n+\frac{1}{2}})+\delta_t e^{n+\frac{1}{2}} \frac{\delta S}{\delta e}^{n + \frac{1}{2}}] d\bx\\
=\int_\Omega (\delta_t \phi^{n+\frac{1}{2}} \frac{\delta S}{\delta \phi}^{n + \frac{1}{2}}+\delta_t e^{n+\frac{1}{2}} \frac{\delta S}{\delta e}^{n + \frac{1}{2}}) d\bx.
\label{Theorem semi-discrete eq1 recast form}
\eena

Notice that
\bena
\nabla s^{n+\frac{1}{2}}=\frac{\delta S}{\delta \phi}^{n+\frac{1}{2}}\nabla \phi^{n+\frac{1}{2}}+\frac{\delta S}{\delta e}^{n+\frac{1}{2}}\nabla e^{n+\frac{1}{2}}+\nabla\cdot(\frac{\partial s}{\partial \nabla \phi}^{n+\frac{1}{2}}\nabla \phi^{n+\frac{1}{2}}).
\label{especial condition s1}
\eena
We have
\bena
\int_\Omega (\delta_t \phi^{n+\frac{1}{2}} \frac{\delta S}{\delta \phi}^{n + \frac{1}{2}}+\delta_t e^{n+\frac{1}{2}} \frac{\delta S}{\delta e}^{n + \frac{1}{2}}) d\bx\\

=\int_\Omega [(-\nabla\cdot\bM\nabla\frac{\delta S}{\delta \phi}^{n + \frac{1}{2}}-\nabla\cdot(\phi^{n+\frac{1}{2}}\tilde{\bv}^{n+\frac{1}{2}}))\frac{\delta S}{\delta \phi}^{n + \frac{1}{2}}- (\frac{p}{T})^{n+\frac{1}{2}}\nabla\cdot \tilde{\bv}^{n+\frac{1}{2}}\\
+\big((\sigma_e^{n+\frac{1}{2}}+2\sqrt{\frac{Pr}{Ra}}\tilde{\bD}^{n+\frac{1}{2}}):\nabla \tilde{\bv}^{n+\frac{1}{2}}+\frac{C_A}{\sqrt{PrRa}}\nabla^2 T^{n+\frac{1}{2}}-\tilde{\bv}^{n+\frac{1}{2}} \cdot \nabla e^{n+\frac{1}{2}}\big)\frac{\delta S}{\delta e}^{n + \frac{1}{2}}] d\bx\\

=\int_\Omega [\bM(\nabla\frac{\delta S}{\delta \phi}^{n+\frac{1}{2}})^2+2\sqrt{\frac{Pr}{Ra}}\frac{1}{T^{n+\frac{1}{2}}}\tilde{\bD}^{n+\frac{1}{2}}:\nabla \tilde{\bv}^{n+\frac{1}{2}}+ \frac{C_A}{\sqrt{PrRa}}\frac{(\nabla T^{n+\frac{1}{2}})^2}{(T^{n+\frac{1}{2}})^2}
+\frac{1}{T^{n+\frac{1}{2}}}\sigma_e^{n+\frac{1}{2}}:\nabla\tilde{\bv}^{n+\frac{1}{2}}\\
+ (\frac{p}{T})^{n+\frac{1}{2}}\epsilon \nabla^2 (\frac{p}{T})^{n+\frac{1}{2}}_t
+[s^{n+\frac{1}{2}}\bI-\frac{\partial s}{\partial \nabla \phi}^{n+\frac{1}{2}}\nabla {\phi}^{n+\frac{1}{2}}-(\frac{\delta S}{\delta \phi}^{n+\frac{1}{2}}\phi^{n+\frac{1}{2}})\bI)]:\nabla\tilde{\bv}^{n+\frac{1}{2}} ]d\bx\\
-\int_{\pr \Omega}  \bn \cdot (s^{n+\frac{1}{2}}\bI-\frac{\partial s}{\partial \nabla \phi}^{n+\frac{1}{2}}\nabla {\phi}^{n+\frac{1}{2}})\cdot \tilde{\bv}^{n+\frac{1}{2}} d\ba
+\frac{C_A}{\sqrt{PrRa}}\int_{\pr \Omega}  \bn \cdot \frac{ \nabla T^{n+\frac{1}{2}}}{T^{n+\frac{1}{2}}} d\ba+\epsilon\int_{\pr \Omega}  \bn \cdot (\frac{p}{T})^{n+\frac{1}{2}}\epsilon \nabla (\frac{p}{T})^{n+\frac{1}{2}}_t d\ba\\
=\int_\Omega [\bM(\nabla\frac{\delta S}{\delta \phi}^{n+\frac{1}{2}})^2+2\sqrt{\frac{Pr}{Ra}}\frac{1}{T^{n+\frac{1}{2}}}\tilde{\bD}^{n+\frac{1}{2}}: \tilde{\bD}^{n+\frac{1}{2}}+ \frac{C_A}{\sqrt{PrRa}}\frac{(\nabla T^{n+\frac{1}{2}})^2}{(T^{n+\frac{1}{2}})^2}]d\bx\\
-\int_\Omega[\frac{\epsilon}{2}(\nabla(\frac{p}{T})^{n+1}+\nabla(\frac{p}{T})^n)\frac{\nabla(\frac{p}{T})^{n+1}-\nabla(\frac{p}{T})^n}{\Delta t} ]d\bx\\
+\frac{C_A}{\sqrt{PrRa}}\int_{\pr \Omega}  \bn \cdot \frac{ \nabla T^{n+\frac{1}{2}}}{T^{n+\frac{1}{2}}} d\ba
+\epsilon\int_{\pr \Omega}  \bn \cdot (\frac{p}{T})^{n+\frac{1}{2}} \nabla (\frac{p}{T})^{n+\frac{1}{2}}_t d\ba,
\label{conclusion of S recast form}
\eena

Then, we obtain
\bena
\frac{\hat{S}^{n+1}-\hat{S}^n}{\Delta t}=\int_\Omega [\bM(\nabla\frac{\delta S}{\delta \phi}^{n+\frac{1}{2}})^2+2\sqrt{\frac{Pr}{Ra}}\frac{1}{T^{n+\frac{1}{2}}}\tilde{\bD}^{n+\frac{1}{2}}: \tilde{\bD}^{n+\frac{1}{2}}+ \frac{C_A}{\sqrt{PrRa}}\frac{(\nabla T^{n+\frac{1}{2}})^2}{(T^{n+\frac{1}{2}})^2}]d\bx\\
~~~~~~~~~~~+\frac{C_A}{\sqrt{PrRa}}\int_{\pr \Omega}  \bn \cdot \frac{ \nabla T^{n+\frac{1}{2}}}{T^{n+\frac{1}{2}}} d\ba
+\epsilon\int_{\pr \Omega}  \bn \cdot (\frac{p}{T})^{n+\frac{1}{2}} \nabla (\frac{p}{T})^{n+\frac{1}{2}}_t d\ba.
\eena
This completes the proof.
\rem{When the entire boundary is isothermal, i.e. $\bn \cdot \nabla T\mid_{\partial \Omega}=0$ and $\bn\cdot \nabla p|_{\partial \Omega}=0$, semi-discrete scheme-III yields a positive entropy production rate:
\bena
\frac{\hat{S}^{n+1}-\hat{S}^n}{\Delta t}=\int_\Omega [\bM(\nabla\frac{\delta S}{\delta \phi}^{n+\frac{1}{2}})^2+2\sqrt{\frac{Pr}{Ra}}\frac{1}{T^{n+\frac{1}{2}}}\tilde{\bD}^{n+\frac{1}{2}}: \tilde{\bD}^{n+\frac{1}{2}}+ \frac{C_A}{\sqrt{PrRa}}\frac{(\nabla T^{n+\frac{1}{2}})^2}{(T^{n+\frac{1}{2}})^2}]d\bx\geq 0.
\eena

The numerical scheme requires $\epsilon=O(\Delta t^k), k \geq 4$ in order to be second order for the incompressible model. This makes the forcing term in the Poisson equation for $p$ large. In the numerical simulations presented next, we use semi-discrete scheme II since it's easier to implement and more robust. As the result, we will only discuss the spatial discretization for scheme II and present the corresponding fully discrete scheme.  }

\rem{If we use BDF2 scheme to discretize the reformulated models, we can obtain a set of second order entropy-rate-preserving numerical algorithms with a slightly modified entropy functionals. We will not elaborate the detail in this paper. Interested readers are referred to \cite{Zhao&L&W&Y,Zhao2016Numerical}. }

\subsection{ Fully-discrete algorithms}

\noindent \indent We discretize semi-discrete scheme II spatially on staggered grid using finite difference methods to arrive at a fully discrete scheme. Then, we show that the fully discrete numerical scheme preserve the properties of the entropy-production-rate and the volume of each fluid phase under suitable boundary conditions. We remark that when the same spatial discretization method is applied to other numerical schemes, the resulting fully discrete schemes share the same properties as those of scheme II. We adopt the notations defined in \cite{SunShouwen} and supply them in the Appendix for completeness.

\begin{sch}[{\bf Fully discrete entropy-production-rate-preserving scheme II}] Given $u^{n}$, $v^{n}$, $\phi^n$, $T^n$ and $p^n$, we update $\tilde{u}^{n+1}$, $\tilde{v}^{n+1}$, $u^{n+1}$, $v^{n+1}$, $\phi^{n+1}$, $T^{n+1}$ and $p^{n+1}$ as follows:
\ben\bea{l}
\textit{Step 1.}\\
\begin{cases}
\{\frac{1}{\Delta t}(\tilde{u}^{n+1}-u^{n})+\frac{1}{2}(\bar{u}^{n+\frac{1}{2}}D_x(a_x\tilde{u}^{n+\frac{1}{2}})+A_x(d_x(\tilde{u}^{n+\frac{1}{2}}\bar{u}^{n+\frac{1}{2}}))\\
+a_y(A_x\bar{v}^{n+\frac{1}{2}}
D_y\tilde{u}^{n+\frac{1}{2}})+d_y(A_y\tilde{u}^{n+\frac{1}{2}}A_x\bar{v}^{n+\frac{1}{2}}))
\\=-D_x p^{n}+
[d_xA_x((-T^{n+\frac{1}{2}}\gamma_1)(d_x\phi^{n+\frac{1}{2}})^2)+d_yA_y((-T^{n+\frac{1}{2}}\gamma_1)(d_x\phi^{n+\frac{1}{2}})(a_yD_ya_x\phi^{n+\frac{1}{2}}))]\\
+\sqrt{\frac{Pr}{Ra}}\Delta_h\tilde{u}^{n+\frac{1}{2}}\}|_{i+\frac{1}{2},j}, i=1,2,\ldots,N_x-1,j=1,2,\ldots,N_y.\\

\\
\{\frac{1}{\Delta t}(\tilde{v}^{n+1}-v^{n})+\frac{1}{2}(a_x(A_y\bar{u}^{n+\frac{1}{2}}
D_x\tilde{v}^{n+\frac{1}{2}})+d_x(A_y\bar{u}^{n+\frac{1}{2}}A_x\tilde{v}^{n+\frac{1}{2}})\\+\bar{v}^{n+\frac{1}{2}}D_y(a_y\tilde{v}^{n+\frac{1}{2}})+A_y(d_y(\tilde{v}^{n+\frac{1}{2}}\bar{v}^{n+\frac{1}{2}})))
\\=-D_y p^{n}+[d_yA_y((-T^{n+\frac{1}{2}}\gamma_1)(d_y\phi^{n+\frac{1}{2}})^2)+d_xA_x((-T^{n+\frac{1}{2}}\gamma_1)(d_y\phi^{n+\frac{1}{2}})(a_xD_xa_y\phi^{n+\frac{1}{2}}))]\\
+\sqrt{\frac{Pr}{Ra}}\Delta_h\tilde{v}^{n+\frac{1}{2}}+A_yT^{n+\frac{1}{2}}\}|_{i,j+\frac{1}{2}}, i=1,2,\ldots,N_x,j=1,2,\ldots,N_y-1.\\
\\
\{\frac{1}{\Delta t}(u^{n+1}-\tilde{u}^{n+1})=-\frac{1}{2}D_x(p^{n+1}-p^n)\}|_{i+\frac{1}{2},j}, i=1,2,\ldots,N_x-1,j=1,2,\ldots,N_y.\\
\\
\{\frac{1}{\Delta t}(v^{n+1}-\tilde{v}^{n+1})=-\frac{1}{2}D_y(p^{n+1}-p^n)\}|_{i,j+\frac{1}{2}}, i=1,2,\ldots,N_x,j=1,2,\ldots,N_y-1.\\
\\
\{d_xu^{n+\frac{1}{2}}+d_yv^{n+\frac{1}{2}} = 0\}|_{i,j},i=1,2,\ldots,N_x,j=1,2,\ldots,N_y.  \\
\end{cases}
\\where ~\tilde{u}^{n+1} \in \mathcal{\varepsilon}^{ew0}_{x\times {y}}, \tilde{v}^{n+1} \in \mathcal{\varepsilon}^{ns0}_{x\times {y}};
\\
\textit{Step 2.}\\
\begin{cases}
\{\delta_t\phi^{n+\frac{1}{2}} +d_x(A_x\bar{\phi}^{n+\frac{1}{2}}u^{n+\frac{1}{2}})+d_y(A_y\bar{\phi}^{n+\frac{1}{2}}v^{n+\frac{1}{2}})\\

=-\nabla_h\cdot\bM\nabla_h(\gamma_1\Delta_h\phi^{n+\frac{1}{2}}-2q^{n+\frac{1}{2}}\bar{q}_\phi^{n+\frac{1}{2}}-2\gamma_2\phi^{n+\frac{1}{2}})\}|_{i,j},
i=1,2,\ldots,N_x, j=1,2,\ldots, N_y.\\
\\
\{\delta_tT^{n+\frac{1}{2}}+d_x(A_x\bar{T}^{n+\frac{1}{2}}u^{n+\frac{1}{2}})+d_y(A_y\bar{T}^{n+\frac{1}{2}}v^{n+\frac{1}{2}})\\

=\frac{1}{C_A}(-T^{n+\frac{1}{2}}\gamma_1)[d_x(A_x\phi^{n+\frac{1}{2}})^2d_xu^{n+\frac{1}{2}}
+d_x(A_x\phi^{n+\frac{1}{2}})d_y(A_y\phi^{n+\frac{1}{2}})(a_yD_ya_xu^{n+\frac{1}{2}}
+a_xD_xa_yv^{n+\frac{1}{2}})\\
+d_y(A_y\phi^{n+\frac{1}{2}})^2d_yv^{n+\frac{1}{2}}]
+\frac{2}{C_A}\sqrt{\frac{Pr}{Ra}}[(d_xu^{n+\frac{1}{2}})^2+\frac{1}{2}(a_xD_xa_yv^{n+\frac{1}{2}}+a_yD_ya_xu^{n+\frac{1}{2}})^2+(d_yv^{n+\frac{1}{2}})^2]\\
+\frac{1}{\sqrt{PrRa}} \nabla_h\cdot\nabla_h T^{n+\frac{1}{2}})\}|_{i,j},
i=1,2,\ldots,N_x, j=1,2,\ldots, N_y.\\
\\
\{\delta_tq^{n+\frac{1}{2}}=\bar{q}^{n+\frac{1}{2}}_\phi\delta_t\phi^{n+\frac{1}{2}}+\bar{q}^{n+\frac{1}{2}}_e\delta_te^{n+\frac{1}{2}}\}|_{i,j},i=1,2,\ldots,N_x, j=1,2,\ldots, N_y.
\\
\end{cases}
\eea
\label{Fully Scheme}
\een
along with boundary conditions as follows:
\bena
u^n, ~D_x\phi^n, ~D_x\frac{\delta S}{\delta \phi}^n, ~D_xT^n \in \mathcal{\varepsilon}^{ew0}_{x\times {y}},~~~
v^n, ~D_y\phi^n, ~D_y\frac{\delta S}{\delta \phi}^n \in \mathcal{\varepsilon}^{ns0}_{x\times {y}}. 
\label{Fully-discrete BC}
\eena
\end{sch}

Next, we prove that the entropy production rate and the volume of each fluid phase are preserved at the fully discrete level.

\begin{thm}
Given boundary conditions (\ref{Fully-discrete BC}), the fully discrete scheme preserves the volume conservation law: $
V^{n+1}_h=V^n_h,
$
where
$
V^n_h=(\phi^{n},1)_2,
$
and the entropy production rate
\bena
\frac{S^{n+1}_{h}-S^n_{h}}{\Delta t}
=(\frac{1}{T^{n+\frac{1}{2}}}, 2\sqrt{\frac{Pr}{Ra}}[(d_xu^{n+\frac{1}{2}})^2+\frac{1}{2}(a_xD_xa_yv^{n+\frac{1}{2}}+a_yD_ya_xu^{n+\frac{1}{2}})^2+(d_yv^{n+\frac{1}{2}})^2])_2\\
+(\bM\nabla_h\vparl{S}{\phi}^{n+\frac{1}{2}},\nabla_h\vparl{S}{\phi}^{n+\frac{1}{2}})_2
+\frac{C_A}{\sqrt{PrRa}}\sum_{i=1}^{N_x-1}\sum_{j=1}^{N_y-1}(\frac{(T_{i+1,j}^{n+\frac{1}{2}}-T_{i,j}^{n+\frac{1}{2}})^2}{T_{i,j}^{n+\frac{1}{2}}T_{i+1,j}^{n+\frac{1}{2}}}+\frac{(T_{i,j+1}^{n+\frac{1}{2}}-T_{i,j}^{n+\frac{1}{2}})^2}{T_{i,j}^{n+\frac{1}{2}}T_{i,j+1}^{n+\frac{1}{2}}})\\
+\frac{C_A}{\sqrt{PrRa}}h(\sum_{i=1}^{N_x}(\delta_yT_{i,N_y+1/2}^{n+\frac{1}{2}}(\frac{1}{T})_{i,N_y}^{n+\frac{1}{2}}-\delta_yT_{i,1/2}^{n+\frac{1}{2}}(\frac{1}{T})_{i,1}^{n+\frac{1}{2}})),
\eena
where
\ben
S^n_h=-\|q^n\|^2_2-\gamma_2\|\phi^n\|^2_2-\gamma_3\|e^n\|^2_2-\frac{\gamma_1}{2}\|\nabla_h\phi^n\|^2_2+(C_0,1)_2.
\een
\end{thm}
\noindent {\bf Proof.}
We denote
\bena
\frac{\delta S}{\delta \phi}^{n+\frac{1}{2}}=\gamma_1\Delta_h\phi^{n+\frac{1}{2}}-2q^{n+\frac{1}{2}}\bar{q}_\phi^{n+\frac{1}{2}}-2\gamma_2\phi^{n+\frac{1}{2}},\\
\frac{\delta S}{\delta e}^{n+\frac{1}{2}}=-2q^{n+\frac{1}{2}}\bar{q}_e^{n+\frac{1}{2}}-2\gamma_3e^{n+\frac{1}{2}}.
\label{Approximate formula}
\eena
Then, we obtain
\bena
\frac{S_h^{n+1}-S_h^n}{\Delta t}

=-\gamma_2(\phi^{n+1}+\phi^n,\frac{\phi^{n+1}-\phi^n}{\Delta t})_2-(q^{n+1}+q^n,\frac{q^{n+1}-q^n}{\Delta t})_2\\

-\frac{\gamma_1}{2}({[D_x\phi^{n+\frac{1}{2}},D_x\delta_t\phi^{n+\frac{1}{2}}]}_{ew}
+{[D_y\phi^{n+\frac{1}{2}},D_y\delta_t\phi^{n+\frac{1}{2}}]}_{ns})-\gamma_3(e^{n+1}+e^n,\frac{e^{n+1}-e^n}{\Delta t})_2\\

=(\delta_t \phi^{n+\frac{1}{2}},\frac{\delta S}{\delta \phi}^{n +\frac{1}{2}})_2+(\delta_t e^{n+\frac{1}{2}},\frac{\delta S}{\delta e}^{n + \frac{1}{2}})_2.
\label{Fully discrete eq}
\eena
Apply the phase field equation and temperature field equation in the $Step~2$ part of Fully discrete EQ Scheme (\ref{Fully Scheme}), we have
\bena
(\delta_t \phi^{n+\frac{1}{2}},\frac{\delta S}{\delta \phi}^{n +\frac{1}{2}})_2=\\
(\bM\nabla_h\vparl{S}{\phi}^{n+\frac{1}{2}},\nabla_h\vparl{S}{\phi}^{n+\frac{1}{2}})_2
-(d_x(A_x\bar{\phi}^{n+\frac{1}{2}}u^{n+\frac{1}{2}})
+d_y(A_y\bar{\phi}^{n+\frac{1}{2}}v^{n+\frac{1}{2}}),\frac{\delta S}{\delta \phi}^{n + \frac{1}{2}})_2,
\label{Fully discrete eq1}
\eena
and
\bena
(\delta_t e^{n+\frac{1}{2}},\frac{\delta S}{\delta e}^{n + \frac{1}{2}})_2
=\\(\frac{1}{T^{n+\frac{1}{2}}},(-T^{n+\frac{1}{2}}\gamma_1)[d_x(A_x\phi^{n+\frac{1}{2}})d_x(A_x\bar{\phi}^{n+\frac{1}{2}})d_xu^{n+\frac{1}{2}}
+d_x(A_x\phi^{n+\frac{1}{2}})d_y(A_y\bar{\phi}^{n+\frac{1}{2}})a_yD_ya_xu^{n+\frac{1}{2}}\\
+d_x(A_x\bar{\phi}^{n+\frac{1}{2}})d_y(A_y\phi^{n+\frac{1}{2}})a_xD_xa_yv^{n+\frac{1}{2}}
+d_y(A_y\bar{\phi}^{n+\frac{1}{2}})d_y(A_y\phi^{n+\frac{1}{2}})d_yv^{n+\frac{1}{2}}])_2\\+(\frac{1}{T^{n+\frac{1}{2}}},2\sqrt{\frac{Pr}{Ra}}[(d_xu^{n+\frac{1}{2}})^2
+\frac{1}{2}(a_xD_xa_yv^{n+\frac{1}{2}}+a_yD_ya_xu^{n+\frac{1}{2}})^2+(d_yv^{n+\frac{1}{2}})^2])_2\\
+(\frac{1}{T^{n+\frac{1}{2}}}, \frac{C_A}{\sqrt{PrRa}} \nabla_h\cdot\nabla_h T^{n+\frac{1}{2}})_2-(d_x(A_x\bar{T}^{n+\frac{1}{2}}u^{n+\frac{1}{2}})+d_y(A_y\bar{T}^{n+\frac{1}{2}}v^{n+\frac{1}{2}}),\frac{1}{T^{n+\frac{1}{2}}})_2.
\label{Fully discrete eq2}
\eena

We calculate
\bena
(d_x(A_x\bar{\phi}^{n+\frac{1}{2}}u^{n+\frac{1}{2}})+d_y(A_y\bar{\phi}^{n+\frac{1}{2}}v^{n+\frac{1}{2}}),\frac{\delta S}{\delta \phi}^{n +\frac{1}{2}})_2\\
+(d_x(A_x\bar{T}^{n+\frac{1}{2}}u^{n+\frac{1}{2}})+d_y(A_y\bar{T}^{n+\frac{1}{2}}v^{n+\frac{1}{2}}),\frac{1}{T^{n+\frac{1}{2}}})_2\\
=(a_x(D_x\bar{\phi}^{n+\frac{1}{2}}u^{n+\frac{1}{2}})+a_y(D_y\bar{\phi}^{n+\frac{1}{2}}v^{n+\frac{1}{2}}),\frac{\delta S}{\delta \phi}^{n +\frac{1}{2}})_2\\
+(a_x(D_x\bar{T}^{n+\frac{1}{2}}u^{n+\frac{1}{2}})+a_y(D_y\bar{T}^{n+\frac{1}{2}}v^{n+\frac{1}{2}}),\frac{1}{T^{n+\frac{1}{2}}})_2\\
=[A_x\frac{\delta S}{\delta \phi}^{n+\frac{1}{2}}D_x\bar{\phi}^{n+\frac{1}{2}}+A_x\frac{1}{T^{n+\frac{1}{2}}}D_x\bar{T}^{n+\frac{1}{2}},u^{n+\frac{1}{2}}]_{ew}\\
+[A_y\frac{\delta S}{\delta \phi}^{n+\frac{1}{2}}D_y\bar{\phi}^{n+\frac{1}{2}}+A_y\frac{1}{T^{n+\frac{1}{2}}}D_y\bar{T}^{n+\frac{1}{2}},v^{n+\frac{1}{2}}]_{ns},
\eena
where the following equalities are used
\bena
d_x(A_x\phi u)=\phi d_xu+a_x(D_x\phi u),~~~d_y(A_y\phi v)=\phi d_yv+a_y(D_y\phi v),\\
d_x(A_xe u)=e d_xu+a_x(D_xe u),~~~d_y(A_ye v)=e d_yv+a_y(D_ye v),~~~d_xu+d_yv=0,\\
(a_x(D_x\phi u),\frac{\delta S}{\delta \phi})_2=[A_x\frac{\delta S}{\delta \phi}D_x\phi,u]_{ew},~~~
(a_y(D_y\phi v),\frac{\delta S}{\delta \phi})_2=[A_y\frac{\delta S}{\delta \phi}D_y\phi,v]_{ns},\\
(a_x(D_xT u),\frac{1}{T})_2=[A_x\frac{1}{T}D_xT,u]_{ew},~~~
(a_y(D_yT v),\frac{1}{T})_2=[A_y\frac{1}{T}D_yT,v]_{ns}.
\eena

Recalling (\ref{especial condition s11}), we can derive
\bena
D_xs_{*}^{n+\frac{1}{2}}=A_x\frac{\delta S}{\delta \phi}^{n+\frac{1}{2}}D_x\bar{\phi}^{n+\frac{1}{2}}+A_x\frac{1}{T^{n+\frac{1}{2}}}D_x\bar{T}^{n+\frac{1}{2}}
-\gamma_1[D_x(d_x(A_x\phi^{n+\frac{1}{2}})d_x(A_x\bar{\phi}^{n+\frac{1}{2}}))\\
~~~~~~~~~~~~~~+D_ya_xa_y(d_x(A_x\bar{\phi}^{n+\frac{1}{2}})d_y(A_y\phi^{n+\frac{1}{2}}))],
\\
D_ys_{*}^{n+\frac{1}{2}}=A_y\frac{\delta S}{\delta \phi}^{n+\frac{1}{2}}D_y\bar{\phi}^{n+\frac{1}{2}}+A_y\frac{1}{T^{n+\frac{1}{2}}}D_y\bar{T}^{n+\frac{1}{2}}
-\gamma_1[D_y(d_y(A_y\bar{\phi}^{n+\frac{1}{2}})d_y(A_y\phi^{n+\frac{1}{2}}))\\
~~~~~~~~~~~~~~+D_xa_ya_x(d_x(A_x\phi^{n+\frac{1}{2}})d_y(A_y\bar{\phi}^{n+\frac{1}{2}}))].
\eena

Then, we arrive at
\bena
[A_x\frac{\delta S}{\delta \phi}^{n+\frac{1}{2}}D_x\bar{\phi}^{n+\frac{1}{2}}+A_x\frac{1}{T^{n+\frac{1}{2}}}D_x\bar{T}^{n+\frac{1}{2}},u^{n+\frac{1}{2}}]_{ew}\\
+[A_y\frac{\delta S}{\delta \phi}^{n+\frac{1}{2}}D_y\bar{\phi}^{n+\frac{1}{2}}+A_y\frac{1}{T^{n+\frac{1}{2}}}D_y\bar{T}^{n+\frac{1}{2}},v^{n+\frac{1}{2}}]_{ns}\\
=-\gamma_1
[(d_x(A_x\phi^{n+\frac{1}{2}})d_x(A_x\bar{\phi}^{n+\frac{1}{2}}),d_xu^{n+\frac{1}{2}})_2+(d_y(A_y\phi^{n+\frac{1}{2}})d_y(A_y\bar{\phi}^{n+\frac{1}{2}}),d_yv^{n+\frac{1}{2}})_2\\
+(d_x(A_x\bar{\phi}^{n+\frac{1}{2}})d_y(A_y\phi^{n+\frac{1}{2}}),a_yD_ya_xu^{n+\frac{1}{2}})_2
+(d_x(A_x\phi^{n+\frac{1}{2}})d_y(A_y\bar{\phi}^{n+\frac{1}{2}}),a_xD_xa_yv^{n+\frac{1}{2}})_2],
\eena
where we have used equalities
\bena
[D_xs_{\ast},u]_{ew}=-(s_{\ast},d_xu)_2, ~~[D_ys_{\ast},v]_{ns}=-(s_{\ast},d_yv)_2,~~d_xu+d_yv=0\\

[D_ya_xa_y\phi,u]_{ew}=-(\phi,a_yD_ya_xu)_2, ~~[D_xa_ya_x\phi,v]_{ns}=-(\phi,a_xD_xa_yv)_2.
\eena
From the above results, we obtain
\bena
(\delta_t \phi^{n+\frac{1}{2}},\frac{\delta S}{\delta \phi}^{n +\frac{1}{2}})_2+(\delta_t e^{n+\frac{1}{2}},\frac{\delta S}{\delta e}^{n + \frac{1}{2}})_2\\
=(\bM\nabla_h\vparl{S}{\phi}^{n+\frac{1}{2}},\nabla_h\vparl{S}{\phi}^{n+\frac{1}{2}})_2+(\frac{1}{T^{n+\frac{1}{2}}},2\sqrt{\frac{Pr}{Ra}}[(d_xu^{n+\frac{1}{2}})^2
+\frac{1}{2}(a_xD_xa_yv^{n+\frac{1}{2}}+a_yD_ya_xu^{n+\frac{1}{2}})^2\\+(d_yv^{n+\frac{1}{2}})^2])_2
+(\frac{1}{T^{n+\frac{1}{2}}}, \frac{C_A}{\sqrt{PrRa}}\Delta_hT^{n+\frac{1}{2}})_2.
\label{spatial result1}
\eena
Next, with inhomogeneous boundary conditions (\ref{inhomogeneous BC1}), we have
\bena
(\frac{1}{T^{n+\frac{1}{2}}}, \frac{C_A}{\sqrt{PrRa}}\Delta_hT^{n+\frac{1}{2}})_2
=\frac{C_A}{\sqrt{PrRa}}\sum_{i=1}^{N_x-1}\sum_{j=1}^{N_y-1}(\frac{(T_{i+1,j}^{n+\frac{1}{2}}-T_{i,j}^{n+\frac{1}{2}})^2}{T_{i,j}^{n+\frac{1}{2}}T_{i+1,j}^{n+\frac{1}{2}}}+\frac{(T_{i,j+1}^{n+\frac{1}{2}}-T_{i,j}^{n+\frac{1}{2}})^2}{T_{i,j}^{n+\frac{1}{2}}T_{i,j+1}^{n+\frac{1}{2}}})\\
+\frac{C_A}{\sqrt{PrRa}}h(\sum_{i=1}^{N_x}(\delta_yT_{i,N_y+1/2}^{n+\frac{1}{2}}(\frac{1}{T})_{i,N_y}^{n+\frac{1}{2}}-\delta_yT_{i,1/2}^{n+\frac{1}{2}}(\frac{1}{T})_{i,1}^{n+\frac{1}{2}})).
\label{spatial result2}
\eena
Finally, it follows from (\ref{spatial result1}) and (\ref{spatial result2}) that
\bena
\frac{S^{n+1}_{h}-S^n_{h}}{\Delta t}=(\frac{1}{T^{n+\frac{1}{2}}}, 2\sqrt{\frac{Pr}{Ra}}[(d_xu^{n+\frac{1}{2}})^2+\frac{1}{2}(a_xD_xa_yv^{n+\frac{1}{2}}+a_yD_ya_xu^{n+\frac{1}{2}})^2+(d_yv^{n+\frac{1}{2}})^2])_2\\
+(\bM\nabla_h\vparl{S}{\phi}^{n+\frac{1}{2}},\nabla_h\vparl{S}{\phi}^{n+\frac{1}{2}})_2
+\frac{C_A}{\sqrt{PrRa}}\sum_{i=1}^{N_x-1}\sum_{j=1}^{N_y-1}(\frac{(T_{i+1,j}^{n+\frac{1}{2}}-T_{i,j}^{n+\frac{1}{2}})^2}{T_{i,j}^{n+\frac{1}{2}}T_{i+1,j}^{n+\frac{1}{2}}}+\frac{(T_{i,j+1}^{n+\frac{1}{2}}-T_{i,j}^{n+\frac{1}{2}})^2}{T_{i,j}^{n+\frac{1}{2}}T_{i,j+1}^{n+\frac{1}{2}}})\\
+\frac{C_A}{\sqrt{PrRa}}h(\sum_{i=1}^{N_x}(\delta_yT_{i,N_y+1/2}^{n+\frac{1}{2}}(\frac{1}{T})_{i,N_y}^{n+\frac{1}{2}}-\delta_yT_{i,1/2}^{n+\frac{1}{2}}(\frac{1}{T})_{i,1}^{n+\frac{1}{2}})).
\eena
Namely, the Fully discrete scheme preserves the entropy production rate.

Meanwhile, computing the discrete inner product of the phase field equation in  $Step~2$ part of (\ref{Fully Scheme}) with constant function 1 follows from (\ref{Fully-discrete BC}) and Lemma 5.2 that we achieve $\frac{V^{n+1}_h-V^n_h}{\Delta t}=0$, i.e., the Fully discrete scheme preserves the volume conservation.

\rem{Under adiabatic boundary conditions  (\ref{Fully-discrete BC}) except for the temperature condition being chosen as
\bena
D_xT^n \in \mathcal{\varepsilon}^{ew0}_{x\times {y}}, D_yT^n \in \mathcal{\varepsilon}^{ns0}_{x\times {y}}.
\label{inhomogeneous BC111}
\eena
 the Fully discrete EQ Scheme yields a positive entropy production rate
\bena
\frac{S^{n+1}_{h}-S^n_{h}}{\Delta t}=(\frac{1}{T^{n+\frac{1}{2}}}, 2\sqrt{\frac{Pr}{Ra}}[(d_xu^{n+\frac{1}{2}})^2+\frac{1}{2}(a_xD_xa_yv^{n+\frac{1}{2}}+a_yD_ya_xu^{n+\frac{1}{2}})^2+(d_yv^{n+\frac{1}{2}})^2])_2\\
+(\bM\nabla_h\vparl{S}{\phi}^{n+\frac{1}{2}},\nabla_h\vparl{S}{\phi}^{n+\frac{1}{2}})_2
+\frac{C_A}{\sqrt{PrRa}}\sum_{i=1}^{N_x-1}\sum_{j=1}^{N_y-1}(\frac{(T_{i+1,j}^{n+\frac{1}{2}}-T_{i,j}^{n+\frac{1}{2}})^2}{T_{i,j}^{n+\frac{1}{2}}T_{i+1,j}^{n+\frac{1}{2}}}+\frac{(T_{i,j+1}^{n+\frac{1}{2}}-T_{i,j}^{n+\frac{1}{2}})^2}{T_{i,j}^{n+\frac{1}{2}}T_{i,j+1}^{n+\frac{1}{2}}})\geq0.
\eena
Hence, the structure-preserving numerical scheme is second order in both spatial and temporal satisfying thermodynamically consistency at the discrete level for proper boundary conditions.}

{The resulting system of algebraic equations is solved using an iterative method assisted by the fast Fourier transform discussed in detail in \cite{SunShouwen}.  }

\subsection{Adaptive time-stepping strategy}

\noindent \indent Phase field dynamics is often dominated by multiple time scales determined by the temporal energy landscape. The hydrodynamics and thermal effect all have their own time scales. The EQ reformulation also introduce its own time scale which is often quite restrictively small. For the evolution of the hydrodynamic variables involve multiple time scales, it's better off for us to use adaptive time steps to ensure solution accuracy and computational efficiency. In the simulations presented next, we implement a time adaptive strategy for time step $\Delta t$ following the work of Zhang and Qiao in \cite{zhang_qiao_2012}:
\bena
\Delta t=max(\Delta t_{min},\frac{\Delta t_{max}}{\sqrt{1+\beta|S'(t)|^2}}),
\label{the time step sizes}
\eena
where $\beta$ is a constant and used to adjust the level of adaptivity, $S(t)$ is the entropy functional of this model. In the implementation of the adaptive time-stepping method, we use the preset smallest time step $\Delta t_{min}$ in the first step, and the following step size is determined by (\ref{the time step sizes}).

\section{Numerical Results and Discussion}

\noindent \indent In this section, we first validate the  convergence rate of the proposed fully discrete scheme (\ref{Fully Scheme}) through  mesh refine tests and show its entropy-production-rate and volume preserving property in simulating some thermally induced hydrodynamical phenomena in a two immiscible viscous fluid system.
In the numerical experiments, we use the initial condition for velocity $\bv$ as follows
\bena
 u(x,y,0)= 0, \quad v(x,y,0)= 0, ~~(x,y)\in\Omega,
 \label{initial conditions of velocity}
\eena
and the model parameter values
\bena
~~~~L_x=L_y=1, ~C_A=1, ~T_M=0.05, ~T_a=1, \\
T_b=0,~\gamma_1=10^{-3}, ~\gamma_2=1, ~\gamma_3=10^{-1}, ~C_0=10^3.
\label{parameter values}
\eena

\subsection{Mesh refinement test}

\noindent \indent In order to conduct a mesh refinement test, we calculate errors by taking the difference between results obtained from successive coarse steps and those of adjacent finer steps. Consequently, we conduct mesh refinement tests for the above numerical fully scheme to demonstrate its second order accuracy numerically. The remaining parameter values are selected as follows
\bena
M=10^{-4}, Pr=10^{2}, Ra=10^{5}.
\label{parameter values1}
\eena
In addition to (\ref{initial conditions of velocity}), the initial state of the temperature field and phase field are given respectively by
\begin {equation}
T(x,y,0)=\left\{
\begin{array}{lr}
T_a,& 0\leq x\leq L_x, y=0,\\
-(T_a-T_b)y+T_a,&  0< x< L_x, 0< y< L_y,\\
T_b,& 0\leq x\leq L_x, y=L_y,
\end{array}
\right.
\label{initial conditions of temperature}
\end{equation}
\begin {equation}
\phi(x,y,0)=\left\{
\begin{array}{lr}
1,& r_1\leq0.2-\delta ~~$or$~~ r_2\leq0.2-\delta,\\
\text{tanh}(\frac{0.2-r_1}{\delta}),& 0.2-\delta<r_1\leq0.2+\delta,\\
\text{tanh}(\frac{0.2-r_2}{\delta}),& 0.2-\delta<r_2\leq0.2+\delta,\\
0,& \text {other},
\end{array}
\right.
\label{initial conditions of phase}
\end{equation}
where $r_1=\sqrt{(x-0.3+\delta)^2+(y-0.5)^2},r_2=\sqrt{(x-0.7+\delta)^2+(y-0.5)^2}$ and $\delta=0.01$.

\begin{figure*}
\centering
\subfigure[Temporal convergence test for $\phi$ and $T$]{
\includegraphics[width=0.475\linewidth]{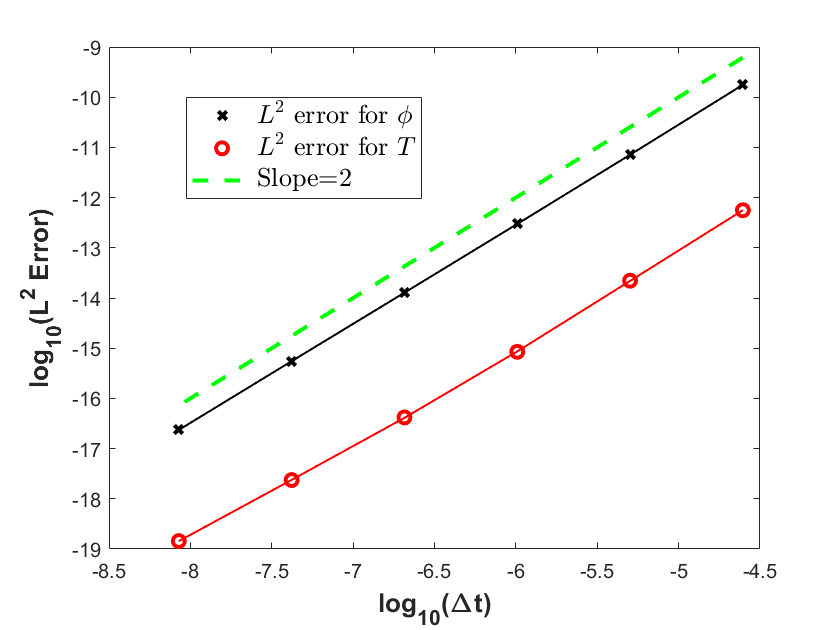}}
\subfigure[Temporal convergence test for $u$ and $v$]{
\includegraphics[width=0.475\linewidth]{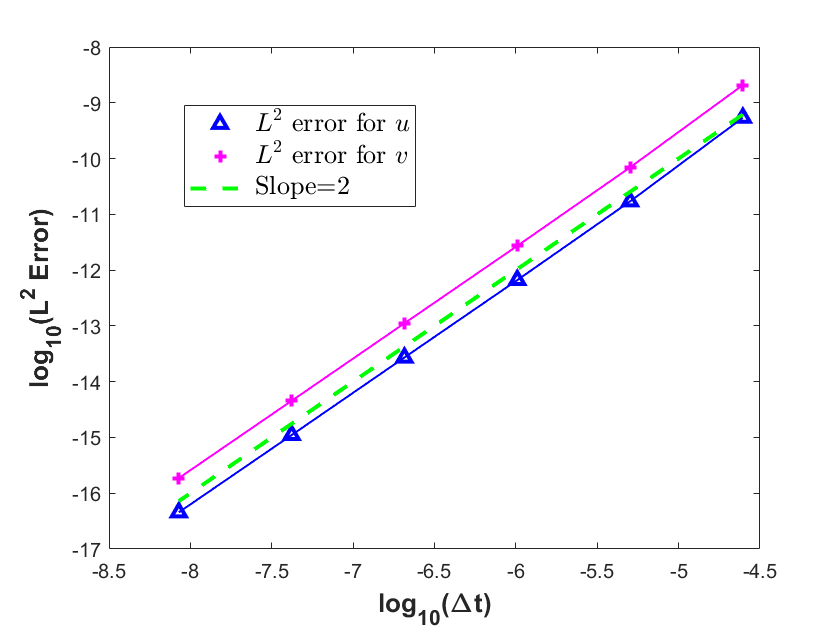}}
\caption{Convergence test in time of scheme (\ref{Fully Scheme}). (a): The error in $L_2$ norm of $\phi$ and $T$. (b): The error in $L_2$ norm of $u$ and $v$. In this simulation, we set the spatial meshes number as $N_x =N_y=64$. The results of (a) and (b) confirm second-order convergence rates in time for all variables, indicating a high level of accuracy in the numerical simulations.}
\label{Figure_convergence_Time}
\end{figure*}

In time step refinement tests, we choose the spatial meshes number $N_x=N_y=64$ and time step $\Delta t=10^{-2}\times\frac{1}{2^{k-1}}, k=1,2,3,\ldots$, respectively. Moreover, we compute the errors at time $t=0.1$, measured in $L^2$ norms of differences of quantity $\phi$, $T$ and $u$, $v$ between consecutive mesh sizes, respectively. The  results are summarized in Figure \ref{Figure_convergence_Time} (a)(b), where the second-order convergence rate in time is demonstrated clearly.

\begin{figure*}
\centering
\subfigure[Spatial convergence test for $\phi$ and $T$]{
\includegraphics[width=0.475\linewidth]{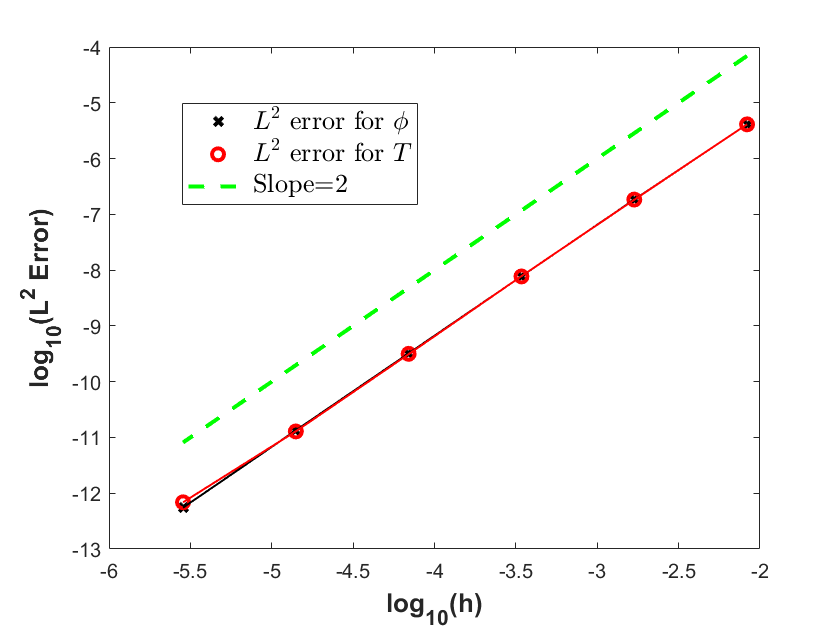}}
\subfigure[Spatial convergence test for $u$ and $v$]{
\includegraphics[width=0.475\linewidth]{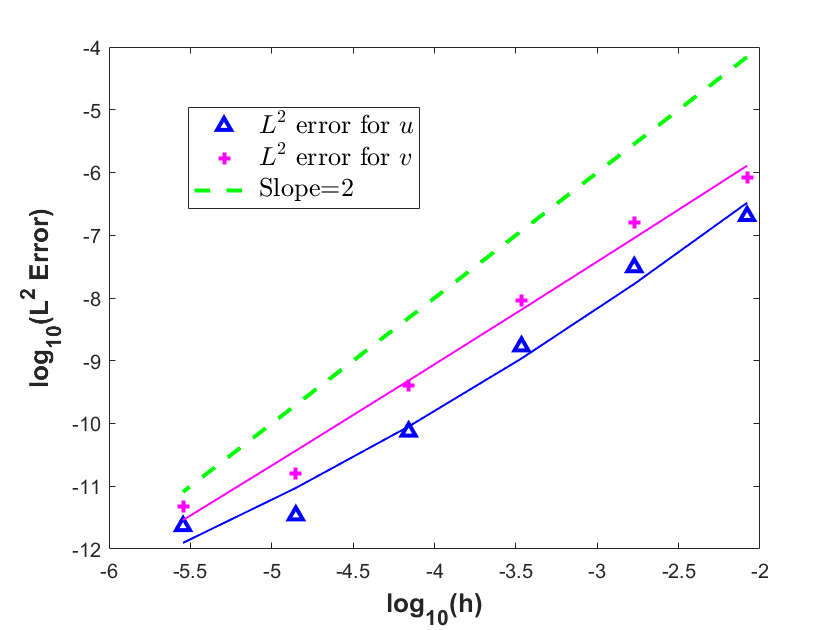}}
\caption{Convergence test in space of scheme (\ref{Fully Scheme}). (a): The error in $L_2$ norm of $\phi$ and $T$. (b): The error in $L_2$ norm of $u$ and $v$. In this simulation, we set the spatial meshes number as $N_x =N_y=64$. The results of (a) and (b) confirm second-order convergence rates in space for all variables, also indicating a high level of accuracy in the numerical simulations.}
\label{Figure_convergence_space}
\end{figure*}


To test the spatial convergence rate, we adopt identical parameter values as those specified in the preceding section and fix the time step size, and use another set of initial conditions
\bena
~~~T(x,y,0)=\frac{1}{2}\sin(\pi x)\sin(\pi y),\\
\phi(x,y,0)=\frac{1}{2}+\frac{1}{2}\cos(\pi x)\cos(\pi y), (x,y)\in\Omega.
\eena

A time step of $\Delta t=1.0\times10^{-2}$ and spatial mesh sizes $N_x=N_y=8\times2^k, k=0,1,2,3,\ldots$ are employed, respectively. We calculate $L^2$ norms of differences of $\phi$, $T$ and $u$, $v$ between consecutive mesh sizes as the error at time $t=1.5$, respectively. The mesh refinement test results are summarized in Figure \ref{Figure_convergence_space} (a)(b), where the second-order convergence rate is clearly established in space.

Next, we use the developed code to study the Rayleigh-B\'{e}nard convection and dynamics of a pair of merging drops in a binary immiscible viscous fluid confined in a rectangular domain and driven by the competing  temperature gradient, gravity and the interfacial force collectively.

\subsection{Rayleigh-B\'{e}nard convection in an immiscible binary viscus fluid}

\begin{figure}[htbp]
  \centering
  \includegraphics[width=.6 \textwidth]{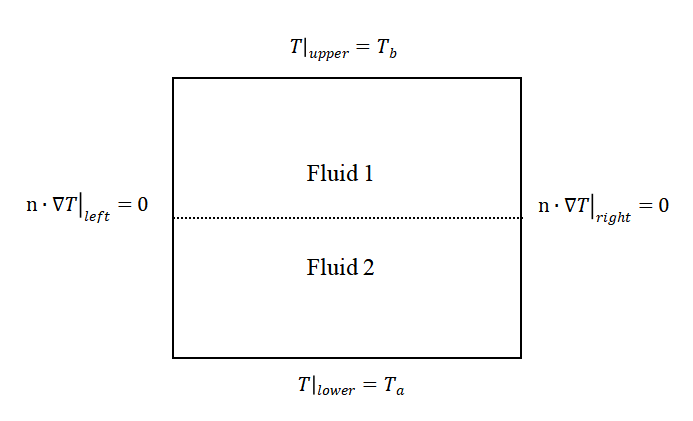}
  \caption{Schematics of the physical domain and  the temperature boundary condition. The velocity boundary condition is zero. } \label{Fig5.1}
  \label{Fig1}
\end{figure}

\noindent \indent In this numerical simulation, we consider the Rayleigh-B\'{e}nard convection of two-layered, superimposed viscous fluids confined in a rectangular domain subject to a transverse temperature gradient at their interface. Initially, the immiscible fluids are placed one on top of the other with a flat interface. The phase variables have homogeneous Neumann boundary conditions while the velocity has homogeneous Dirichlet boundary conditions.
We allow heat exchanges with the outside to maintain a constant temperature at the top and bottom boundaries while the left and right boundary conditions for the temperature are set as homogeneous Neumann, i.e., adiabatic (see Figure \ref{Fig1}), and the velocity is zero.

\begin{figure}[htbp]
  \centering
  \subfigure{
\includegraphics[width=0.32\linewidth]{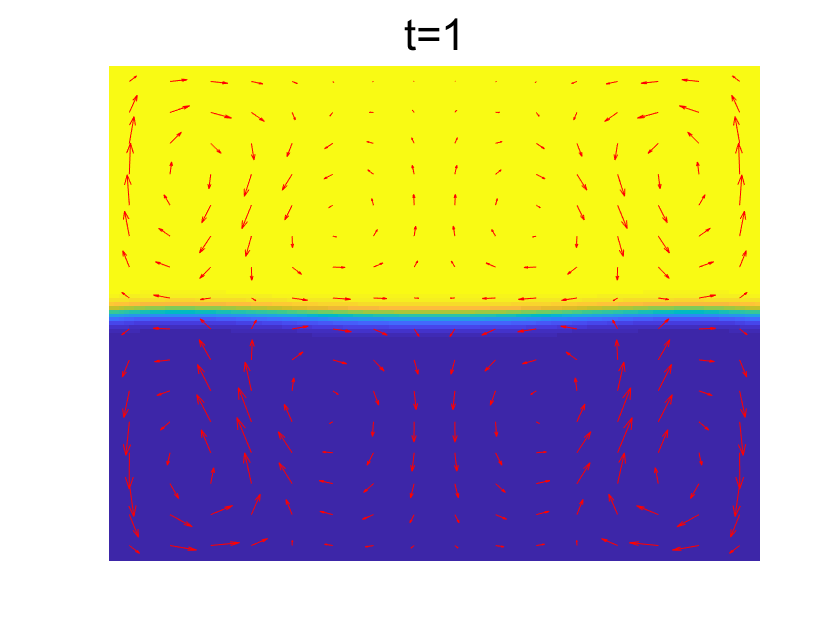}}
\subfigure{
\includegraphics[width=0.32\linewidth]{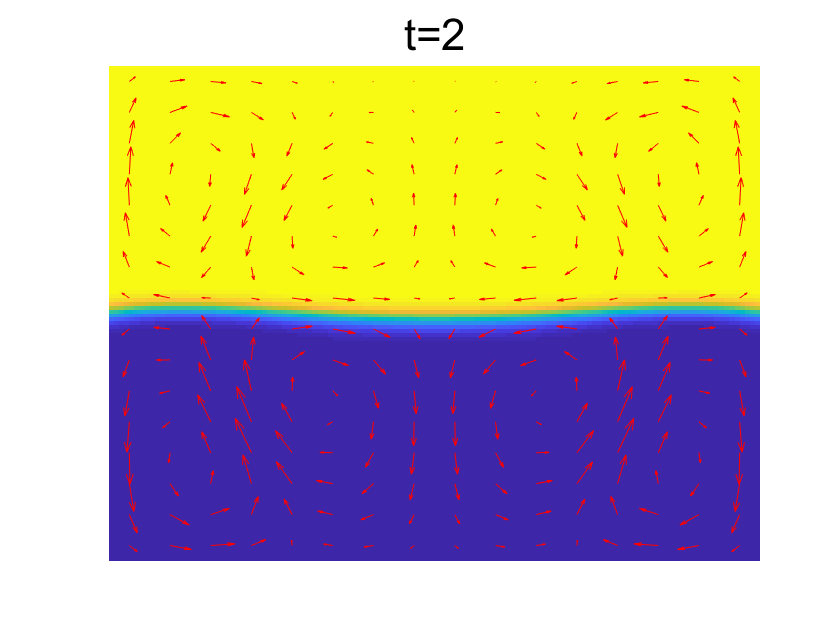}}
\subfigure{
\includegraphics[width=0.32\linewidth]{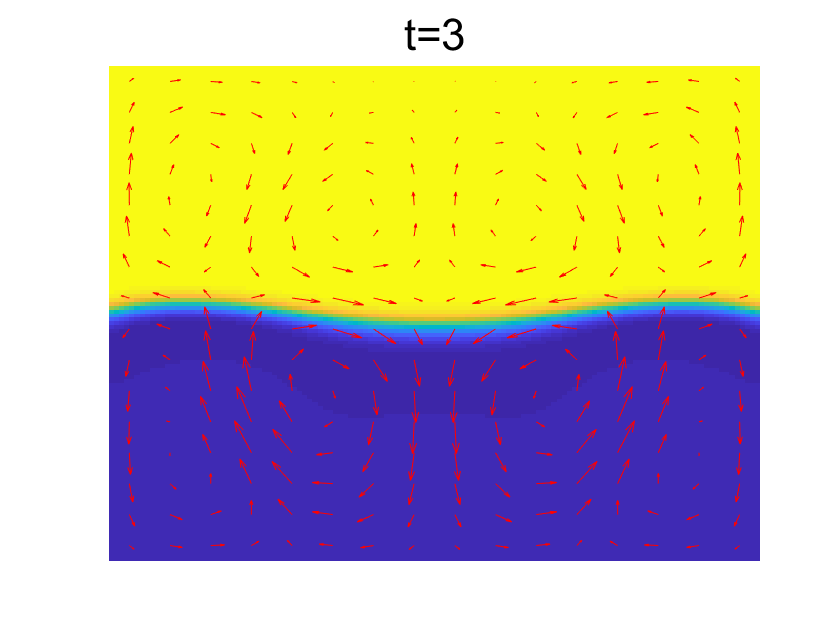}}
\caption{Snapshots of the phase field and velocity field at $t=1, 2, 3$, respectively. The results shows, at the beginning, there are four small roll cells on the upper and lower sides near the interface, respectively. The results are the same as those in Ref.\cite{QingmingChang}.}
  \label{Fig5.3combineref}
\end{figure}

At first, we would like to  compare the simulation of our model  with existing experimental numerical results to validate the new model. In the simulation, initial conditions of the velocity field and the temperature field are given by (\ref{initial conditions of velocity}) and (\ref{initial conditions of temperature}), respectively, and the initial condition of the phase variable is given by
\bena
\phi(x,y,0)=\frac{1}{2}+\frac{1}{2}\tanh(\frac{y-0.5}{\epsilon}), 0 \leq y \leq L_y,
\label{the initial phase field1}
\eena

where $\epsilon$ represents the thickness of the diffuse interface. Next, we establish parameter values $\epsilon=0.02$, $Ra=8.0\times10^{4}$ and $Pr=7.1$, other model parameter values are the same as in (\ref{parameter values}). In the simulation, we use  $N_x=N_y=128$ mesh points in space, $L_x=L_y=2$, the max and min  adaptive time step $\Delta t_{max}=1$, $\Delta t_{min}=1.0\times10^{-2}$, respectively, and solve the initial-boundary value problem up to $t=3$.
From the numerical result depicted in Figure \ref{Fig5.3combineref}, we observe that at the beginning of the simulation, four roll cells above and four below the interface form in the velocity field and the evolution of fluid convection is primarily caused by the thermally induced buoyancy force. The interface deformation is mainly because of the thermal induced fluid flow. This numerical result is consistent with the results in \cite{QingmingChang}.

\begin{figure}[htbp]
  \centering
  \subfigure{
\includegraphics[width=0.32\linewidth]{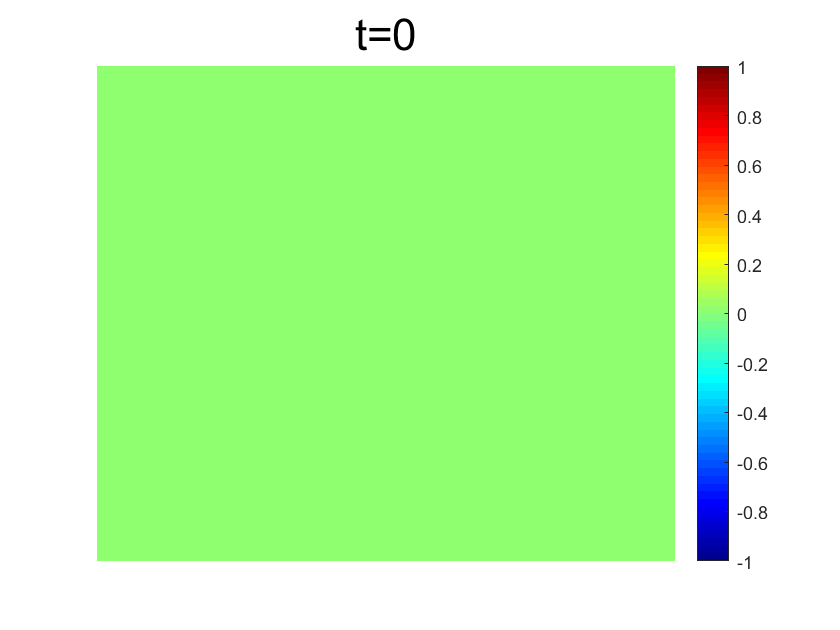}}
\subfigure{
\includegraphics[width=0.32\linewidth]{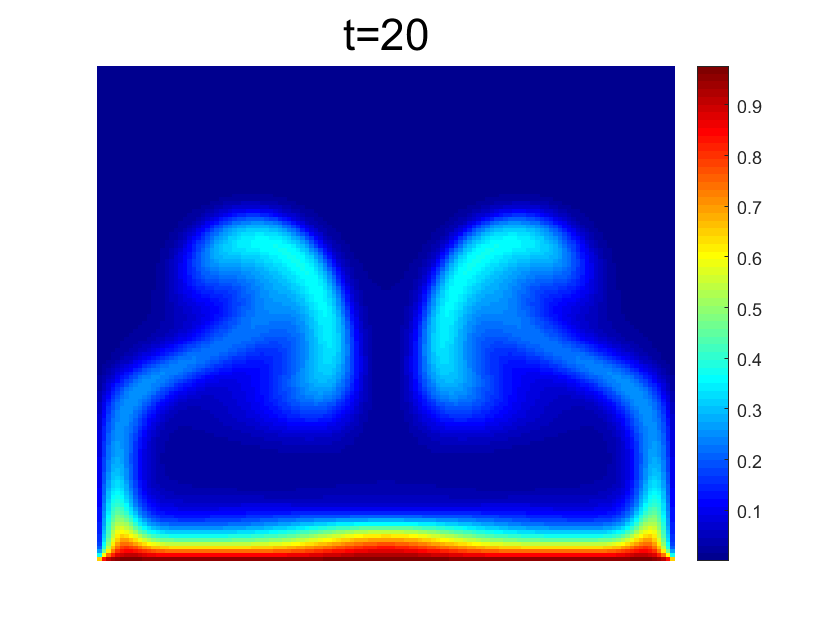}}
\subfigure{
\includegraphics[width=0.32\linewidth]{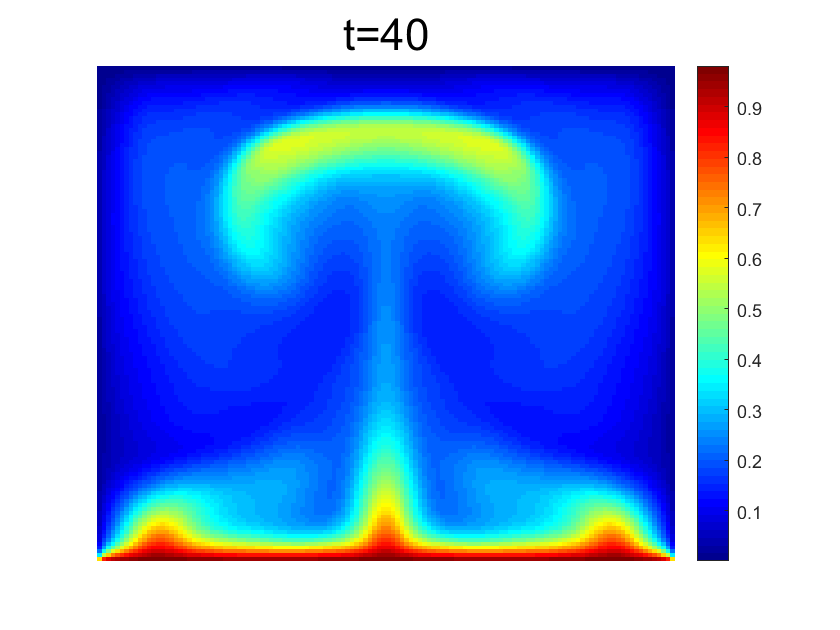}}
\subfigure{
\includegraphics[width=0.32\linewidth]{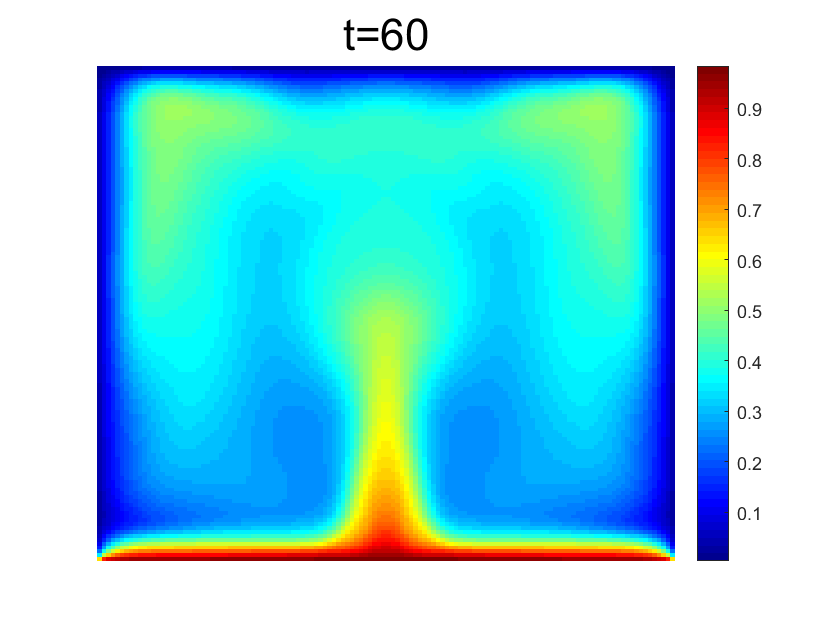}}
\subfigure{
\includegraphics[width=0.32\linewidth]{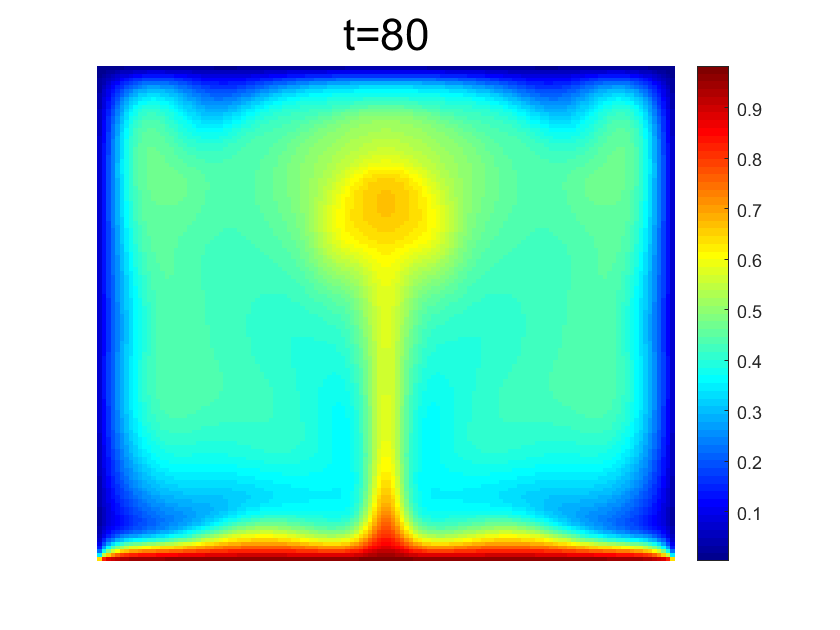}}
\subfigure{
\includegraphics[width=0.32\linewidth]{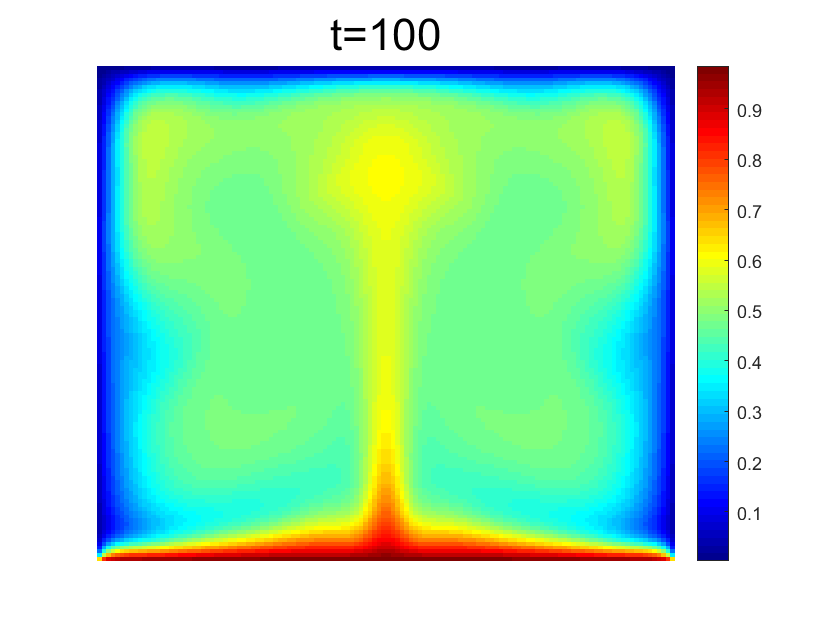}}
\subfigure{
\includegraphics[width=0.32\linewidth]{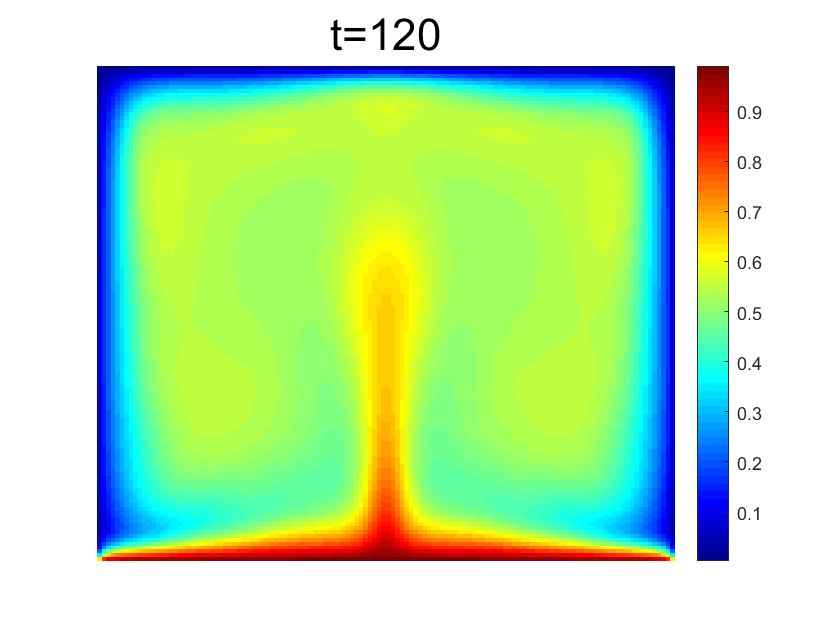}}
\subfigure{
\includegraphics[width=0.32\linewidth]{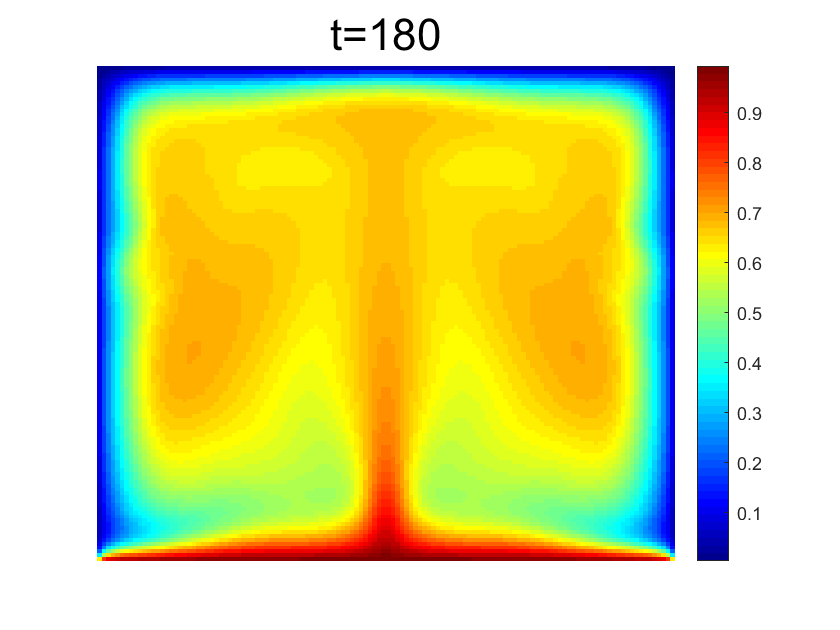}}
\subfigure{
\includegraphics[width=0.32\linewidth]{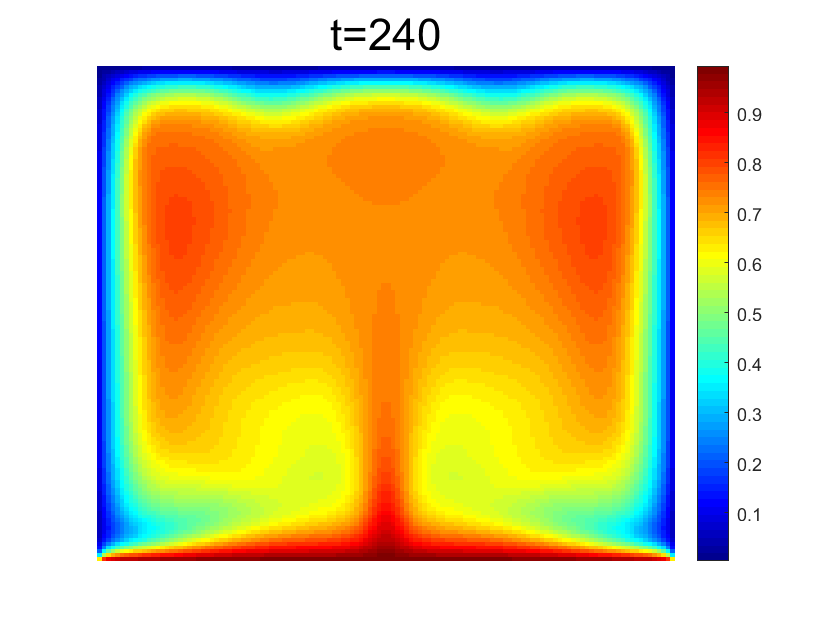}}
  \caption{Snapshots of the temperature field at $t=0, 20, 40, 60, 80, 100, 120, 180, 240$, respectively. The computational is heated up in the middle while the cool areas are concentrated near the top and the insulated lateral boundaries. }
  \label{Fig5.2}
\end{figure}

\begin{figure}[htbp]
  \centering
  \subfigure{
\includegraphics[width=0.32\linewidth]{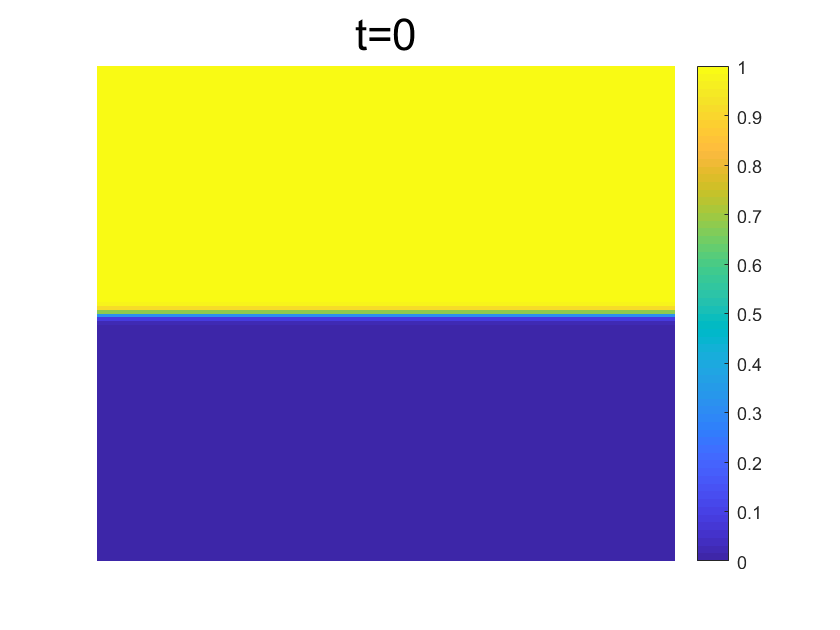}}
\subfigure{
\includegraphics[width=0.32\linewidth]{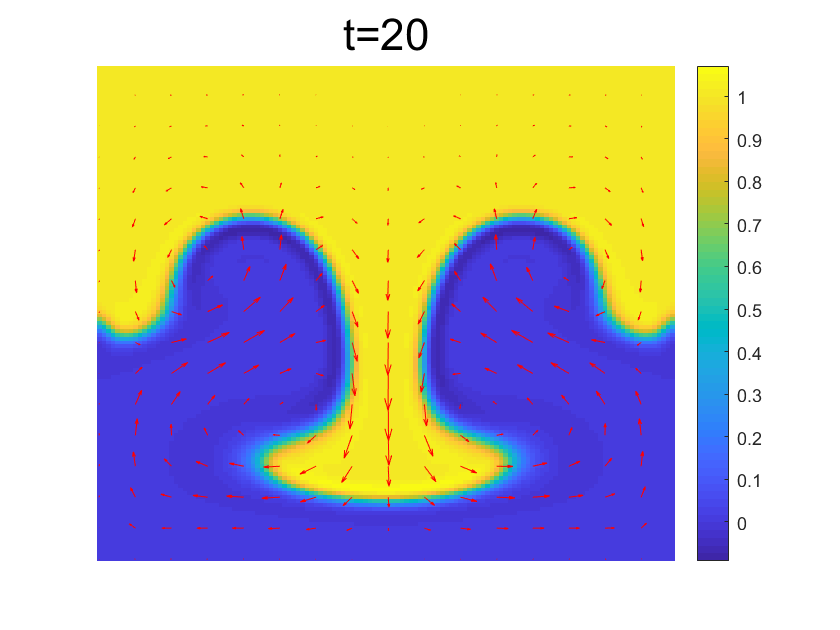}}
\subfigure{
\includegraphics[width=0.32\linewidth]{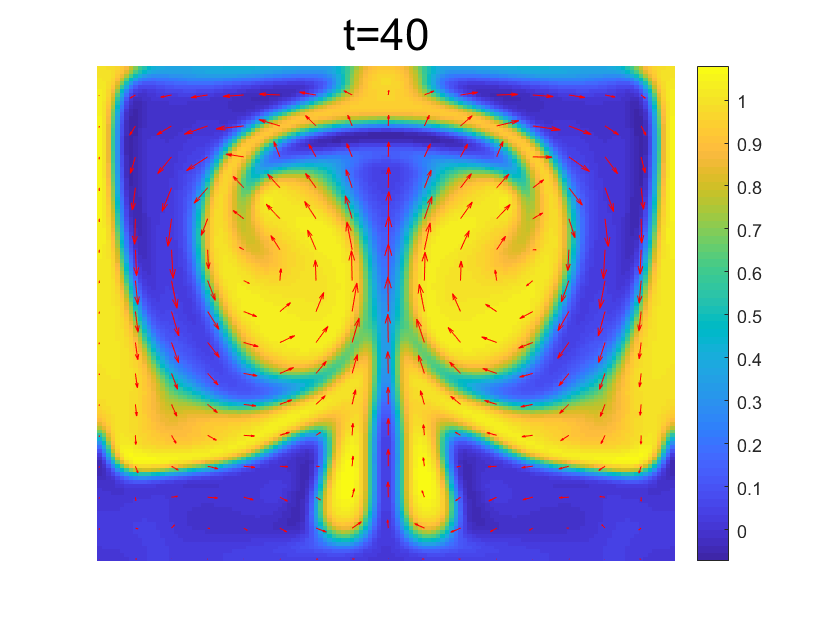}}
\subfigure{
\includegraphics[width=0.32\linewidth]{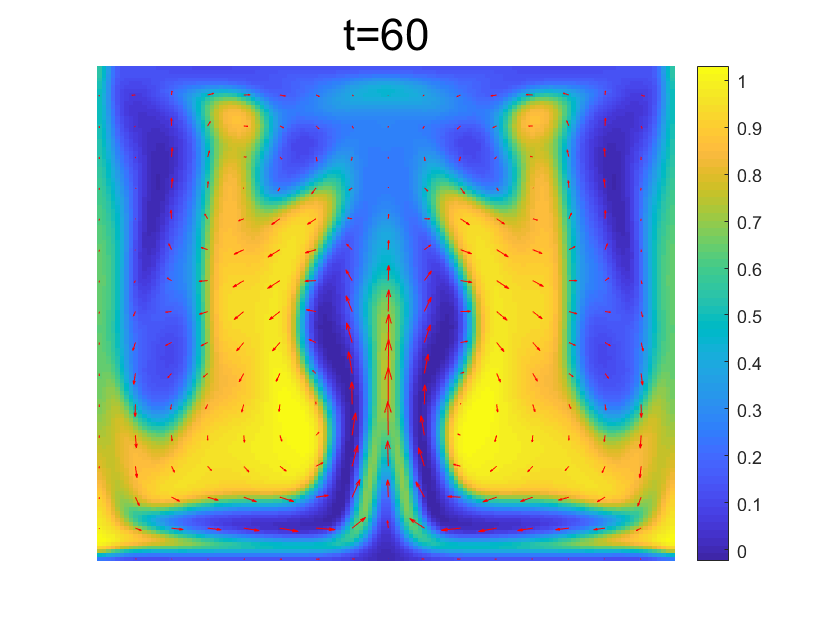}}
\subfigure{
\includegraphics[width=0.32\linewidth]{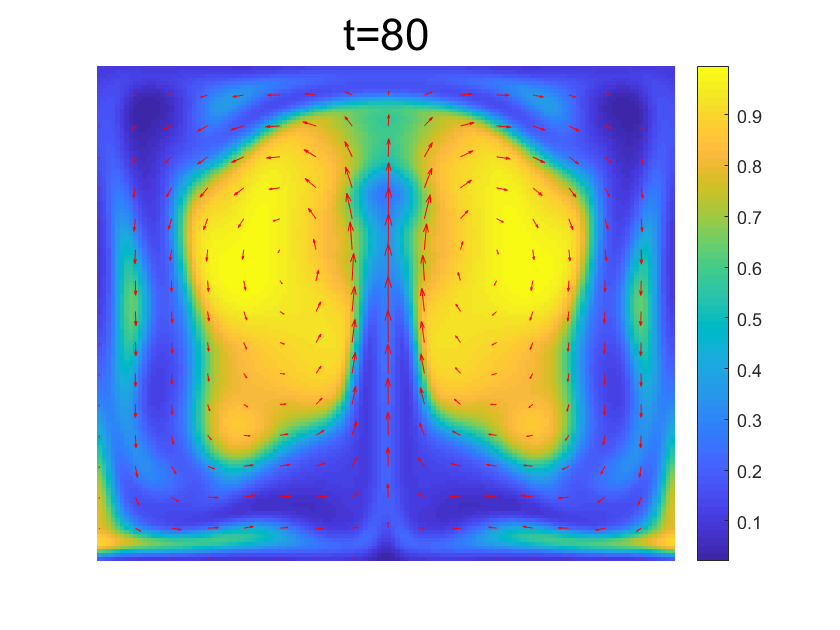}}
\subfigure{
\includegraphics[width=0.32\linewidth]{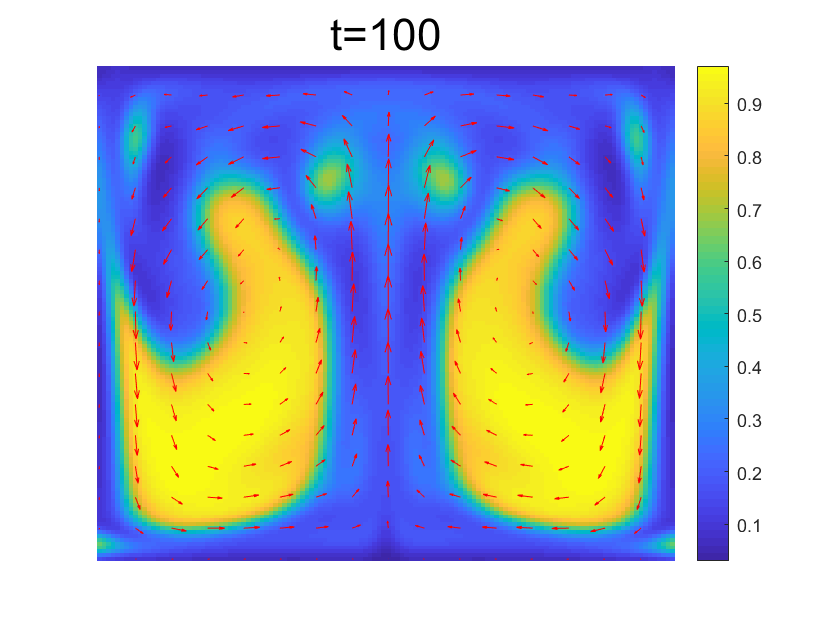}}
\subfigure{
\includegraphics[width=0.32\linewidth]{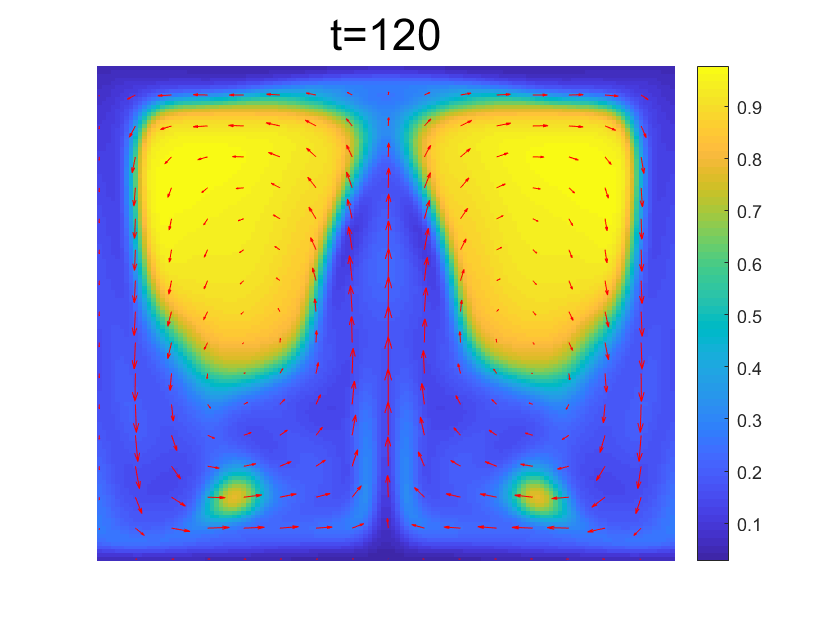}}
\subfigure{
\includegraphics[width=0.32\linewidth]{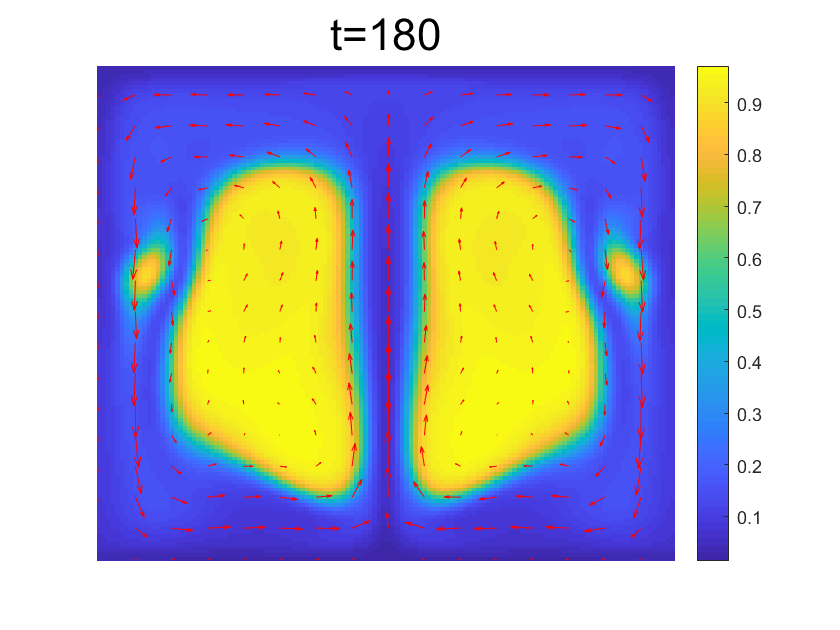}}
\subfigure{
\includegraphics[width=0.32\linewidth]{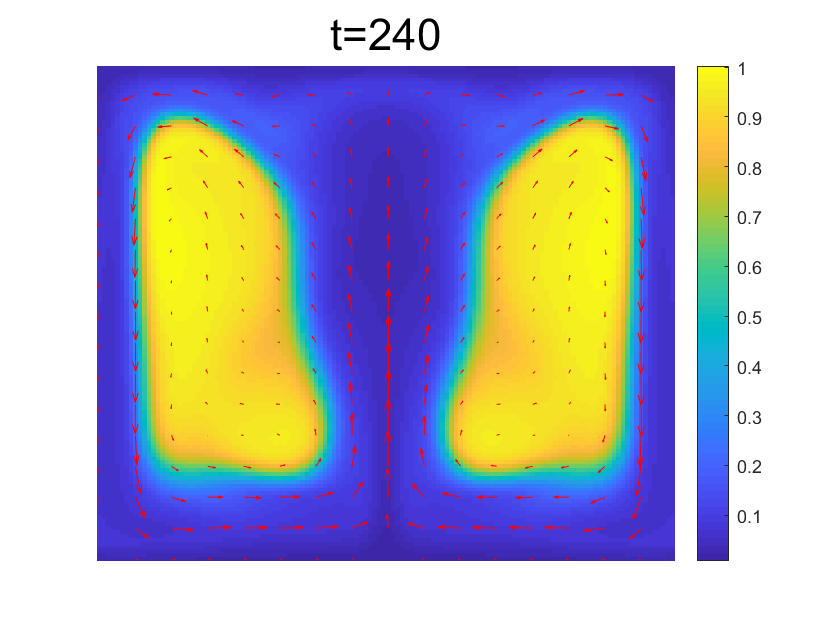}}
  \caption{Snapshots of the phase field and velocity field at $t=0, 20, 40, 60, 80, 100, 120, 180, 240$, respectively. The results indicate that as the fluid at the bottom heats up, its density decreases, so buoyant forces push the less-dense fluid up towards the cooler end of the container. Meanwhile, the cooler fluid at the top is denser, so it sinks due to gravity to displace the warmer fluid. The interface deformation is mainly because of the thermal induced fluid flow. The maximum of $|\bv|$ at $t=20, 40, 60, 80, 100, 120, 180, 240$ are $3.5611e-01, 3.1431e-01, 2.8915e-01, 3.7954e-01, 3.0357e-01, 2.5127e-01, 1.8731e-01, 1.6531e-01$, respectively. Initially, two small roll cells are present in the vicinity of the interface at t = 20. As the fluid at the bottom heats up, the temperature gradient in the system persists, and roll cells also changes gradually. The roll cells cause the fluid motion to disrupt the fluid interfaces. As time goes by, the mixing of the two fluids intensifies near the drops of fluid 1. }
  \label{Fig5.3}
\end{figure}

After benchmarking the numerical results at the onset of interfacial instability for a short time, we conduct a long time simulation up to $t=240.$
Figure \ref{Fig5.2} and Figure \ref{Fig5.3} depict the temperature field, the phase field and the velocity field, respectively, at a few selected time points in the long time simulation. We notice that the  temperature field keeps changing under the constant heating from the bottom  creating a heat flow across the entire domain. Initially, there is no significant change in the temperature field when the velocity is small and the interface changes slowly. However, owing to both convection and continuous heating, the heat flow  coalesces the small roll cells separated by the fluid interface into a pair of large, circular fluid flow patterns, resembling roll cells, in the domain. For instance, at $t=20$ in Figure \ref{Fig5.3}, there exist two large roll cells, wherein the velocity field rotates in opposite directions within each cell throughout the domain. As time goes by, the roll cells persist driving the interface apart and move the heat and fluid mass quickly to other parts of the domain where the temperature is low. So, the fluid and heat transport intensifies over the entire domain as time goes by.

Figure \ref{Fig5.2} depicts snapshots of the numerical simulations illustrating the temperature field at $t=0$, $20$, $40$, $60$, $80$, $100$, $120$, $180$, $240$, respectively. Figure \ref{Fig5.3} displays the phase field and velocity field snapshots at $t=20, 40, 60, 80, 100, 120, 180, 240$, respectively. In the process of simulation, we observe that the interface between the two fluids is deformed severely over time and mixing takes place. This is also verified by the drastic redistribution of the biphasic materials in Figure \ref{Fig5.3}. At the end of the simulation at $t=240$, phase A is seen to be given in two large deformed drops accompanied by a few satellites while phase A and B mix in a quite large region adjacent to the drops in variable degrees.

\begin{figure*}
\centering
\subfigure[Entropy $S$.]{
\includegraphics[width=0.321\linewidth]{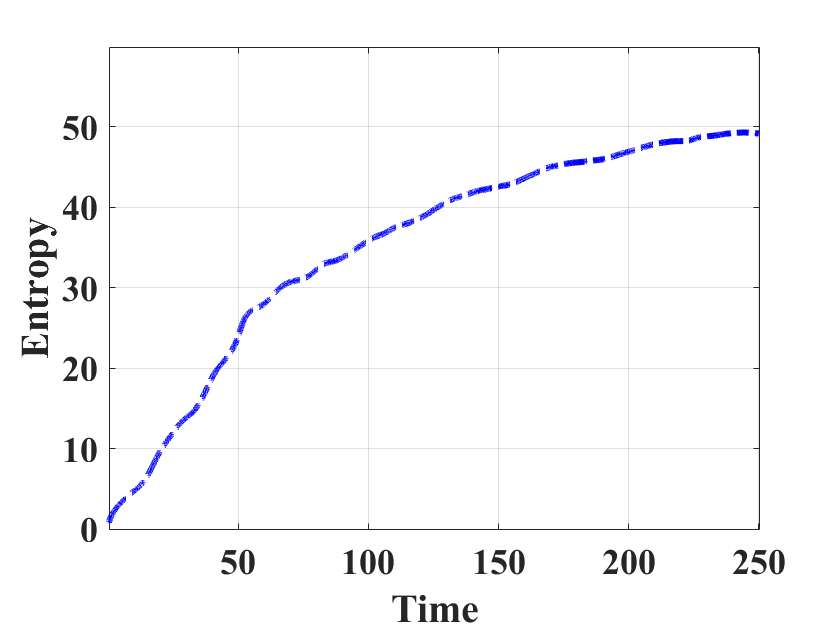}}
\subfigure[Volume $V$.]{
\includegraphics[width=0.321\linewidth]{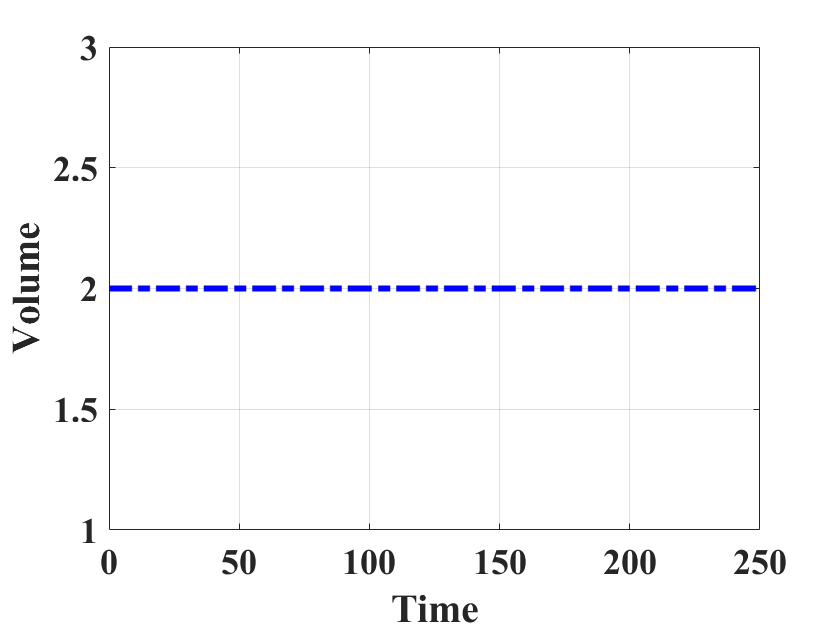}}
\subfigure[Adaptive time step.]{
\includegraphics[width=0.321\linewidth]{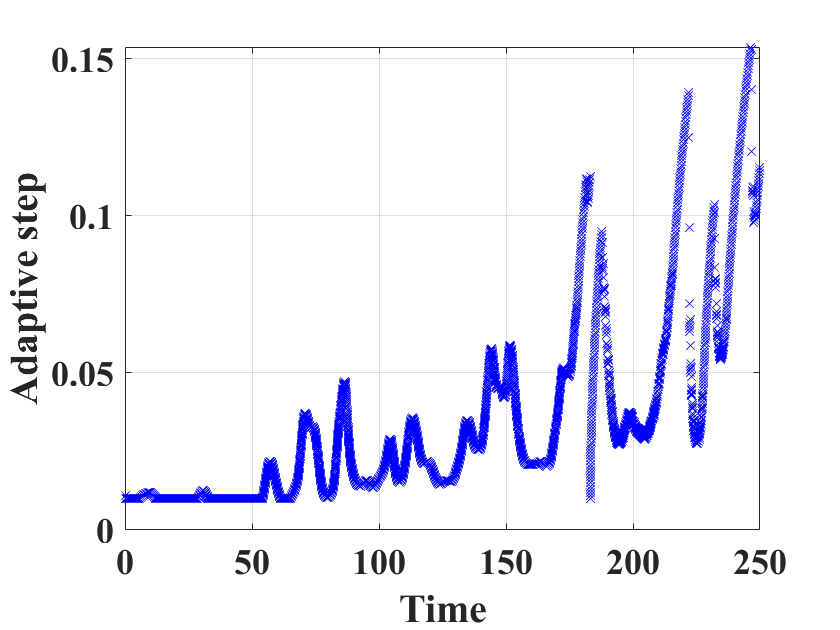}}
\caption{Evolution of the entropy, volume of a fluid phase and adaptive time step over time.
(a): Entropy $S$.  (b): Phase volume $V$. (c): The adaptive time step. It demonstrates that the scheme preserves the positive entropy production rate and the volume of each phase in the long time simulation, which implicitly indicates the computation is accurate.} \label{Fig5.4}
\end{figure*}

These numerical results demonstrate that the Rayleigh-B\'{e}nard convection as a result of the buoyancy-driven and temperature-gradient driven effect in a container  is a good mixer. As the fluid at the bottom heats up, its density decreases, so buoyant forces push the less-dense fluid up towards the cooler end of the container. Meanwhile, the cooler fluid at the top is denser, so it sinks and displaces the warmer fluid. As time goes by, the material distribution and temperature distribution shift so that the fluid phase 1 consolidates into drops  while the fluid phase 2 are pushed to the boundary. The fluid of phase 1 seems to be trapped in the roll cells. Phase separation is maintained in spatially inhomogeneous domains with some parts partially mixed however.
Finally, Figure \ref{Fig5.4} confirms that the numerical scheme preserves both the volume and the positive entropy production rate and shows the adaptive time steps against time in the long time simulation. The time step eases up in general over time.

\subsection{Drop dynamics under the influence of gravity, interfacial force, and temperature gradient}

\noindent \indent To further investigate the competition among the thermal effect, gravity, and interfacial effect, in the hydrodynamics of the nonisothermal binary fluid system in a Raleigh-B\'{e}nard cell, we consider two fluid B droplets with a radius of 1/5 suspended in fluid A initially.

\begin{figure}
   \centering
  \subfigure{
\includegraphics[width=0.32\linewidth]{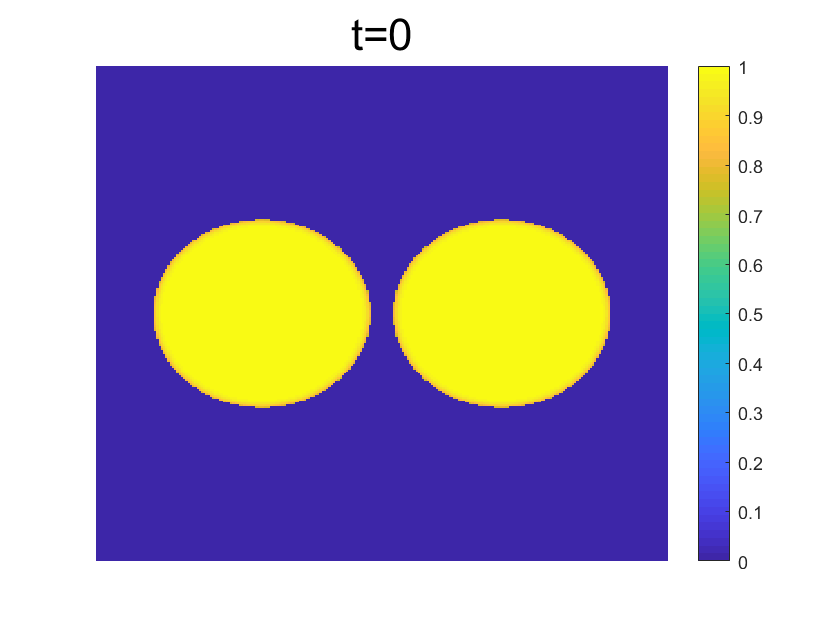}}
\subfigure{
\includegraphics[width=0.32\linewidth]{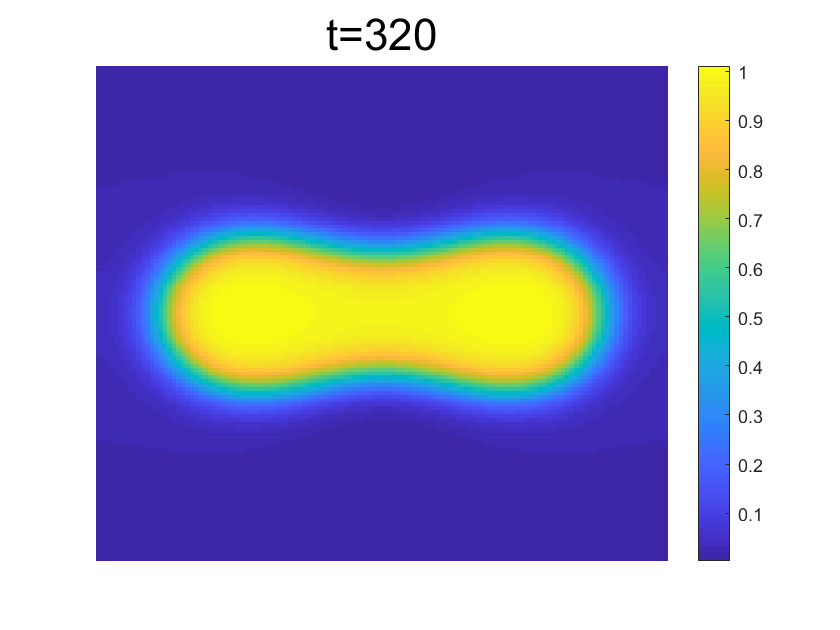}}
\subfigure{
\includegraphics[width=0.32\linewidth]{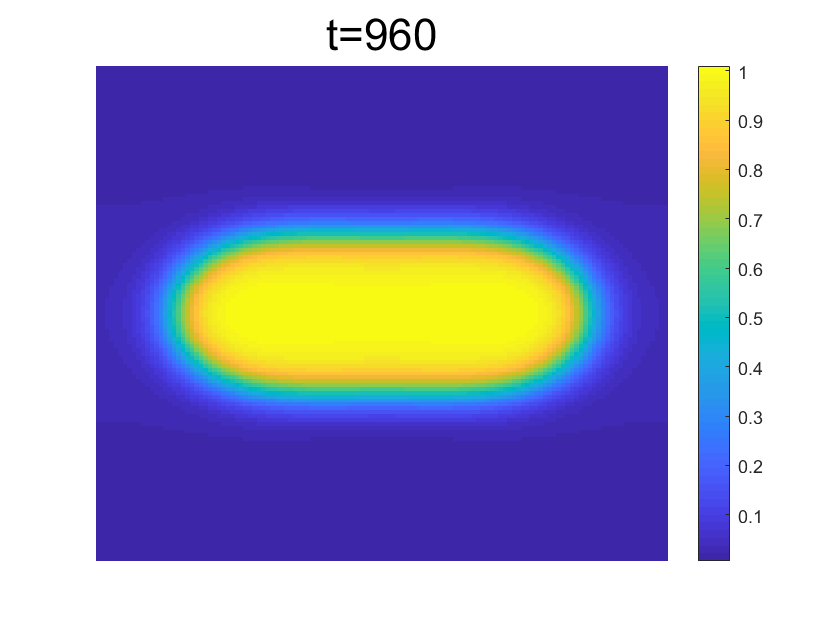}}
\subfigure{
\includegraphics[width=0.32\linewidth]{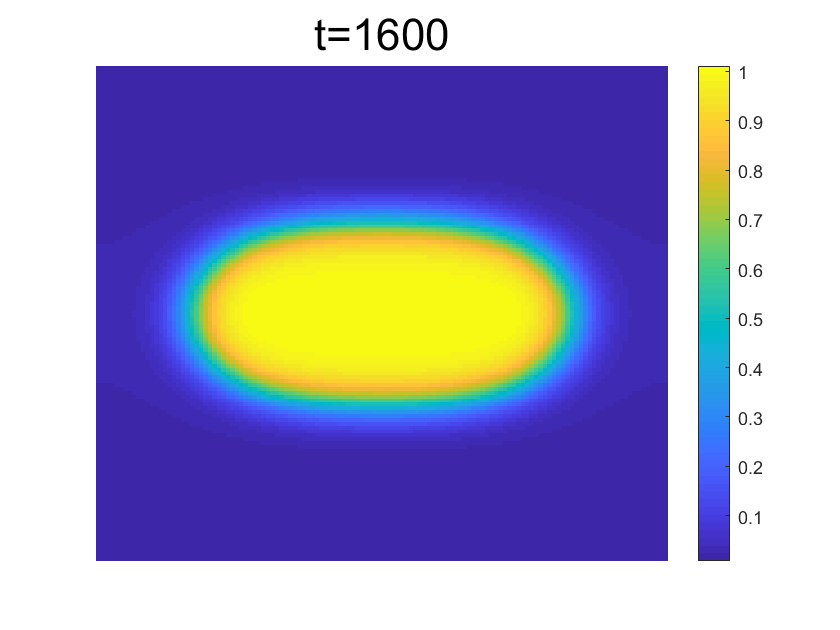}}
\subfigure{
\includegraphics[width=0.32\linewidth]{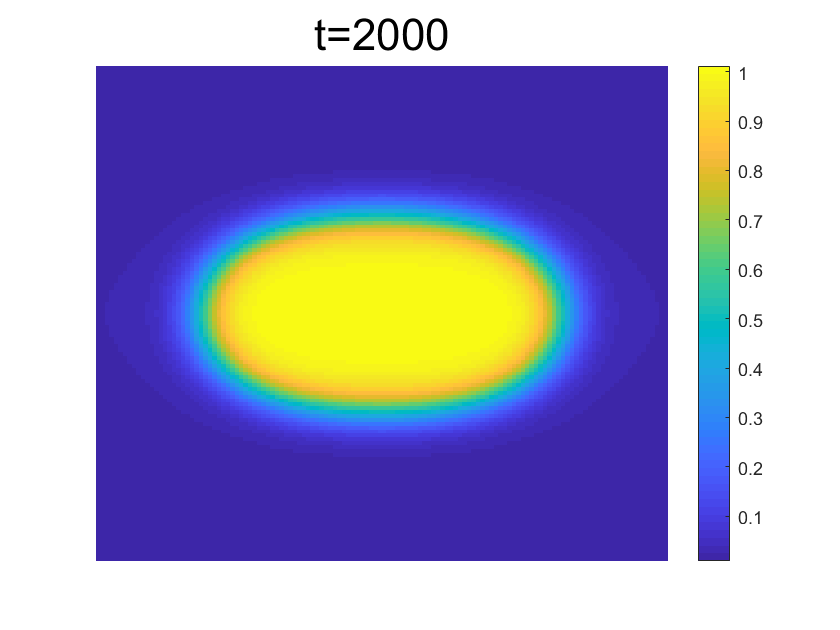}}
\subfigure{
\includegraphics[width=0.32\linewidth]{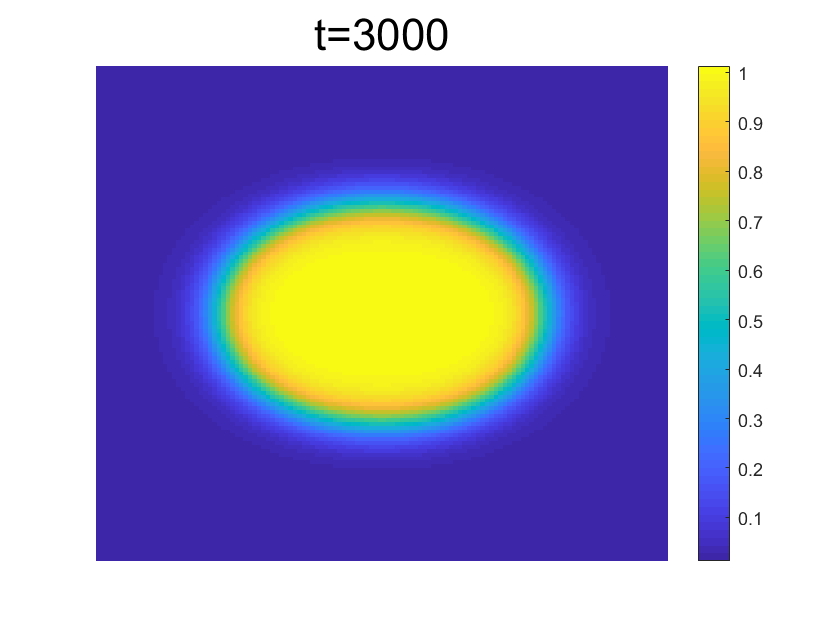}}
  \caption{Snapshots of the phase field at $t=0, 320, 960, 1600, 2000, 3000$ in an isothermal environment, respectively. As time goes on, the two initial droplets merge into a singular, larger droplet. }
  \label{Fig5.16}
\end{figure}

We impose the initial velocity field as (\ref{initial conditions of velocity}), the initial condition of the phase variable as (\ref{initial conditions of phase}), and the parameter values used in the previous simulation are replicated.
We conduct two numerical simulations here. Firstly, we  consider an isothermal system with the temperature of the system as a constant
\bena
~~~T(x,y,t)\equiv T_b, ~(x,y)\in\Omega, t\in[0,+\infty).
\eena
Under the isothermal condition, we employ a spatial meshes of $N_x=N_y=128$ and a temporal step size of $\Delta t=1.0\times10^{-1}$, solving the problem until t=3000. The dynamic process of  drop merging is shown in Figure \ref{Fig5.16}. The constant temperature field  never interfere with the hydrodynamics during drop merging.

\begin{figure}[htbp]
  \centering
  \subfigure{
\includegraphics[width=0.32\linewidth]{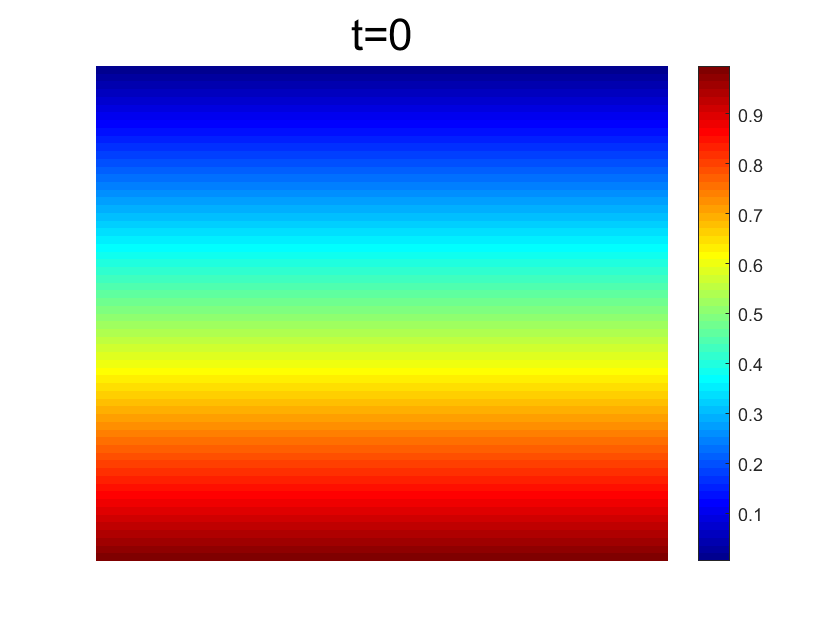}}
\subfigure{
\includegraphics[width=0.32\linewidth]{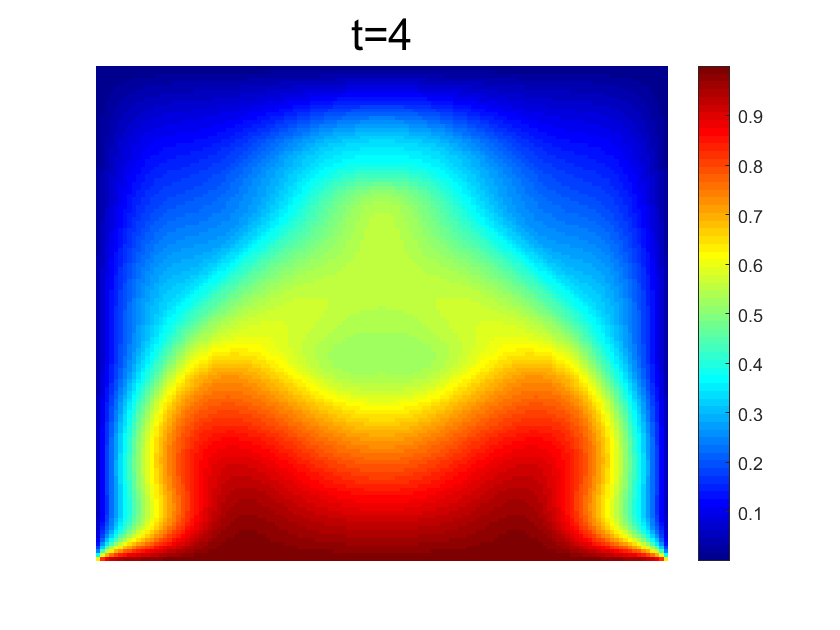}}
\subfigure{
\includegraphics[width=0.32\linewidth]{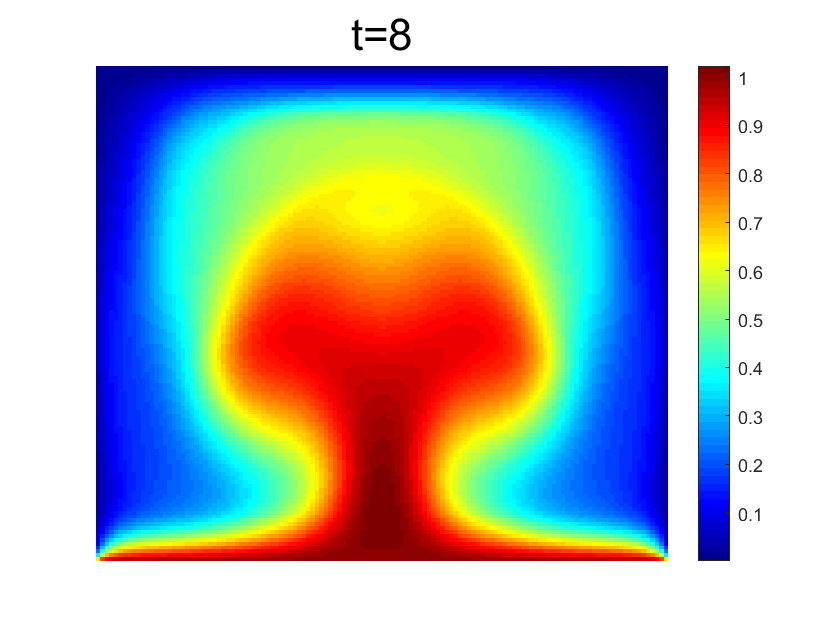}}
\subfigure{
\includegraphics[width=0.32\linewidth]{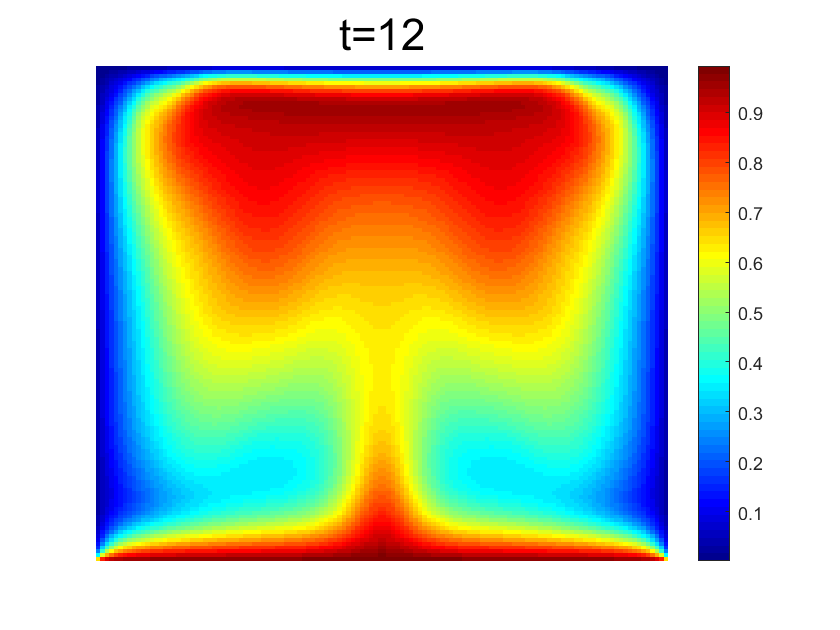}}
\subfigure{
\includegraphics[width=0.32\linewidth]{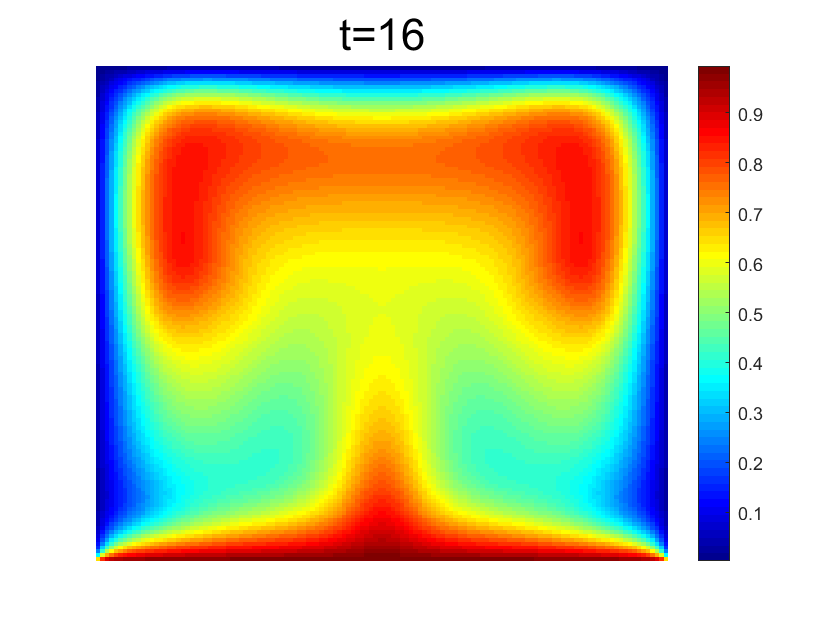}}
\subfigure{
\includegraphics[width=0.32\linewidth]{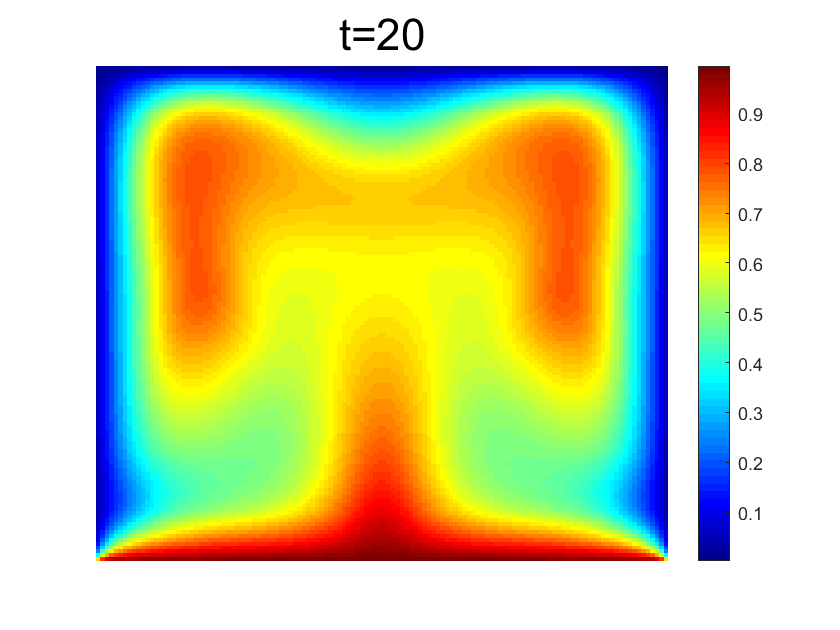}}
\subfigure{
\includegraphics[width=0.32\linewidth]{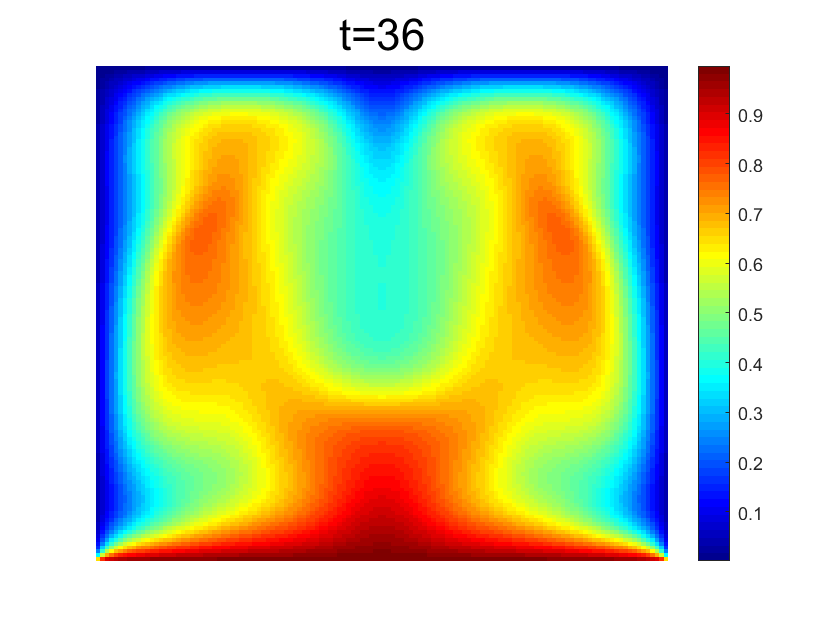}}
\subfigure{
\includegraphics[width=0.32\linewidth]{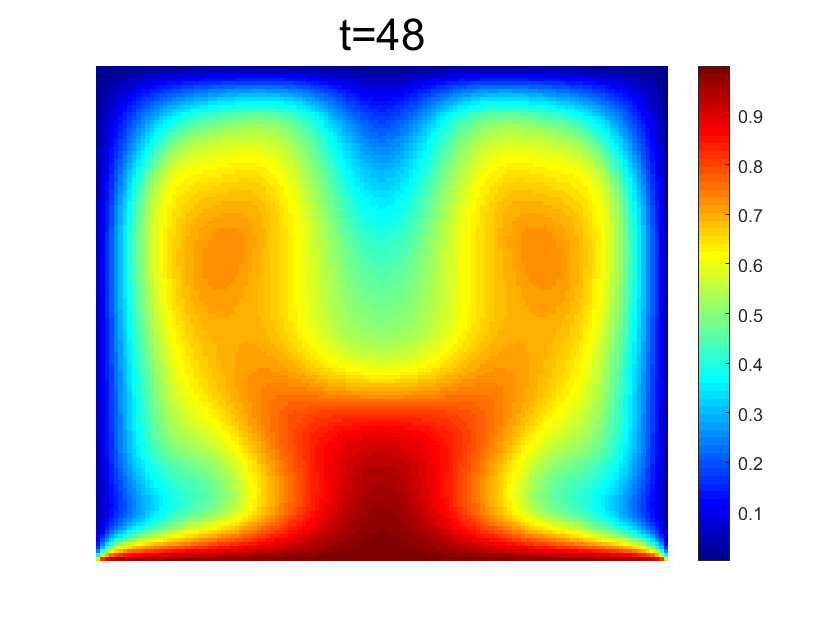}}
\subfigure{
\includegraphics[width=0.32\linewidth]{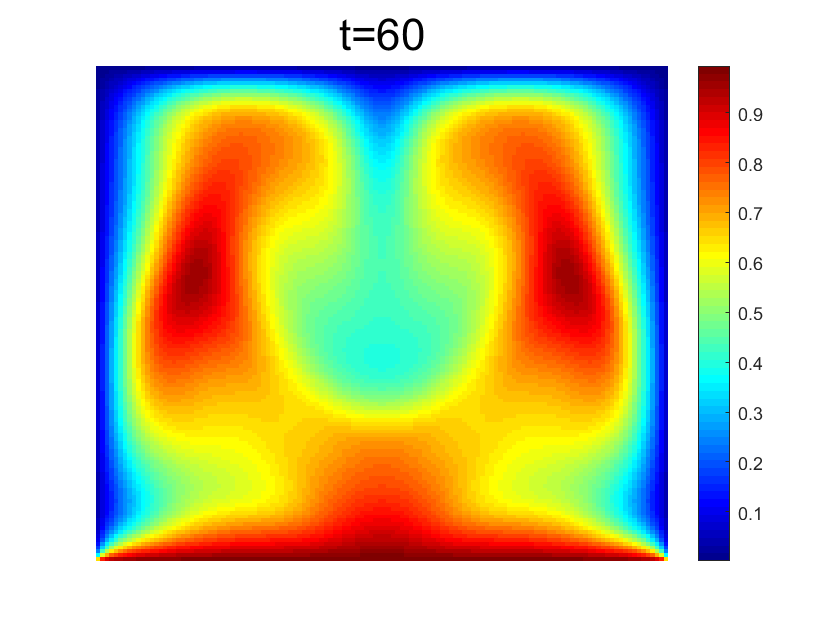}}
  \caption{Snapshots of the temperature field at selected times $t=0, 4, 8, 12, 16, 20, 36, 48, 60$, respectively. The higher temperature regions form a pitchfork shaped domain.}
  \label{Fig5.14}
\end{figure}

\begin{figure}[htbp]
  \centering
  \subfigure{
\includegraphics[width=0.32\linewidth]{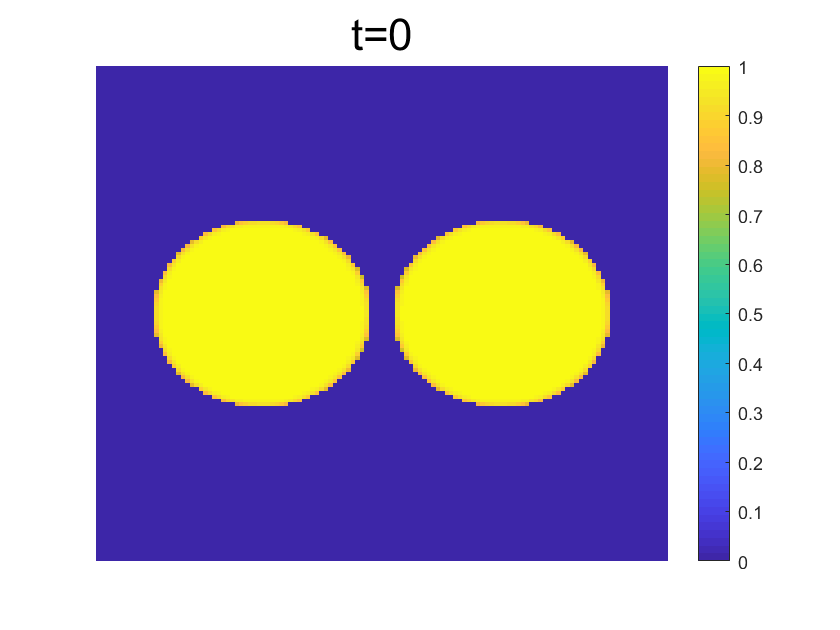}}
\subfigure{
\includegraphics[width=0.32\linewidth]{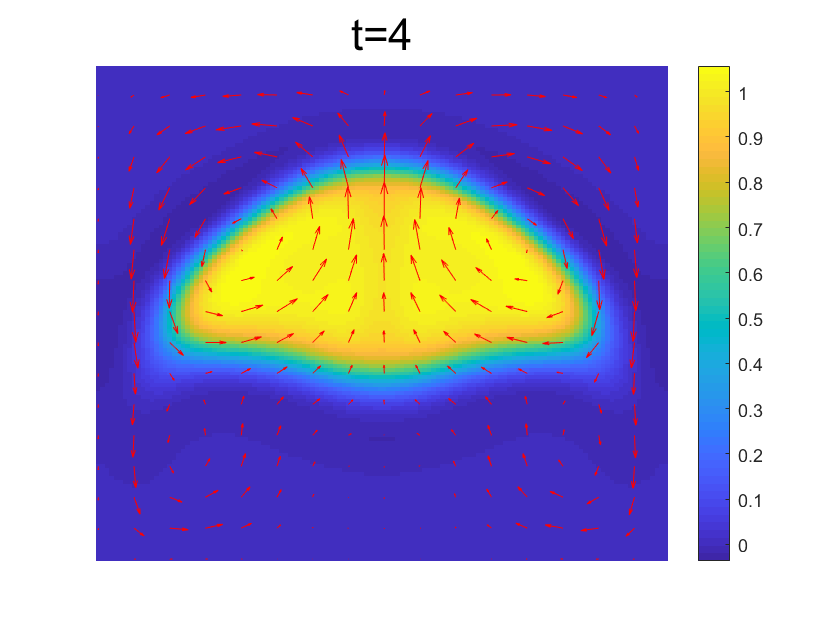}}
\subfigure{
\includegraphics[width=0.32\linewidth]{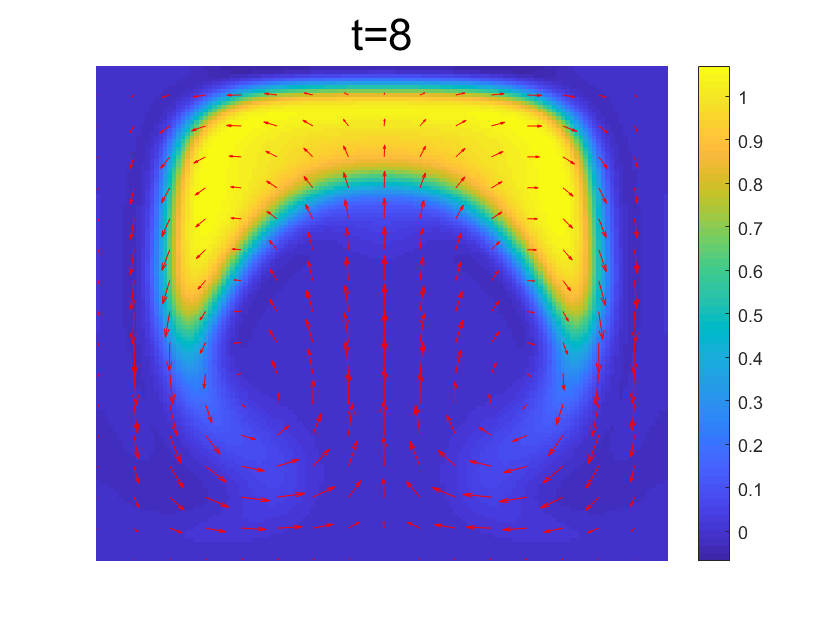}}
\subfigure{
\includegraphics[width=0.32\linewidth]{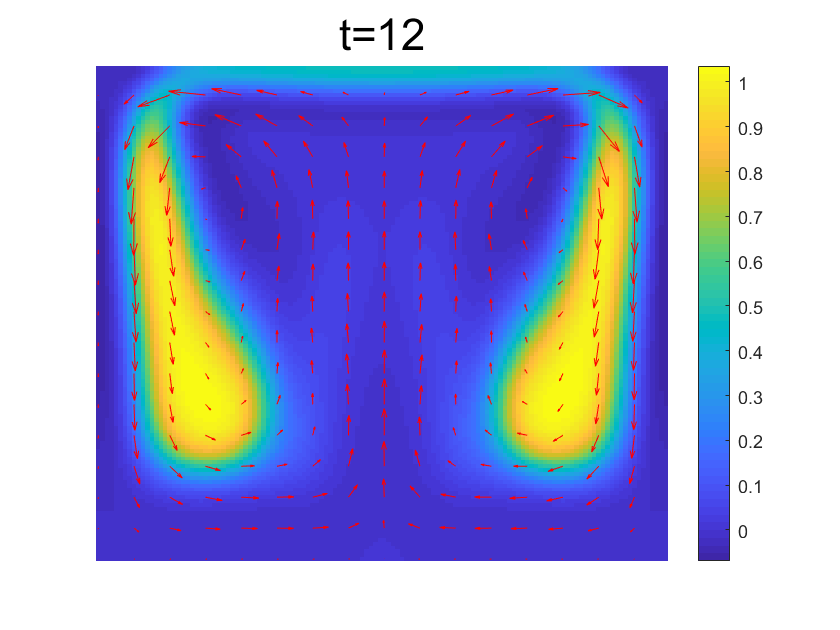}}
\subfigure{
\includegraphics[width=0.32\linewidth]{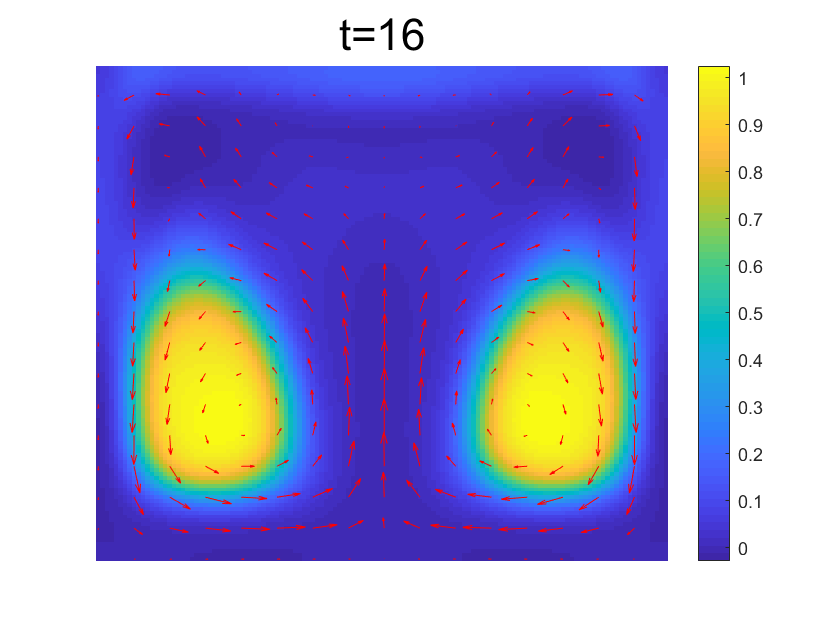}}
\subfigure{
\includegraphics[width=0.32\linewidth]{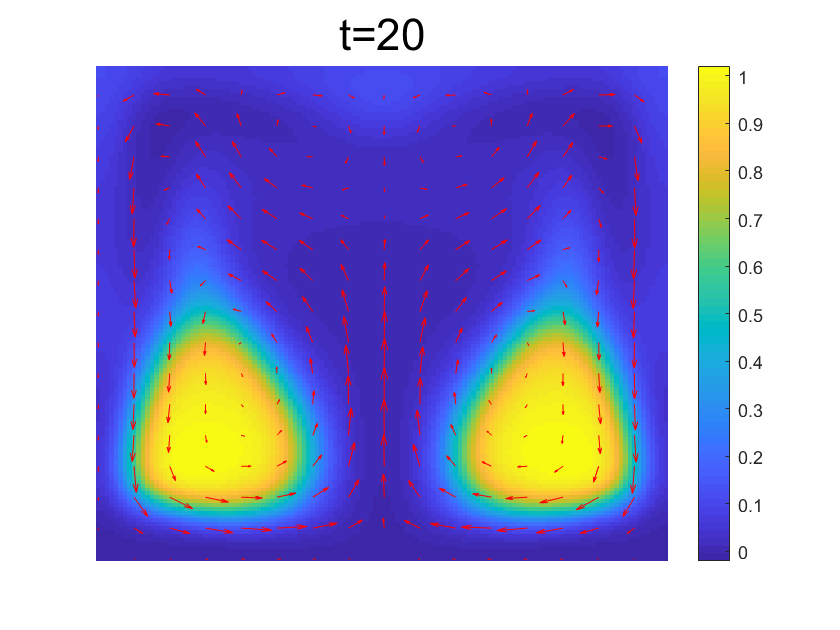}}
\subfigure{
\includegraphics[width=0.32\linewidth]{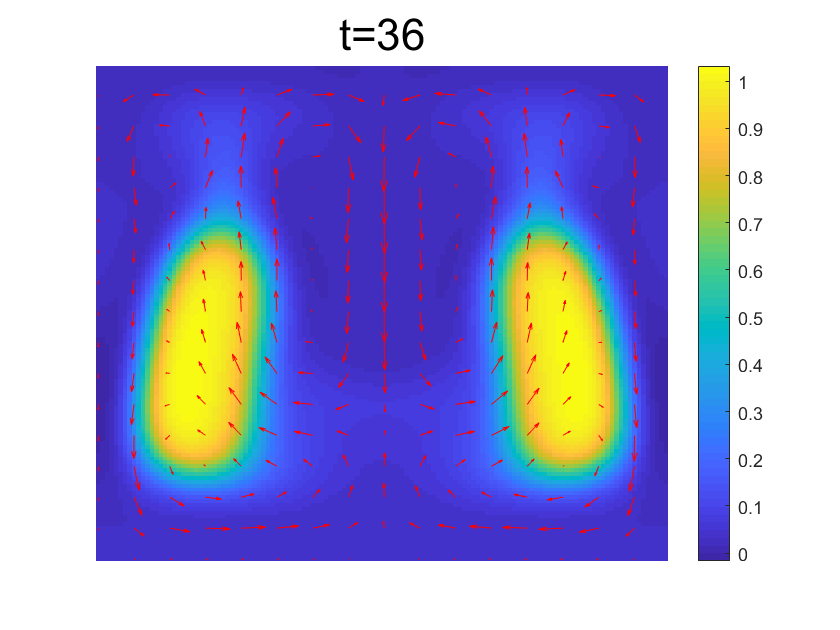}}
\subfigure{
\includegraphics[width=0.32\linewidth]{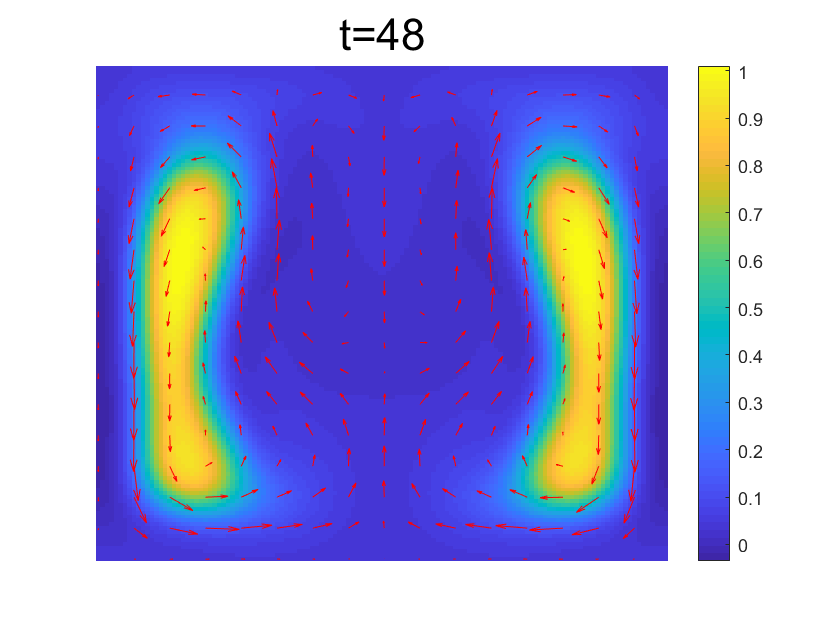}}
\subfigure{
\includegraphics[width=0.32\linewidth]{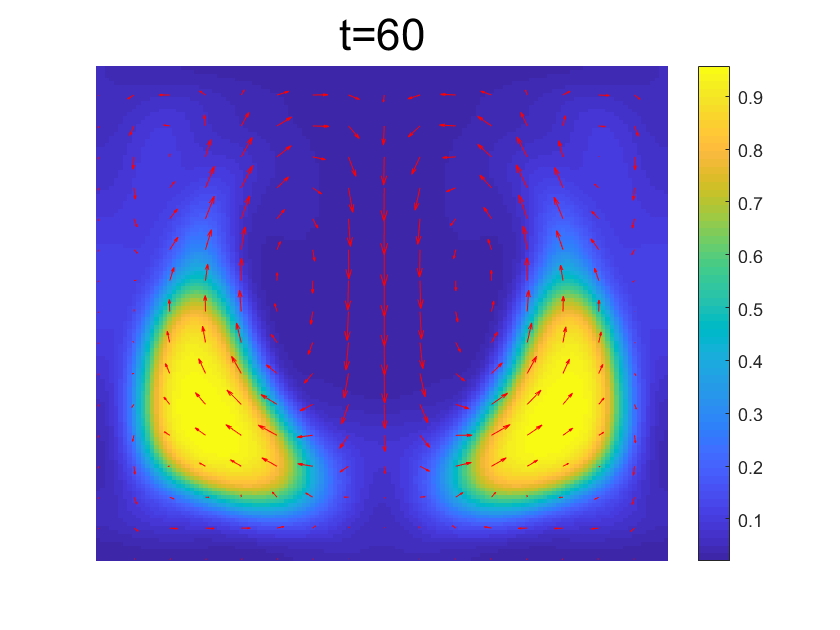}}
  \caption{Snapshots of the phase field and velocity field at $t=0, 4, 8, 12, 16, 20, 36, 48, 60$, respectively. At the beginning of the simulation, there is some mixing occurring at the interface region due to transverse fluid flow motion, resulting in a blurred interface. Meanwhile, as the fluid at the bottom heats up, the buoyant forces push the merging drops towards the cooler end of the container. And the merging drops eventually separate again from the sides of the container at the end of the simulation. And the maximum of $|\bv|$ at $t=4, 8, 12, 16, 20, 36, 48, 60$ are $1.1929e-01, 2.6183e-01, 1.4836e-01, 0.8475e-01, 0.7781e-01, 0.7363e-01, 0.8161e-01, 1.4738e-01$, respectively.
  At the initial moment, there are four small roll cells in the vicinity of interface shown as $t=4$. As the fluid at the bottom heats up, the $\bv$ increases and consequently, the number of roll cells decreases. It documents that as the fluid at the bottom heats up, so buoyant forces push the merging drops up towards the cooler end of the container. Meanwhile, the cooler fluid at the top sinks and displaces the warmer fluid, which leads to the formation of circulating roll cells.}
  \label{Fig5.15}
\end{figure}

\begin{figure*}
\centering
\subfigure[Entropy $S$.]{
\includegraphics[width=0.32\linewidth]{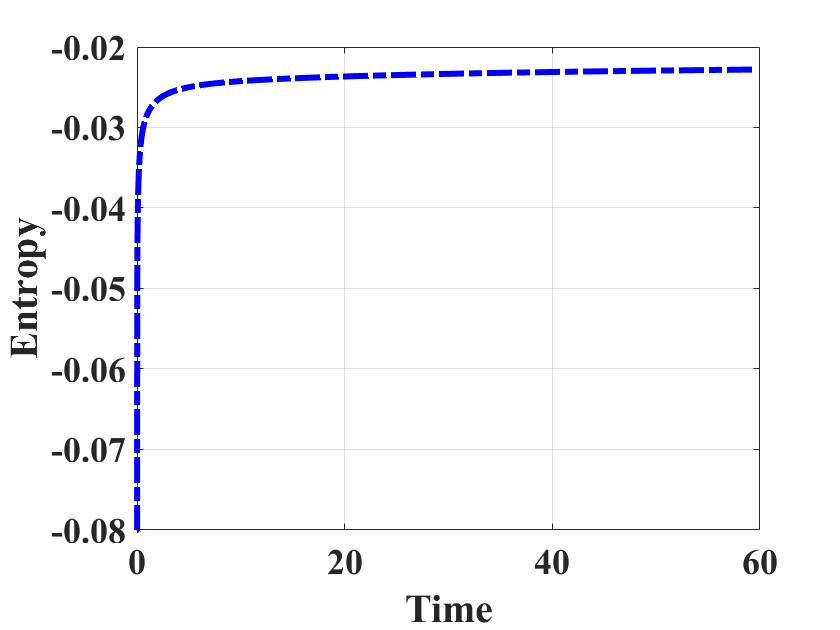}}
\subfigure[Volume $V$.]{
\includegraphics[width=0.32\linewidth]{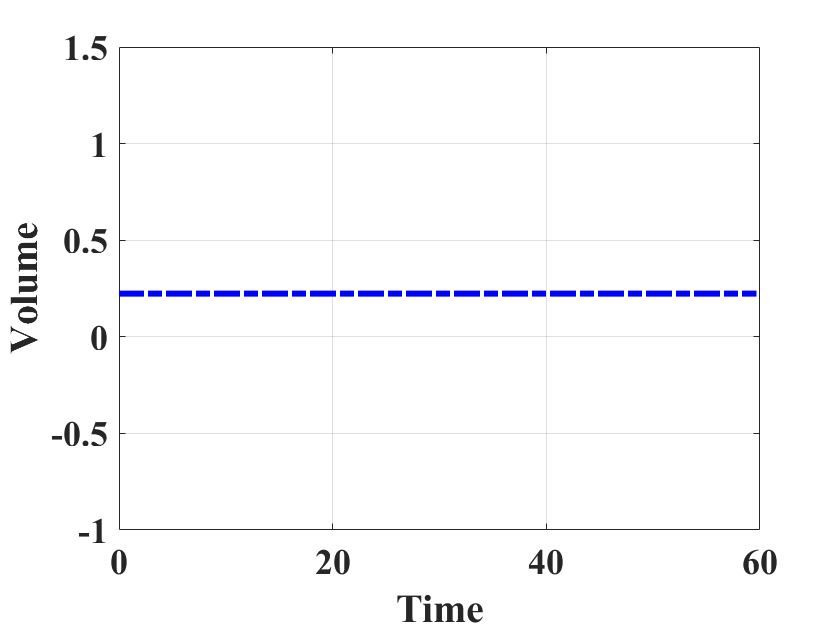}}
\subfigure[Adaptive time step.]{
\includegraphics[width=0.32\linewidth]{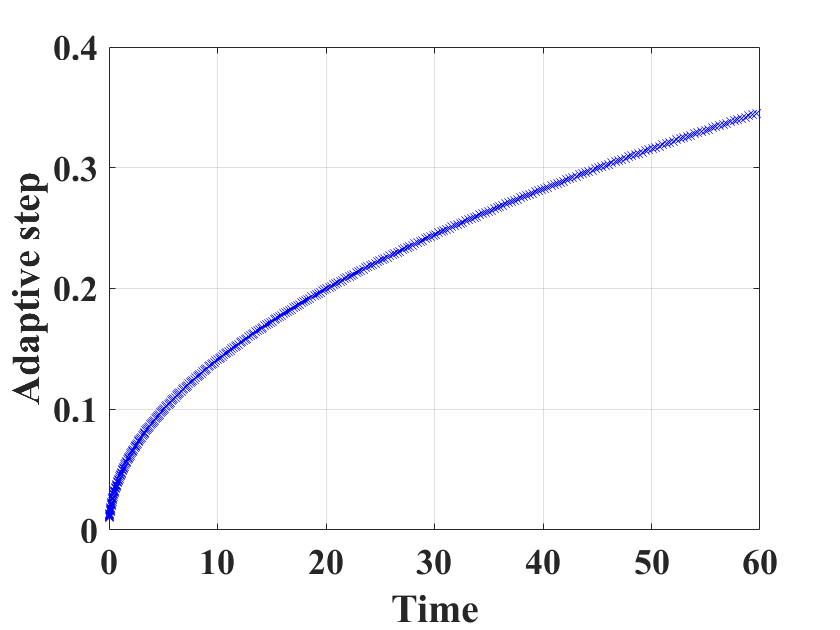}}
\caption{Evolution of the entropy, volume of a fluid phase and adaptive time step over time.
(a): The  entropy $S$.  (b): The phase volume $V$. (c): The adaptive time step. It demonstrates that the scheme maintains the positive entropy production rate and the volume of each fluid phase.} \label{Fig5.17}
\end{figure*}

Secondly in contrast,  we simulate dynamics of the two  drop system with a temperature gradient induced by the imposed boundary temperature difference, where initial conditions for the velocity field, temperature field, and phase variable are specified by (\ref{initial conditions of velocity}), (\ref{initial conditions of temperature}) and (\ref{initial conditions of phase}), respectively. In this simulation, we employ spatial meshes $N_x=N_y=128$ and  max/min adaptive time step $\Delta t_{max}=1$/$\Delta t_{min}=1.0\times10^{-2}$ to solve the problem up to $t=60$. Figure \ref{Fig5.14} depicts the solution of the temperature at $t=0, 4, 8, 12, 16, 20, 36, 48, 60$, respectively. The corresponding velocity field are shown in Figure \ref{Fig5.15}.
Owing to the Rayleigh B\'{e}nard convection, the buoyant force pushes the fluid upwards and eventually disrupt the coalescing dynamics during the merging of the two drops. Due to the formation of the roll cells or rotational flows in the domain,  the coalesced fluid A regions are disrupted into two separate, deformed drops at $t=16$. This phenomenon is shown in Figure \ref{Fig5.15}. As time goes by, the two drops sit below the centers of the two roll cells. They are occasionally stretched thin and then recover to thick drops. Mixing takes place in the neighborhood of the two drops. The hydrodynamics shown in this simulation differ significantly from the corresponding isothermal case, revealing the disruptive impact of the thermal flow to the hydrodynamics.
Figure \ref{Fig5.17} shows the evolution of the entropy, the volume of a fluid phase and the adaptive time step over time. The time step eases up quite significantly as time elapses.

\section{Conclusion}

\noindent \indent
We have developed a new hydrodynamic model for studying the Rayleigh-B'{e}nard convection in two-phase fluids, which is thermodynamically consistent and accounts for non-isothermal effects, gravity, and interfacial forces in incompressible binary viscous fluids. To simulate this system, we have devised a set of second-order numerical algorithms that preserve volume and entropy-production rate. We have numerically validated the convergence rate and structure-preserving properties of one fully discrete scheme.
Using an adaptive time-stepping implementation of the scheme, we have presented a couple of numerical examples that showcase the collective effect of thermal transport, gravity, and interfacial force in the Rayleigh-B\'{e}nard convection of two-layered viscous fluids in a rectangular container with specified boundary conditions. Additionally, we have demonstrated the dynamics of merging drops in an immiscible binary viscous fluid system subject to competing effects of gravity, thermal transport, and interfacial force. Our simulations indicate that the non-isothermal effect imposed by a temperature gradient across the boundary has a significant influence on the hydrodynamics of the binary fluid system.
Overall, our models, associated structure-preserving schemes, and simulation tools can be applied to various real-world scenarios involving multiphasic fluid flows where non-isothermal effects are important.

\section*{Acknowledgements}

\noindent \indent Shouwen Sun's work is partially supported by Key Scientific Research Project of Colleges and Universities in Henan Province, China (No.22A110018) and by National Natural Science Foundation of China (No.12101387).


\bibliographystyle{plain}
\bibliography{reference}

\section*{Appendix: Definitions and Notations}
\noindent \indent We summarize the notations used in the spatial discretization and some useful lemmas here for completeness, which are defined in \cite{SunShouwen}.  We set the computational domain as $\Omega=[0,L_x]\times[0,L_y]$ with $L_x=h_x N_x, L_y=h_y N_y$, where $N_x, N_y$ are positive integers and   $h_x, h_y$ are  mesh sizes. We define the following sets for various  grid points:
$$E_x=\{x_{i+\frac{1}{2}}|i=0,1,\ldots,N_x\}, C_x=\{x_{i}|i=1,2,\ldots,N_x\}, C_{\bar{x}}=\{x_{i}|i=0,1,\ldots,N_x+1\},$$
$$E_y=\{y_{j+\frac{1}{2}}|j=0,1,\ldots,N_y\}, C_y=\{y_{j}|j=1,2,\ldots,N_y\}, C_{\bar{y}}=\{y_{j}|j=0,1,\ldots,N_y+1\},$$
where $x_l=(l-\frac{1}{2})h_x, y_l=(l-\frac{1}{2})h_y$, $l$ can take on integer or half-integer values. The elements of $E_x, E_y$ are called edge-centered points, the elements of $C_x, C_y, C_{\bar{x}}, C_{\bar{y}}$ are called cell-centered points and the two points belonging to
$C_{\bar{x}}\backslash C_x$ are called ghost points. In this paper, we chose $h_x=h_y=h$ for simplicity.

We define the following discrete function spaces
$$\mathcal{C}_{x\times y}=\{\phi: C_x \times C_y\rightarrow \mathbb{R}\}, \mathcal{C}_{\bar{x}\times y}=\{\phi: C_{\bar{x}} \times C_y\rightarrow \mathbb{R}\}, \mathcal{C}_{x\times \bar{y}}=\{\phi: C_x \times C_{\bar{y}}\rightarrow \mathbb{R}\},$$
$$\mathcal{C}_{{\bar{x}}\times \bar{y}}=\{\phi: C_{\bar{x}} \times C_{\bar{y}}\rightarrow \mathbb{R}\}, \mathcal{\varepsilon}^{ew}_{x\times y}=\{u: E_x \times C_y\rightarrow \mathbb{R}\}, \mathcal{\varepsilon}^{ew}_{x\times {\bar{y}}}=\{u: E_x \times C_{\bar{y}}\rightarrow \mathbb{R}\},$$
$$\mathcal{\varepsilon}^{ns}_{x\times y}=\{v: C_x \times E_y\rightarrow \mathbb{R}\}, \mathcal{\varepsilon}^{ns}_{{\bar{x}}\times y}=\{v: C_{\bar{x}} \times E_{\bar{y}}\rightarrow \mathbb{R}\}, \mathcal{\nu}_{x\times y}=\{f: E_x \times E_y\rightarrow \mathbb{R}\},$$
where the functions in $\mathcal{C}_{x\times y}, \mathcal{C}_{\bar{x}\times y}, \mathcal{C}_{x\times \bar{y}}, \mathcal{C}_{{\bar{x}}\times \bar{y}}$ are called cell centered discrete functions,  the functions in $\mathcal{\varepsilon}^{ew}_{x\times y}, \mathcal{\varepsilon}^{ew}_{x\times {\bar{y}}}$, $\mathcal{\varepsilon}^{ns}_{x\times y}, \mathcal{\varepsilon}^{ns}_{{\bar{x}}\times y}$ are called east-west and north-south edge centered discrete functions and the functions in $\mathcal{\nu}_{x\times y}$ are called vertex centered discrete functions, respectively.

\noindent \indent Firstly, in order to define the operator symbols, we assume $\phi, \psi$ are cell centered functions, $u, r$ are east-west edge centered functions, $v, w$ are north-south edge centered functions and $f, g$ are vertex centered functions. Namely, $\phi, \psi \in \mathcal{C}_{x\times y}\cup \mathcal{C}_{\bar{x}\times y}\cup\mathcal{C}_{x\times \bar{y}}\cup\mathcal{C}_{{\bar{x}}\times \bar{y}}$, $u, r \in \mathcal{\varepsilon}^{ew}_{x\times y}\cup \mathcal{\varepsilon}^{ew}_{x\times {\bar{y}}}$, $v, w\in \mathcal{\varepsilon}^{ns}_{x\times y}\cup \mathcal{\varepsilon}^{ns}_{{\bar{x}}\times y}$, $f, g \in \mathcal{\nu}_{x\times y}$.

Secondly, we define the east-west-edge-to-center average and difference operators as $a_x, d_x$,
\bena
a_xu_{i,j}:=\frac{1}{2}(u_{i+\frac{1}{2},j}+u_{i-\frac{1}{2},j}),~~~ d_xu_{i,j}:=\frac{1}{h_x}(u_{i+\frac{1}{2},j}-u_{i-\frac{1}{2},j}),\\
a_xf_{i,j+\frac{1}{2}}:=\frac{1}{2}(f_{i+\frac{1}{2},j+\frac{1}{2}}+f_{i-\frac{1}{2},j+\frac{1}{2}}),~~~ d_xf_{i,j+\frac{1}{2}}:=\frac{1}{h_x}(f_{i+\frac{1}{2},j+\frac{1}{2}}-f_{i-\frac{1}{2},j+\frac{1}{2}}).
\eena
The north-south-edge-to-center average and difference operators are defined as $a_y, d_y$,
\bena
a_yv_{i,j}:=\frac{1}{2}(v_{i,j+\frac{1}{2}}+u_{i,j-\frac{1}{2}}),~~~ d_yv_{i,j}:=\frac{1}{h_y}(v_{i,j+\frac{1}{2}}-v_{i,j-\frac{1}{2}}),\\
a_yf_{i+\frac{1}{2},j}:=\frac{1}{2}(f_{i+\frac{1}{2},j+\frac{1}{2}}+f_{i+\frac{1}{2},j-\frac{1}{2}}),~~~ d_yf_{i+\frac{1}{2},j}:=\frac{1}{h_y}(f_{i+\frac{1}{2},j+\frac{1}{2}}-f_{i+\frac{1}{2},j-\frac{1}{2}}).
\eena
The center-to-east-west-edge average and difference operators are defined as $A_x, D_x$,
\bena
A_x\phi_{i+\frac{1}{2},j}:=\frac{1}{2}(\phi_{i+1,j}+\phi_{i,j}),~~~ D_x\phi_{i+\frac{1}{2},j}:=\frac{1}{h_x}(\phi_{i+1,j}-\phi_{i,j}),\\
A_xv_{i+\frac{1}{2},j+\frac{1}{2}}:=\frac{1}{2}(v_{i+1,j+\frac{1}{2}}+v_{i,j+\frac{1}{2}}),~~~ D_xv_{i+\frac{1}{2},j+\frac{1}{2}}:=\frac{1}{h_x}(v_{i+1,j+\frac{1}{2}}-v_{i,j+\frac{1}{2}}).
\eena
The center-to-north-south-edge average and difference operators are defined as $A_y, D_y$,
\bena
A_y\phi_{i,j+\frac{1}{2}}:=\frac{1}{2}(\phi_{i,j+1}+\phi_{i,j}),~~~ D_y\phi_{i,j+\frac{1}{2}}:=\frac{1}{h_y}(\phi_{i,j+1}-\phi_{i,j}),\\
A_yu_{i+\frac{1}{2},j+\frac{1}{2}}:=\frac{1}{2}(u_{i+\frac{1}{2},j+1}+u_{i+\frac{1}{2},j}),~~~ D_yu_{i+\frac{1}{2},j+\frac{1}{2}}:=\frac{1}{h_y}(u_{i+\frac{1}{2},j+1}-u_{i+\frac{1}{2},j}).
\eena

In this paper, the cell centered functions $\phi\in \mathcal{C}_{{\bar{x}}\times \bar{y}}$ is said to satisfy homogeneous Neumann boundary conditions if and only if
\bena
\phi_{0,j}=\phi_{1,j},~~~ \phi_{N_x,j}=\phi_{N_x+1,j},~~~ j=1,2,\ldots,N_y; \\
\phi_{i,0}=\phi_{i,1},~~~ \phi_{i,N_y}=\phi_{i,N_y+1},~~~ i=0,1,\ldots,N_x+1.\label{BC Ne}
\eena
In addition, the edge centered functions $u\in \mathcal{\varepsilon}^{ew}_{x\times {\bar{y}}}, v\in \mathcal{\varepsilon}^{ns}_{{\bar{x}}\times y}$ are said to satisfy homogeneous Dirichlet boundary conditions if and only if
\bena
u_{\frac{1}{2},j}=u_{N_x+\frac{1}{2},j}=0, ~~~j=1, 2, \ldots, N_y,\\
A_yu_{i+\frac{1}{2},\frac{1}{2}}=A_yu_{i+\frac{1}{2},N_y+\frac{1}{2}}=0, ~~~i=0,1, 2, \ldots, N_x,\\
v_{i,\frac{1}{2}}=v_{i,N_y+\frac{1}{2}}=0, ~~~i=1, 2, \ldots, N_x,\\
A_xv_{\frac{1}{2},j+\frac{1}{2}}=A_xv_{N_x+\frac{1}{2},j+\frac{1}{2}}=0, ~~~j=0,1, 2, \ldots, N_y.\label{BC Di}
\eena

The discrete Laplacian operator $\Delta_h: \mathcal{\varepsilon}^{ew}_{x\times {\bar{y}}} \cup \mathcal{\varepsilon}^{ns}_{{\bar{x}}\times y}\cup \mathcal{C}_{{\bar{x}}\times \bar{y}}\longrightarrow \mathcal{\varepsilon}^{ew}_{x\times {y}} \cup \mathcal{\varepsilon}^{ns}_{{x}\times y}\cup \mathcal{C}_{{x}\times y}$ is defined as follows
\ben
\Delta_h u=D_x(d_xu)+d_y(D_yu), ~\Delta_h v=D_x(d_xv)+d_y(D_yv), ~\Delta_h \phi=D_x(d_x\phi)+d_y(D_y\phi).
\een
In addition, we define the corresponding inner products $(\cdot,\cdot)$ and norms $\|\cdot\|$ as follows:
\bena
(\phi,\psi)_2:=h_xh_y\sum_{i=1}^{N_x}\sum_{j=1}^{N_y} \phi_{i,j}\psi_{i,j},\\
{[u,r]}_{ew}:=(a_x(ur),1)_2=\frac{1}{2}h_xh_y\sum_{i=1}^{N_x}\sum_{j=1}^{N_y}(u_{i+\frac{1}{2},j}r_{i+\frac{1}{2},j}+u_{i-\frac{1}{2},j}r_{i-\frac{1}{2},j}),\\
{[v,w]}_{ns}:=(a_y(vw),1)_2=\frac{1}{2}h_xh_y\sum_{i=1}^{N_x}\sum_{j=1}^{N_y}(v_{i,j+\frac{1}{2}}w_{i,j+\frac{1}{2}}+v_{i,j-\frac{1}{2}}w_{i,j-\frac{1}{2}}),\\
(f,g)_{vc}:=(a_x(a_y(fg)),1)_2, (\nabla_h \phi, \nabla_h \psi):=[D_x\phi, D_x\psi]_{ew}+[D_y\phi, D_y\psi]_{ns},
\eena
where $\psi$ is a cell centered functions and satisfy homogeneous Neumann boundary conditions.
\bena
\|\phi\|_2:=(\phi,\phi)^{\frac{1}{2}}_2,~~~\|u\|_{ew}:=[u,u]^{\frac{1}{2}}_{ew},~~~\|v\|_{ns}:=[v,v]^{\frac{1}{2}}_{ns},~~~\|f\|_{vc}:=(f,f)^{\frac{1}{2}}_{vc}.
\eena
For $\phi \in \mathcal{C}_{x\times y}\cup \mathcal{C}_{\bar{x}\times y}\cup\mathcal{C}_{x\times \bar{y}}\cup\mathcal{C}_{{\bar{x}}\times \bar{y}}$ we define $\|\nabla\phi\|_2$ as
\bena
\|\nabla\phi\|_2^2:=\|D_x\phi\|^2_{ew}+\|D_y\phi\|_{ns}^2.
\eena
Given the edge-centered velocity vector $\bv=(u,v), u\in\mathcal{\varepsilon}^{ew}_{x\times{\bar{y}}}, v\in\mathcal{\varepsilon}^{ns}_{{\bar{x}}\times y}$, we define $\|\bv\|_2, \|\nabla\bv\|_2$ as
\bena
\|\bv\|_2^2:=\|u\|^2_{ew}+\|v\|^2_{ns},~~~\|\nabla\bv\|_2^2:=\|d_xu\|^2_2+\|D_yu\|^2_{vc}+\|D_xv\|^2_{vc}+\|d_yv\|^2_2,\\
\|\bD\|_2^2:=\|d_xu\|^2_2+\frac{1}{2}\|D_yu\|^2_{vc}+\frac{1}{2}\|D_xv\|^2_{vc}+(D_yu, D_xv)_{vc}+\|d_yv\|^2_2.\\
\eena
where $\bD=\frac{1}{2}(\nabla\bv+\nabla\bv^T)$. Next, we present some useful lemmas to be used in the proof of the theorems in the next section.
\begin{lem}
Given $\phi, \psi\in\mathcal{C}_{{\bar{x}}\times \bar{y}}$ satisfying the discrete homogeneous Neumann boundary condition, the following summation by parts formula is valid
\bena
-(\Delta_h \phi,\psi)_2=(\nabla_h \phi, \nabla_h \psi)_2.
\eena
\end{lem}
\begin{lem}
For $\phi\in\mathcal{C}_{{\bar{x}}\times \bar{y}}$ satisfying the discrete homogeneous Neumann boundary condition, $\bv=(u,v), u\in\mathcal{\varepsilon}^{ew}_{x\times{\bar{y}}}, v\in\mathcal{\varepsilon}^{ns}_{{\bar{x}}\times y}$ satisfying the homogeneous Dirichlet boundary condition, the following summation by parts formulae are valid
\bena
{[A_x\phi,u]}_{ew}=(\phi,a_xu)_2,~~~{[A_y\phi,v]}_{ns}=(\phi,a_yv)_2,\\
{[D_x\phi,u]}_{ew}=-(\phi,d_xu)_2,~~~{[D_y\phi,v]}_{ns}=-(\phi,d_yv)_2.
\eena
\end{lem}
\begin{lem}
Given $f\in\mathcal{\nu}_{x\times y}$ satisfying the discrete homogeneous Dirichlet boundary condition and $u\in\mathcal{\varepsilon}^{ew}_{x\times{\bar{y}}}, v\in\mathcal{\varepsilon}^{ns}_{{\bar{x}}\times y}$, the following formulae are valid
\bena
{[a_yf,u]}_{ew}=(f,A_yu)_{vc},~~~{[a_xf,v]}_{ns}=(f,A_xv)_{vc}.
\eena
\end{lem}
\begin{lem}
Given $u\in\mathcal{\varepsilon}^{ew}_{x\times{\bar{y}}}, v\in\mathcal{\varepsilon}^{ns}_{{\bar{x}}\times y}$ satisfying the discrete homogeneous Dirichlet boundary condition and $f\in\mathcal{\nu}_{x\times y}$, the following formulae are valid
\bena
{[d_yf,u]}_{ew}=-(f,D_yu)_{vc},~~~{[d_xf,v]}_{ns}=-(f,D_xv)_{vc}.
\eena
\end{lem}

\end{document}